\documentclass{aa}  

\usepackage{hyperref}
\usepackage{graphicx}
\usepackage{txfonts}
\usepackage{caption}
\usepackage{subcaption}
\usepackage{booktabs}
\usepackage{multirow}
\usepackage{multicol}
\usepackage{makecell}
\newcommand{\ra}[1]{\renewcommand{\arraystretch}{#1}}
\usepackage{lscape}
\usepackage{ulem}
\usepackage[dvipsnames]{xcolor}
\usepackage{bold-extra}
\usepackage{verbatim}

\begin{document}

   \title{From VIPERS to SDSS: Unveiling galaxy spectra evolution over 9 Gyr through unsupervised machine-learning}
\author{J. Dubois\inst{1}
   	\and
   	M. Siudek\inst{2}
   	\and
   	D. Fraix-Burnet\inst{1}
   	\and
   	J. Moultaka\inst{3}
   }
   
   \institute{Univ. Grenoble Alpes, CNRS, IPAG, Grenoble, France \\ \email{Julien Dubois julienhm.dubois@gmail.com, didier.fraix-burnet@univ-grenoble-alpes.fr}
   \and Institute of Space Sciences (ICE, CSIC), Campus UAB, Carrerde Can Magrans, s/n, 08193 Barcelona, Spain \email{siudek@ice.csic.es}
   	\and   IRAP, Université de Toulouse, CNRS, CNES, UPS, 14, avenue Edouard Belin, F-31400 Toulouse, France\\ \email{jihane.moultaka@irap.omp.eu}\\
   }

   \date{Received month day, year; accepted month day, year}

 
  \abstract
    {} 
     {This study aims to trace the chronological evolution of galaxy spectra over cosmic time. Focusing on the VIPERS dataset, we seek to understand the diverse population of galaxies within narrow redshift bins, comparing our findings with the previously mapped diversity of SDSS galaxies.}
      {We use Fisher-EM, an unsupervised subspace model-based classification algorithm to classify a dataset of $79~224$ galaxy spectra from the VIMOS Public Extragalactic Redshift Survey (VIPERS). The dataset was divided into 26 samples by bins of redshift ranging from $z \sim 0.4$ to $z \sim 1.2$, which were classified independently.
    Classes of subsequent bins were linked through the k-Nearest Neighbour method to create a chronological tree of classes at different epochs.}
    {Based on the optical spectra, three main chronological galaxy branches have emerged: (i) red passive, ((ii) blue star-forming, and (iii) very blue, possibly associated with AGN activity. Each of the branches differentiates into sub-branches discriminating finer properties such as D4000 break, colour, star-formation rate, and stellar masses and/or disappear with cosmic time. Notably, these classes align remarkably well with the branches identified in a previous SDSS analysis, indicating a robust and consistent classification across datasets. The chronological "tree" constructed from VIPERS data provides valuable insights into the temporal evolution of these spectral classes.}
    {The synergy between VIPERS and SDSS datasets enhances our understanding of the evolutionary pathways of galaxy spectra. The remarkable correspondence between independently derived branches in both datasets underscores the reliability of our unsupervised machine-learning approach. The three sub-trees show complex branching structures highlighting different physical and evolutionary behaviours. This study contributes to the broader comprehension of galaxy evolution by providing a chronologically organised framework for interpreting optical spectra within specific redshift ranges.}
 

   \maketitle
%

\section{Introduction}
The vastness of the Universe and the sheer number of galaxies it contains makes it necessary to group those showing similar characteristics and study the groups' properties as a whole rather than each object individually. Such an approach is also encouraged in the modern context by the skyrocketing data collection capacity, which will reach unprecedented highs in the coming years with missions such as EUCLID \citep{racca_euclid_2016} or LSST \citep{lsst_science_collaboration_science-driven_2017}. As more and more data is being collected, automated classification procedures are becoming a necessity, whether it be as a mean to study galaxies, or on a more practical aspect, as a tool to automatically and efficiently label and store new observations within large surveys.

Galaxies showcase a wide variety of shapes and structures which were initially used to propose the first classification system by \citet{hubble_extragalactic_1926}, distinguishing elliptical, lenticular, spiral and irregular galaxies. It was later refined by \citet{de_vaucouleurs_classification_1959} and \citet{sandage_hubble_1961}, introducing further discrimination. Morphological classification of galaxies is still commonly used nowadays \citep[e.g.][]{baillard_efigi_2011, Cheng2023, Huertas-Company2023} because it correlates well with certain physical properties such as the colour \citep[e.g.][]{Moutard2016May-a, Siudek2022} or stellar content \citep[e.g.][]{kennicutt_spectrophotometric_1992}. Nonetheless, such classifications remain somewhat subjective and are not directly inferred from the physical characteristics of the galaxies. In addition, morphological classifications are mostly limited to the local Universe where galaxies’ shapes and structures are well resolved and defined. 
Beyond the local Universe, only the Hubble Space Telescope and recently launched James Webb Space Telescope are able to provide images with sufficient spatial resolution to resolve the structure of small high-redshift galaxies~\citep[e.g.][]{Huertas-Company2015,Huertas-Company2023JWST}. 
In addition to morphology, photometric properties and spectral features have proven to be useful attributes to separate different populations of galaxies. The Baldwin-Phillips-Terlevich (BPT) diagram \citep{baldwin_classification_1981} is widely used to distinguish star-forming galaxies, Seyfert galaxies, and low-ionisation nuclear emission-line regions (LINERS) using [O III], [N II], H$\alpha$ and H$\beta$ line ratios. However, the application of such traditional diagnostics is limited at redshifts $z>0.4$ since optical emission lines such as H$\alpha$ and [NII] are shifted out in the near-infrared. Other diagnostic diagrams such as the Mass-Excitation diagram \citep{juneau_new_2011} have been introduced as an alternative for intermediate redshifts. Rest-frame colour magnitudes are also used to separate red passive, “green valley” galaxies and blue star-forming galaxies in colour-colour diagrams such as the (NUV-r)-(r-K) (hereafter NUVrK) introduced in \citet{arnouts_encoding_2013}. But such classification methods can hardly distinguish more than a few classes and only make use of a few numbers of parameters, which cannot realistically reflect all the diversity of galaxies in the Universe~\citep[e.g.][]{Siudek2018Sep, Siudek2022proc}.
   
An alternative approach is to use raw spectra, which encompass a great amount of information related to the intrinsic physical characteristics of galaxies, as the basis for classification. \citet{kennicutt_spectrophotometric_1992} builds an atlas of galaxy spectra, but does not introduce a new classification scheme per se and uses morphological types instead. In \citet{dobos_high_2012}, they build a template of mean galaxy spectra, which has the advantage of increasing the signal-to-noise ratio and thus revealing structures that could be undetectable on individual spectra. However, they do not build their classification off of the raw spectra, and instead use a colour, morphology, and nuclear activity classification from the work of \citet{lee_nature_2008}, and extend it to further separate spectral lines strengths. Similarly, \citet{wang_spectral_2018} propose an atlas of composite spectra of galaxies, but base their classification on spectral features and diagnostic diagrams. In fact, basing a classification off of raw spectra is a challenging task due to a large number of parameters (i.e. the monochromatic fluxes), which requires the use of advanced machine-learning techniques and statistical methods specialised is performing such tasks \citep{Fraix-Burnet2015}.

Supervised classification techniques are used in a wide variety of applications in astrophysics \citep[see][for a review]{Huertas-Company2023}. However, these techniques require learning labelled databases, and therefore, a priori knowledge about the expected classes, which we do not have for galaxy spectra. Unsupervised methods, on the other hand, are more suited for this task. They are designed to look for discriminant patterns in the data and find clusters without prior knowledge and expectation. The k-means algorithm \citep{macqueen_methods_1967} is perhaps one of the most popular unsupervised classification techniques, which sorts the objects using centroids and distance minimization. \citet{sanchez_almeida_automatic_2010} used the k-means algorithm to classify galaxy spectra from the Sloan Digital Sky Survey (SDSS), but \citet{de_clustering_2016} argue that the method is not suited for high-dimensional clustering as it suffers from the curse of dimensionality \citep{bellman_dynamic_1957}. This issue was avoided in \citet{Marchetti2013} by combining a principal component analysis (PCA) for dimension reduction and the k-means algorithm to obtain a spectral classification of galaxies from the VIMOS Public Extragalactic Redshift Survey (VIPERS). However, the PCA extracts axes of maximum variance and does not necessarily form a discriminant subspace with its components \citep{Chang1983}. 
In addition, as in all unsupervised approaches, the number of clusters in the k-means algorithm must be given as an input and should be chosen according to objective criteria \citep[as in][]{de_clustering_2016}.

The Fisher-EM algorithm \citep{Bouveyron2012Jan} deals with all the issues above: it is based on a Mixture Gaussian Model (GMM), in a subspace optimised for the clustering, using an objective criterion to decide altogether the best statistical model, the latent discriminant subspace and the number of clusters.
This method was shown to be very efficient with high-dimensional data and outperforms most commonly used methods, including k-means \citep{Bouveyron2012Jan}. In astrophysics, Fisher-EM was used by \citet{siudek_vimos_2018} to classify a dataset of 50~000 galaxies and broad-line active galactic nuclei (AGNs) on the basis of 12 photometric magnitudes and spectroscopic redshift. They obtained 11 classes of galaxies with distinct properties, and an additional class of AGNs. The same approach is able to recreate the spectral diversity using only photometric information~\citep{siudek_vimos_2018} and is capable of revealing populations hardly distinguishable based on the standard methods~\citep{Lisiecki2023, Siudek2023rednuggets, Siudek2023AGN}or study synergies between galaxy populations between two cosmic epochs~\citep{Turner2021}. Fisher-EM revealed its power in high-dimension on the spectral classification of a dataset of 700~000 local galaxies from the SDSS \citep{Fraix-Burnet2021} leading to 86 classes with little dispersion in most classes and the identification of weird/noisy/misclassified spectra in some specific classes. As shown in \citet{dubois_unsupervised_2022} using simulated galaxy spectra, Fisher-EM is capable of retrieving classes of galaxies sharing similar physical properties on the sole basis of the spectra and in a fully data-driven manner.

The present work aims at extending the classification of galaxy spectra of \citet{Fraix-Burnet2021} to higher redshifts using the VIPERS database, similarly to \citet{Marchetti2013} but with a larger dataset and the more sophisticated Fisher-EM algorithm.

This paper is organised as follows. In Sect.~\ref{section:data} we present the data. In Sect.~\ref{section:method}, we describe the classification algorithm Fisher-EM as well as the data preparation procedure. In Sect.~\ref{section:results}, we present the classification results and study the evolution of the classes throughout cosmic time. In Sect.~\ref{section:discussion}, we discuss the limitations of this work and finally, the summary is given in Sect.~\ref{section:conclusion}. Throughout the paper, we assume the cosmological parameters with $H_0=70$~km~s$^{-1}$, $\Omega_{M}=0.3$, $\Omega_{\Lambda}=0.7$.

\section{Data}
\label{section:data}
This study uses the public data release 2 (PDR-2) of VIPERS, a spectroscopic survey carried at the ESO 8m Very Large Telescope targetting galaxies from the CFHTLS Wide Photometric survey \citep{Scodeggio2018}. The original dataset contains 86 775 spectra of objects brighter than 22.5 mag in the z\textasciitilde 1 vicinity.  The spectra cover wavelengths from 5500A to 9500A, and were observed at a resolution of R=220 within the W1 and W4 fields, covering an area of \textasciitilde 24 deg$^2$ \citep[for further details on the survey, see][]{Garilli2014,Guzzo2014,Marchetti2017}. They were shifted back to their rest-frame and resampled using the same method as in \citet{de_clustering_2016} and \citet{Fraix-Burnet2021}, which consists in doubling the sampling rate beforehand to preserve spectral features. The large range of redshifts results in a common rest-frame wavelength range from 3900 to 4300 \AA. All spectra were normalised by their rest-frame average value between $4150$\r{A} and $4250$\r{A}, a spectral region with no prominent features and common to the entire dataset.

In addition to the spectroscopic measurements, we use the same physical parameters as used in \citet{Siudek2018Sep} for the interpretation of the results. They include spectral features (4000\r{A} break, hereafter D4000, emission-line measurements), observed magnitudes from the VIPERS Multi-Lambda Survey \citep{Moutard2016May-a}, absolute magnitudes, stellar masses and star formation rates (SFRs) derived from spectral energy distribution (SED) fitting with the code "Le Phare" \citep{Arnouts1999Dec,Ilbert2006Oct}. For further information on the SED fitting process, please refer to \citet{Moutard2016May-b}.

In this work, we exclude stars and limit the study to galaxies observed in the redshift range $0.4<z<1.2$. We also exclude galaxies whose spectroscopic redshift measurements are not secure, i.e. obtained from a single or weak features for which the photometric redshift is absent or incompatible\footnote{Flags 1.1, 1.2, 9.1, and 9.2. See \url{http://vipers.inaf.it/data/pdr2/catalogs/PDR2_SPECTRO_TABLES.html}} \citep{Scodeggio2018}. The remaining galaxies have a spectroscopic redshift confidence level of either 99\% or \textasciitilde 90\% but within at least $2\sigma$ of the photometric redshift measurement. The resulting subset contains 79224 galaxies with a median redshift of 0.7.

\section{Method}
\label{section:method}

\subsection{Masks and redshift bins}
\label{section:method:masks}

The VIPERS observations are ground-based, and the raw spectra are therefore contaminated by skylines. The VIPERS-DR2 provides cleaned spectra, from which skylines have been subtracted. These cleaned spectra are masked and reconstructed in regions where the sky residuals are above an empirically chosen threshold \citep{Marchetti2017}. The spectra reconstruction is done using a principal component analysis (PCA) and is not recommended for scientific analysis as it is not capable of accurately reconstructing spectral features other than continuum. We also note that the reconstructed chunks visually stand out from the rest of the spectra by their lack of noise, as well as the sporadic presence of some peculiar features. Such particularities could be interpreted by the classification algorithm as discriminative features, although they are non-physical and irrelevant to the study. To prevent this, we decided to clip these masked regions off in the clustering analysis. However, the classification algorithm requires that all spectra must be sampled identically, meaning that if one spectrum has a masked portion, this mask has to be transposed to all other spectra too. To minimise the share of masked wavelengths, it is thus necessary to split the dataset into smaller samples\footnote{In this paper, \textit{dataset} refers to all the VIPERS spectra, and \textit{sample} refers to the spectra of a single bin of redshift.} defined by bins of redshift. This splitting is also made necessary to increase the overlap between spectra (see Sect.~\ref{section:data}), and has the interesting benefit of allowing an epoch-by-epoch analysis and study of how galaxies change over cosmic time. 

We chose to define the redshift bins such that the cosmic-time step between two consecutive bins is constant. The smaller the time step is, the less information is lost from masking and de-redshifting. But decreasing the time step also decreases the number of galaxies in each bin. As such, a compromise must be found between conserving as much information as possible and having large enough bins to sample properly the diversity and proportion of the populations of galaxies. The latter is difficult to assess without a priori knowledge of the expected populations of galaxies. However, for each bin the sample size is also constrained from an algorithmic standpoint. In fact, most generative clustering methods require the sample size $n$ to be greater than the number of sampled wavelength $p$ \citep[for an overview of this problem, see Chap.~18 of][]{Hastie}. While Fisher-EM can overcome that limit, we chose to use it as a lower bound for the sample size. Under the constraints of minimising the constant time-step between successive bins and having $n>p$ for each sample, the optimum was found to be 26 bins separated by a time-step of 163 Myr (Table~\ref{table:bins}). Despite our efforts to limit the loss of information due to masking, the mask-rate remains close to 50\%, meaning that roughly half of the spectra are masked before being classified. However, we show in Sect.~\ref{section:results:classifications} that our method manages to consistently overcome this limitation. Because the VIPERS dataset is not uniformly distributed in redshift, the sample size varies by one order of magnitude between the smallest (bin 1 with 491 galaxies) and biggest bin (bin 11 with 5504 galaxies).

\subsection{The Fisher-EM algorithm}
\label{section:method:FEM}

We use the unsupervised clustering algorithm Fisher-EM \citep{Bouveyron2012Jan} to classify the galaxy spectra. This algorithm is particularly well suited for high-dimensional data, as it outperforms most common unsupervised classifiers that suffer from the curse of dimensionality. Fisher-EM models the data with a Gaussian mixture model (GMM), which is designed on the assumption that the data can be represented as resulting from a total probability density function $f$ that can be written as the weighted sum of $K$ multivariate Gaussian probability density functions $\Phi$ (eq.~\ref{eq:GMM}). 
\begin{equation}
    f(K, \theta) = \sum_{i=1}^K \pi_i \Phi(\mu_i, \Sigma_i)
    \label{eq:GMM}
\end{equation}
with $\theta$ the parameters to be estimated, i.e. for each cluster $i$: its weight $\pi_i$, its mean $\mu_i$ and its covariance matrix $\Sigma_i$. 

The GMM parameters are commonly estimated using the expectation-maximisation (EM) algorithm, but it does not perform well with high dimensional data. Fisher-EM offers a workaround to this issue by providing a dimension reduction step to the regular EM algorithm, where the data is projected onto a discriminative latent subspace of dimension $d=K-1$. It assumes the existence of $d$ latent variables in which most of the discriminative information is encoded, such that the observed data Y and latent data X are linked by a linear transformation (eq.~\ref{eq:projection}).
\begin{equation}
    Y = UX + \epsilon
    \label{eq:projection}
\end{equation}
with $U$ the projection matrix, and $\epsilon$ is a noise term assumed to be Gaussian and centred around zero. This approach not only reduces the dimension by a great amount (two orders of magnitude in this study, $p \approx 10^3$ and $d \approx 10$) but also increases the separability of the classes much more effectively than a regular PCA. This is achieved by optimising the projection matrix $U$ such that it maximizes the ratio of between-class variance to within-class variance, thus making the clusters compact and distinct from each other.

As such, Fisher-EM combines GMM and dimension reduction into a single model that is called a discriminative latent mixture (DLM) model. Its parameters are estimated with a 3-step loop: (i) in the E-step, the membership probabilities of all the observations are computed given the current estimation of $\theta$; (ii) the F-step computes the projection matrix $U$ of the discriminative latent subspace; and (iii) the M-step updates the estimation of the model parameters $\theta$ given the new membership probabilities computed in the E-step and the new projection matrix $U$ computed in the F-step.

\subsection{Model selection}

There are 12 variations of the DLM model, which are obtained by adding more or less constraints on the degrees of freedom of the model: AB, ABk, AjB, AjBk, AkB, AkBk, AkjB, AkjBk, DB, DBk, DkB, DkBk. Each model has its own specificities and may fit better certain datasets. For example, the DkBk model is the most complex, with no constraints on the DLM parameters. The DkB model assumes the noise is identical for each cluster, and the DBk model assumes that the covariance matrix is the same across all clusters. The combination of both previous assumptions results in the DB model. Similarly, assuming that the covariance matrix is diagonal leads to the AkjBk model, and adding an isotropic condition on the covariance matrix leads to the AkBk model. A detailed mathematical description of all DLMs can be found in \citet{Bouveyron2012Jan}.

The free parameters (i.e. the DLM model, its parameters $\theta$, and the number of clusters $K$) are optimised by maximising the integrated completed likelihood (ICL), a penalized variation of the regular likelihood suited for high dimensional clustering \citep{Biernacki2000Jul}. We find that the best forming DLM model on the VIPERS spectra for all samples is DkBk, and we therefore use it for all the classifications in this study. We note that this model is the most complex available in Fisher-EM, and offers a cluster-dependent modelling  of the variance and noise. On the contrary, the optimal number of clusters differs from one sample to another.

\subsection{Classification procedure}
\label{section:method:masks:classification}
Each sample was individually classified using Fisher-EM. For all the bins, the number of classes (that we will call the main classes) was rather low, with significant dispersion, so that we performed a subclassification to increase the discrimination between spectra. 
This is justified by the fact that the classes are found in a discriminative latent subspace specifically optimised for the sample under study. The larger and more diverse the latter is, the harder it is to find a projection matrix that takes into account all the discriminative features of the spectra. Thus, it only separates the populations of spectra with the most obvious differences within the sample \citep[see e.g.][]{Fraix-Burnet2021}. With subclassification, a more specific latent subspace is found for each main class, and a much more refined classification is made possible. We found that it was not necessary to subclassify all the main classes, as some of them were smaller with lesser dispersion. As such, only the classes larger than $5\%$ of the sample size were subclassified.

\section{Results}
\label{section:results}
\subsection{Classification of the samples}
\label{section:results:classifications}

\begin{table*}
    \centering
    \ra{1.3}
    \caption{The characteristics of the subsamples: their range of redshift and epoch, their rest-frame spectral range, the masking rate, sample size, and number of classes yielded. The last line gives the fraction of classes in each bin for which the  k-NN votes are larger than 0.5 (see Sect.~\ref{section:results:tree} for details).}
    \begin{tabular}{l llllllllllllllllllllllllll}
    \hline\hline
    Bin index   & 1 & 2 & 3 & 4 & 5 & 6 & 7 & 8 & 9 & 10 & 11 & 12 & 13\\
    \hline
    Lower redshift    & 0.40 & 0.42 & 0.44 & 0.46 & 0.48 & 0.51 & 0.53 & 0.55 & 0.58 & 0.60 & 0.63 & 0.65 & 0.68 \\
    Upper redshift    & 0.42 & 0.44 & 0.46 & 0.48 & 0.51 & 0.53 & 0.55 & 0.58 & 0.60 & 0.63 & 0.65 & 0.68 & 0.71 \\
    Lower Epoch (Gyr) & 9.24 & 9.07 & 8.91 & 8.75 & 8.59 & 8.43 & 8.26 & 8.10 & 7.94 & 7.77 & 7.61 & 7.44 & 7.28 \\
    Upper Epoch (Gyr) & 9.40 & 9.24 & 9.07 & 8.91 & 8.75 & 8.59 & 8.43 & 8.26 & 8.10 & 7.94 & 7.77 & 7.61 & 7.44 \\
    Lower wavelength ($\mathring{A}$) & 3939 & 3884 & 3829 & 3774 & 3718 & 3663 & 3608 & 3553 & 3498 & 3443 & 3388 & 3333 & 3278\\
    Upper wavelength ($\mathring{A}$) & 6680 & 6585 & 6490 & 6395 & 6301 & 6206 & 6111 & 6016 & 5922 & 5827 & 5732 & 5637 & 5543\\
    Masking rate (\%) & 42 & 41 & 52 & 42 & 45 & 47 & 43 & 49 & 55 & 44 & 53 & 44 & 45 \\
    Sample size & 491 & 1134 & 1579 & 2084 & 2682 & 3444 & 3119 & 3754 & 3996 & 4978 & 5504 & 4763 & 4984 \\
    Number of classes & 20 & 28 & 30 & 8 & 27 & 31 & 33 & 34 & 27 & 29 & 40 & 40 & 40 \\
    k-NN vote fraction & 0.7 & 0.9 & 1 & 0.72 & 0.74 & 0.85 & 0.92 & 0.83 & 0.86 & 0.87 & 0.74 & 0.71 & 0.75 \vspace{0.2cm}\\
    \hline\hline
    Bin index   & 14 & 15 & 16 & 17 & 18 & 19 & 20 & 21 & 22 & 23 & 24 & 25 & 26\\
    \hline
    Lower redshift    & 0.71 & 0.74 & 0.77 & 0.80 & 0.84 & 0.87 & 0.91 & 0.94 & 0.98 & 1.02 & 1.06 & 1.11 & 1.15  \\
    Upper redshift    & 0.74 & 0.77 & 0.80 & 0.84 & 0.87 & 0.91 & 0.94 & 0.98 & 1.02 & 1.06 & 1.11 & 1.15 & 1.2 \\
    Lower Epoch (Gyr) & 7.12 & 6.95 & 6.79 & 6.62 & 6.46 & 6.30 & 6.14 & 5.97 & 5.81 & 5.65 & 5.49 & 5.33 & 5.17  \\
    Upper Epoch (Gyr) & 7.28 & 7.12 & 6.95 & 6.79 & 6.62 & 6.46 & 6.30 & 6.14 & 5.97 & 5.81 & 5.65 & 5.49 & 5.33  \\
    Lower wavelength ($\mathring{A}$) & 3223 & 3168 & 3112 & 3057 & 3002 & 2947 & 2892 & 2837 & 2782 & 2727 & 2672 & 2617 & 2562\\
    Upper wavelength ($\mathring{A}$) & 5448 & 5353 & 5258 & 5164 & 5069 & 4974 & 4879 & 4785 & 4690 & 4595 & 4500 & 4406 & 4311\\
    Masking rate (\%) & 51 & 51 & 54 & 52 & 52 & 53 & 57 & 47 & 47 & 51 & 48 & 59 & 55\\
    Sample size & 4277 & 5258 & 4466 & 3947 & 3546 & 4088 & 3217 & 2346 & 1775 & 1343 & 1233 & 663 & 553 \\
    Number of classes & 35 & 43 & 38 & 28 & 30 & 26 & 26 & 27 & 25 & 27 & 21 & 20 & 21\\
    k-NN vote fraction & 0.9 & 0.85 & 0.83 & 0.89 & 0.82 & 0.76 & 0.81 & 0.7 & 0.88 & 0.97 & 0.68 & 0.85 & -- \vspace{0.2cm}\\
    \hline
    \end{tabular}
    \label{table:bins}
\end{table*}

The classification procedure described in Sect.~\ref{section:method:masks:classification} yields from 20 to 43 classes per sample, with the exception of 8 classes (Table~\ref{table:bins}) for the 4th bin where the location of the masks completely covers up the H$_\gamma$ and H$_\beta$ lines, as well as partially the OIII line, hence significantly reducing the discriminative information available within the spectra of that sample.
Indeed, the most influential factor for the number of classes is found to be the sample size, with a Pearson correlation factor of $r=0.75$. The larger the sample size, the more diversity in the spectra from a statistical standpoint, leading to additional classes. 
The spectral coverage is another defining factor for the number of classes, as it dictates the characteristic features visible in the spectra. However, because of the spectral overlap of the samples as well as the intrinsic redundancy of information within spectra, it is found not to affect the number of classes significantly ($r=-0.1$). The masks, while they can be problematic when masking out important emission or absorption lines, do not affect much the classifications ($r=0.05$) indicating again that there are enough redundancies and correlations between data points.

The resulting classifications can be visualised by the mean spectra of the classes, which give us a first insight into the kind of galaxies that each class separates. 
The results for all the samples can be found in Appendix~\ref{appendix:spectra}. We here detail the first sample (bin z=0.41) and its 20 classes (Fig.~\ref{fig:vipers_classes_bin1}) for illustration. Aside from a few outlier classes, the spectra are rather well distributed among the classes, and the within-class dispersions are relatively low in most cases, indicating that the classes do, in fact, efficiently gather similar spectra. In this sample, the H$_\beta$ and OIII lines are present and play a significant discriminative role in the classification process. On a side note, computing the mean spectra greatly improves the S/N, and reveals smaller features like H$_\delta$ (see the mean spectrum C2 on the right bottom panel in Fig.~\ref{fig:vipers_classes_bin1}. Essentially three main types of classes emerge visually: red passive galaxies (e.g. classes A and D), blue galaxies with intense emission lines (e.g. class B and C), and outliers (the less populated classes characterised by spectra with low S/N or frigging e.g. class A1). Within each of these three visual categories, further segregation arises, mostly based on continuum slope and line intensities. We can note, however, that the dispersion in line intensities is generally higher than in the continuum.

The same kind of analysis performed on the 25 other samples (Appendix~\ref{appendix:spectra}) confirms that the classifications properly gather galaxies sharing similar spectral features with satisfying dispersion given the S/N of the data. In particular, the classes separate the slope of the continuum as well as the relative importance of the prominent lines and the Balmer break. However, the discriminative features gradually evolve as we progress through the bins due to the change in the rest-frame spectral window.

To go one step further in understanding the results, we can look at the distribution of various properties within the classes. Some of these properties, as described in Sect.~\ref{section:data}, are directly derived from the VIPERS spectra, e.g. lines equivalent width, or Balmer break. It can be expected that these properties could be well separated in the classes due to the information being straightforwardly available within the spectroscopic data on which the classification is based. Some other properties however originate from additional data, different spectral regions, or complex inference processes. Typically, spectral lines in galaxies are often used as diagnostics to identify the source of ionisation, hence separating AGNs from star-forming galaxies. The BPT diagram is perhaps the most commonly used AGN diagnostic tool \citep{baldwin_classification_1981}, but it makes use of the NII and H$\alpha$ lines, which are unfortunately out of scope within the VIPERS spectra. This is a common issue that affects higher redshift galaxies, and alternative diagnostic tools have therefore been developed. Among them, the Mass-Excitation \citep[MEx;][]{juneau_new_2011} diagram, which uses the OIII and H$_\beta$ lines equivalent width, and the stellar mass. This diagram was usable with the VIPERS data up to $z=0.9$, after which the OIII line is no longer available. As a result, MEx diagrams were established for 19 out of the 26 bins (Appendix~\ref{appendix:MEx}). For instance, the MEx diagram for the first bin at $z=0.41$ (Fig.~\ref{fig:vipers_MEx_bin1}) shows that the sample is mostly bimodal, and is distributed in the star-forming and intermediate region, with very little AGNs. We can see that most of the classes isolate either galaxies from one region or the other. However, most of the classes in this sample which end up in the intermediate region show no obvious sign of emission on their mean spectrum. Still, this indicates a successful segregation of the stellar mass despite the fact the spectra were normalised. 

Several other diagrams could be investigated (e.g. colour-colour, colour-magnitude) to assess the relevance of the classification from a physical standpoint, but we would remain severely limited by the fact that in total, 754 classes for the whole set of redshift bins were yielded. One of the initial motivation of classifying galaxies is to simplify the process of studying their general properties. It does not replace a thorough study of one object, but it brings a general understanding of many galaxies at once. However, so far, with a total of 754 classes, this goal is not exactly fulfilled. 
In the next section we propose another approach to synthetise our results.

\subsection{Links between classes of successive epochs}
\label{section:results:tree}

\begin{figure*}    
        \includegraphics[width=1.2\linewidth]{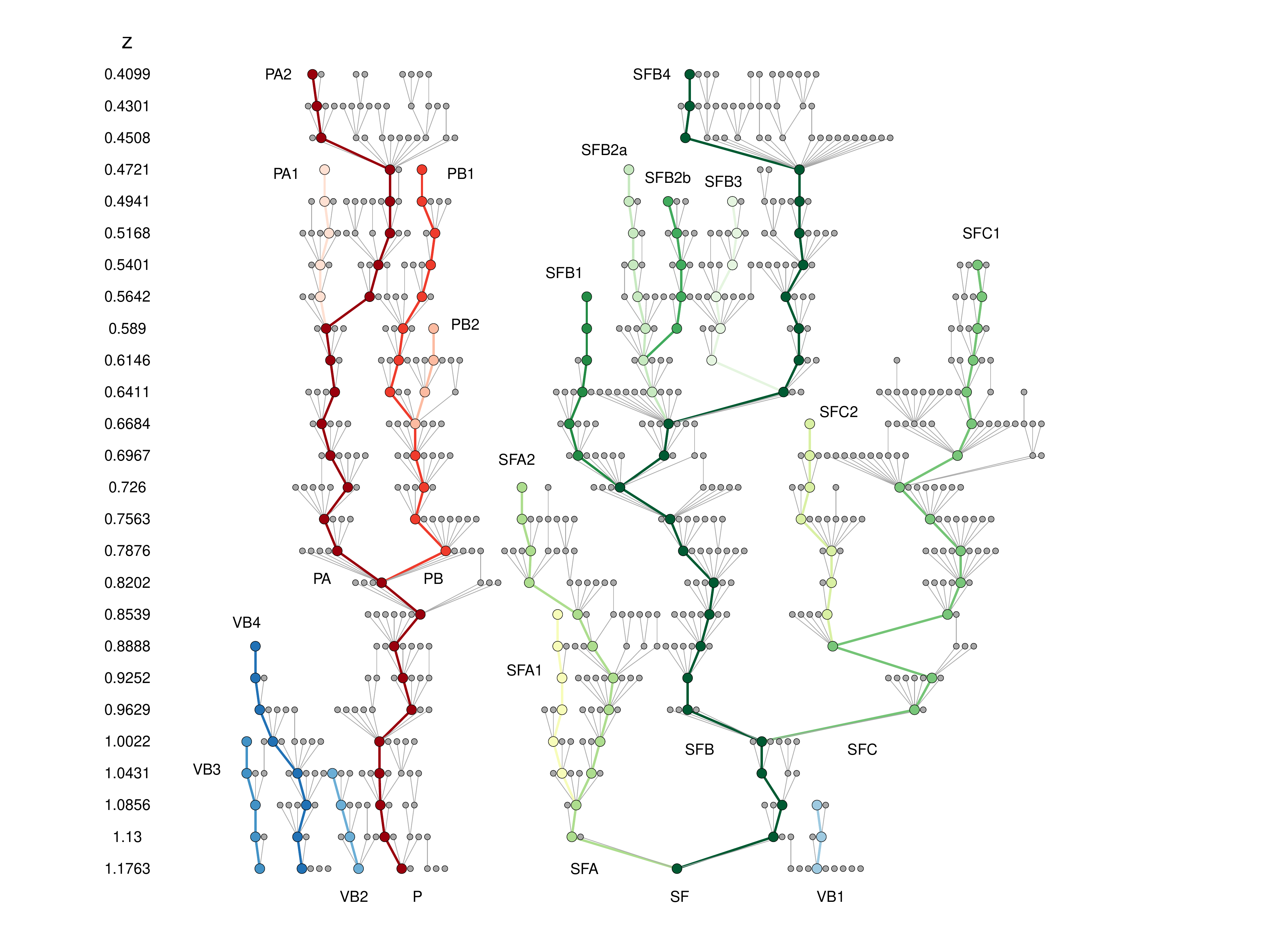}
    \caption{This tree-like structure highlights the links between the galaxy classes from a redshift
of z = 1.2 down to z = 0.4. The black line is the complete classification tree. Each vertical step in the tree corresponds to a certain epoch, linearly sampled from 4 Gyr after the Big Bang (bottom of the tree) to 9 Gyr
(top of the tree). Each node represents a class, and similarity links from epoch to epoch
were retrieved using k-NN (Sect.~\ref{section:results:tree}). The background colors highlight the structures with a common ancestor. VB, P and SF, while the colored branches are examples of evolutionary studies detailed in this paper. The nomenclature is explained in the text. }
    \label{fig:tree}
\end{figure*}

\begin{figure}
    \centering
    \includegraphics[width=\hsize]{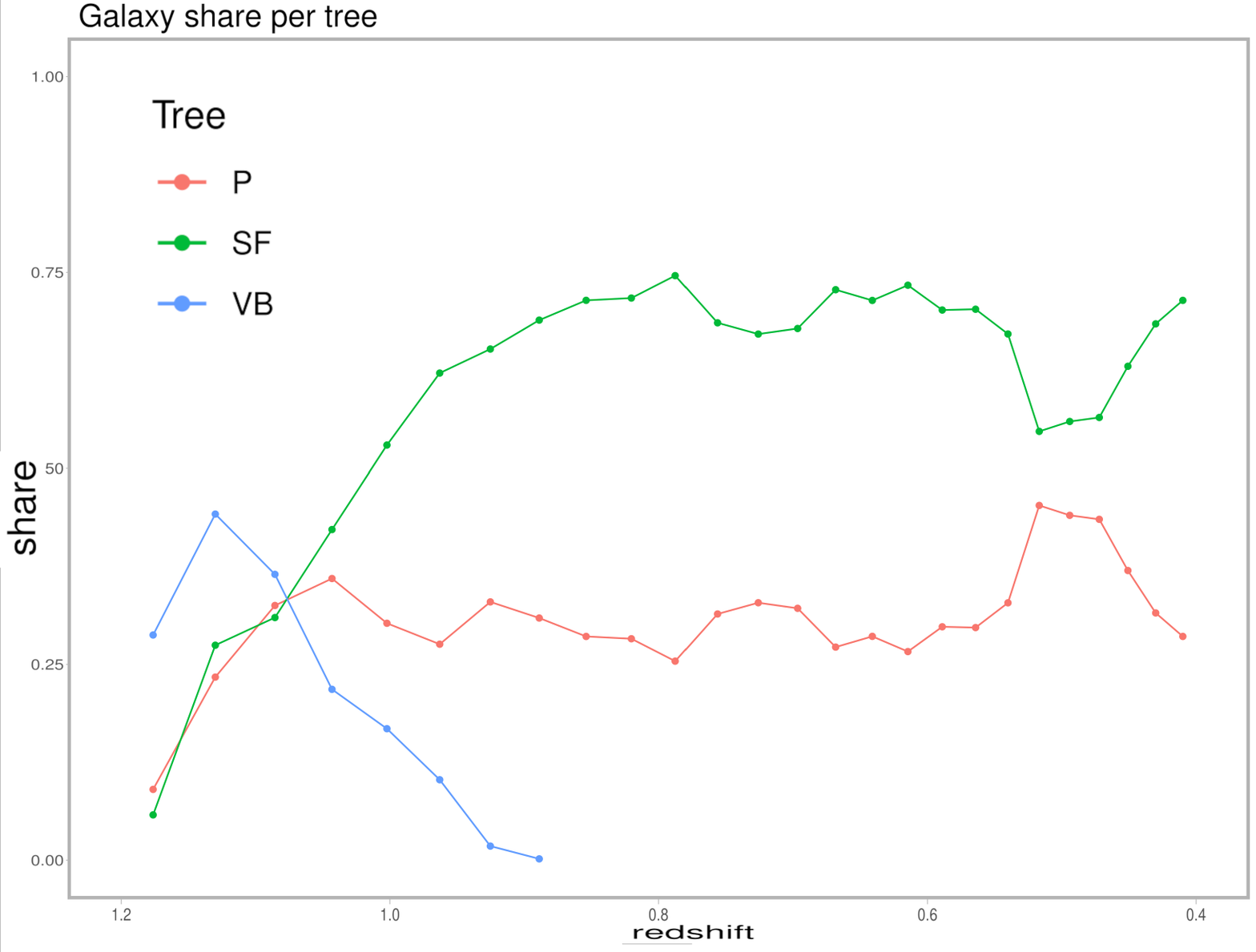}
    \caption{Share of the VIPERS galaxies among the P, SF and VB trees as a function of redshift. The value is normalised to exclude the few classes that do not belong in either of the three trees in the first two bins. }
    \label{fig:tree_share}
\end{figure}

\begin{figure*}
    \centering
    \includegraphics[width=0.99\textwidth]{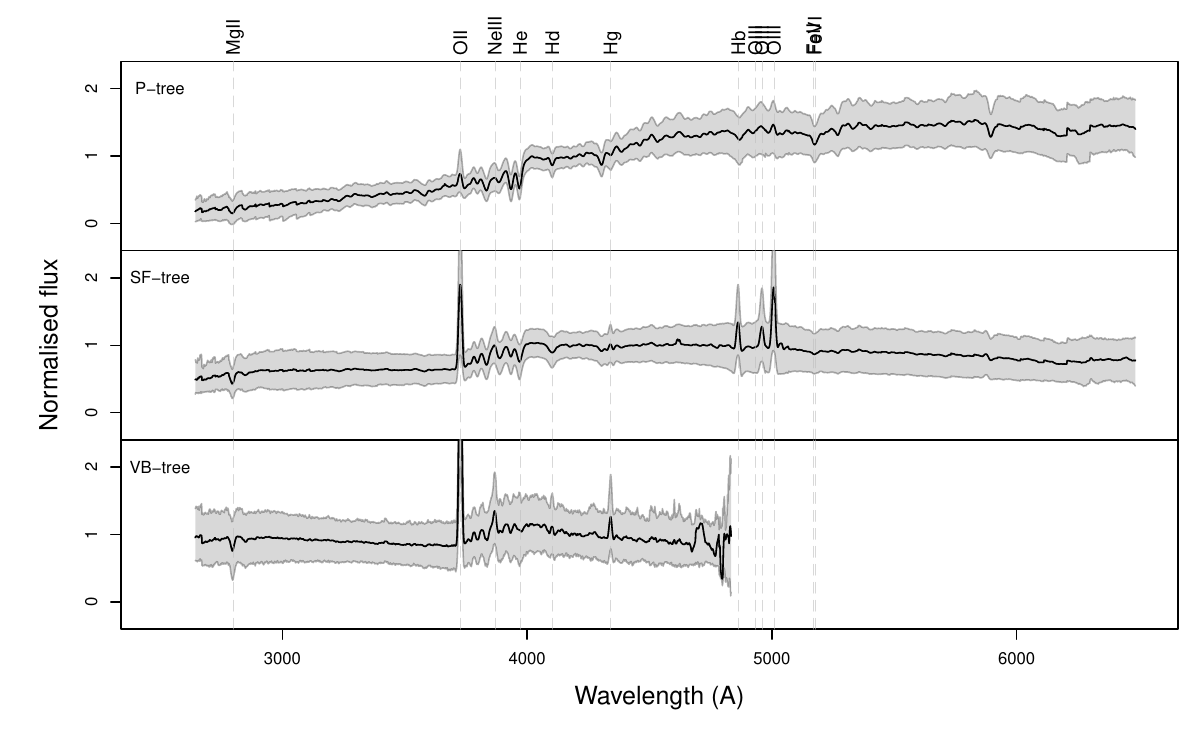}
    \caption{Mean spectra (black) of the P-tree (top), SF-tree (middle), and VB-tree (bottom). The dispersion (10\% and 90\% quantiles) are shown in gray. Some observed emission and absorption lines are highlighted with vertical dashed lines, and the corresponding source is shown at the top. Note that all three panels have the same vertical scale, and that OII line in the bottom panel is cropped out to focus on the continuum and the dispersion. The mean spectrum of the VB-tree is limited to a narrower spectral range than the two other trees, since it only includes galaxies of redshift $z>0.9$.}
    \label{fig:subtree_spectra}
\end{figure*}

\begin{figure}
    \centering
    \includegraphics[width=\linewidth]{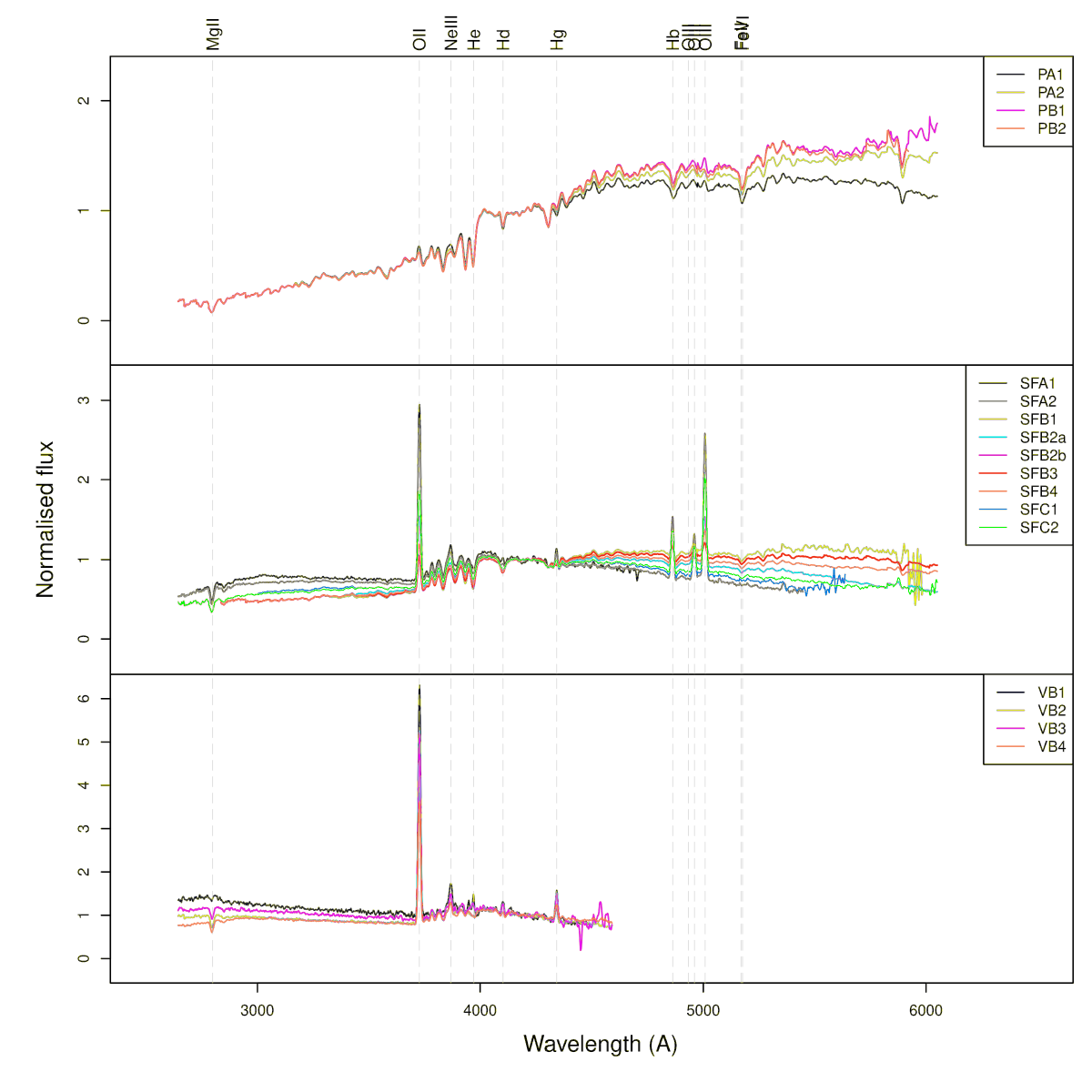}
    \caption{Mean spectra of the branches within the P-tree (top), SF-tree (middle), and VB-tree (bottom). The branch nomenclature can be visualised in Fig.~\ref{fig:tree}. The mean spectra were computed from the very top of the branch down to the common root, e.g. the mean spectrum of PA1 includes the PA and P sections. The dispersion is not shown here for the sake of clarity. Some observed emission and absorption lines are highlighted with vertical dashed lines, and the corresponding source is shown at the top. Note that the vertical scale varies from one panel to another to fully include the emission lines within the plot.}
  \label{fig:stacked_spectra}
\end{figure}

\begin{figure}
    \includegraphics[width=0.5\textwidth]{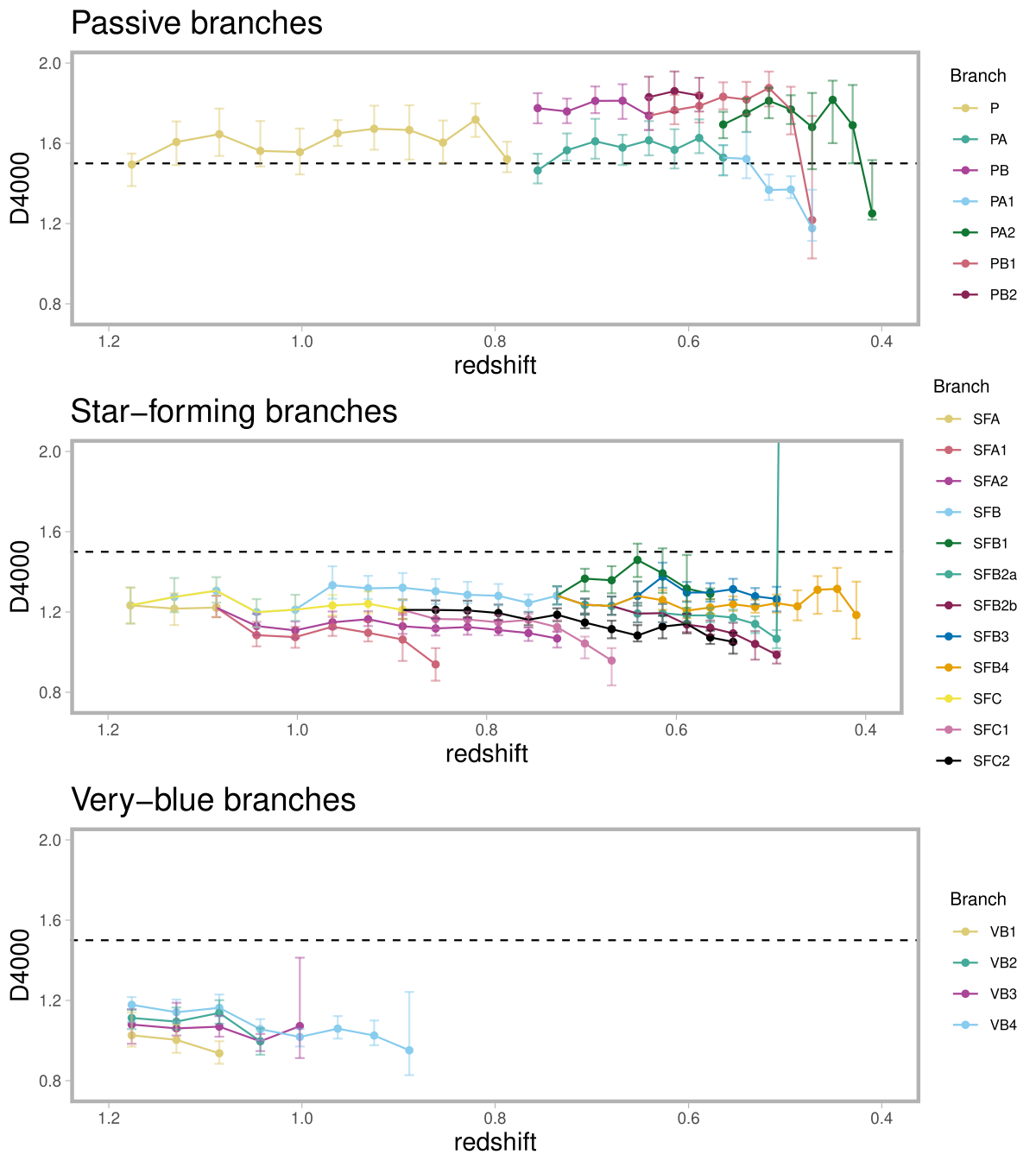}
    \caption{The distribution of the median values of D4000 among the classes in the P, SF and VB branches. The error bars show the 25\% and 75\% quantiles, and the dashed line the separation threshold between passive and star-forming galaxies from \citet{Kauffmann2003May}.}
    \label{fig:D4000_3main}
\end{figure}

\begin{figure}
    \includegraphics[width=0.5\textwidth]{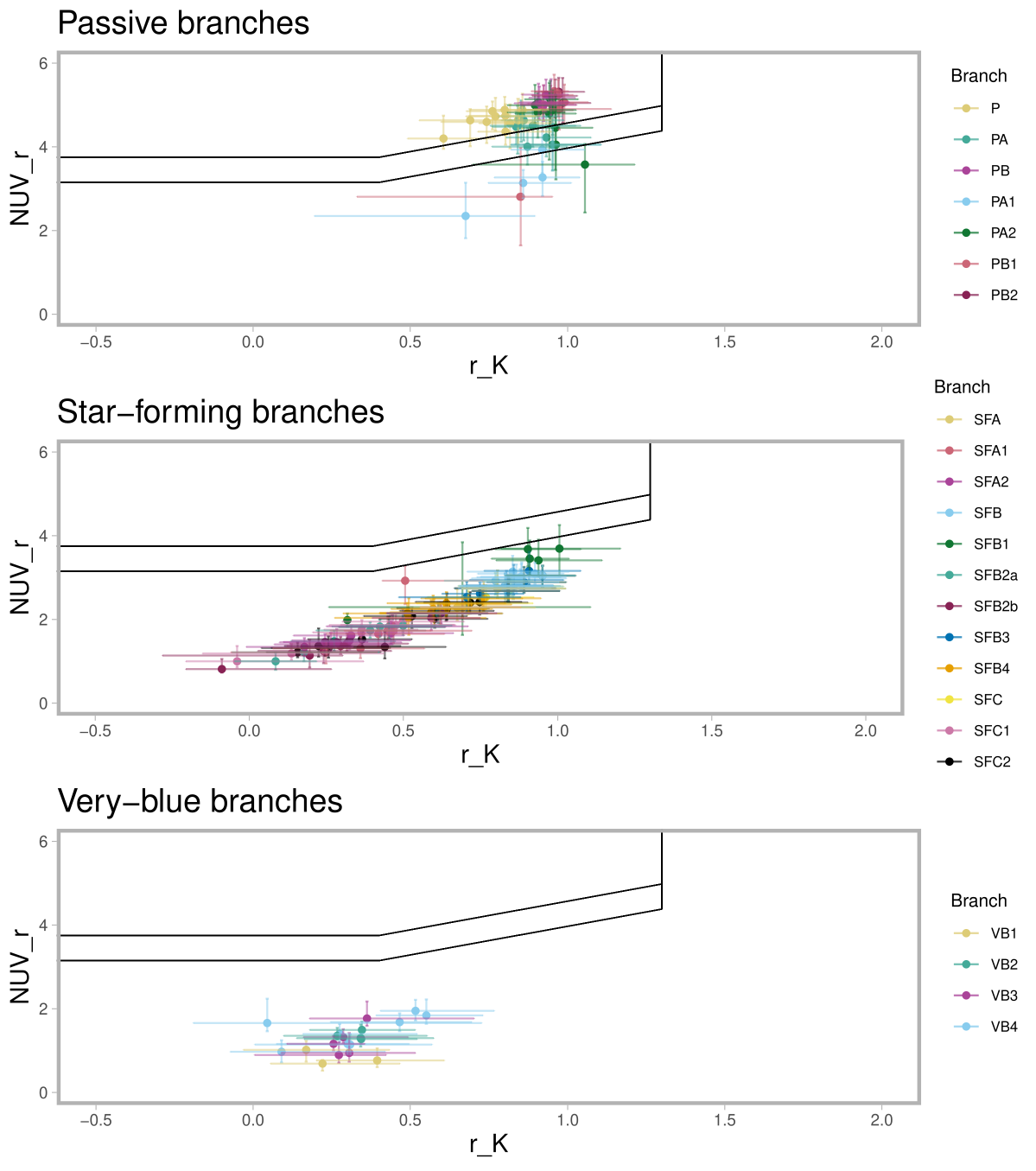}
    \caption{The distribution of the classes of P, SF and VB branches in the NUVrK diagram. According to the classification of \citet{Davidzon2016}, passive galaxies lie above the upper line, galaxies with intense star formation lie below the lower line, the intermediate zone contains galaxies having low sSFR.}
    \label{fig:NUVrK}
\end{figure}

\begin{figure}
    \includegraphics[width=0.5\textwidth]{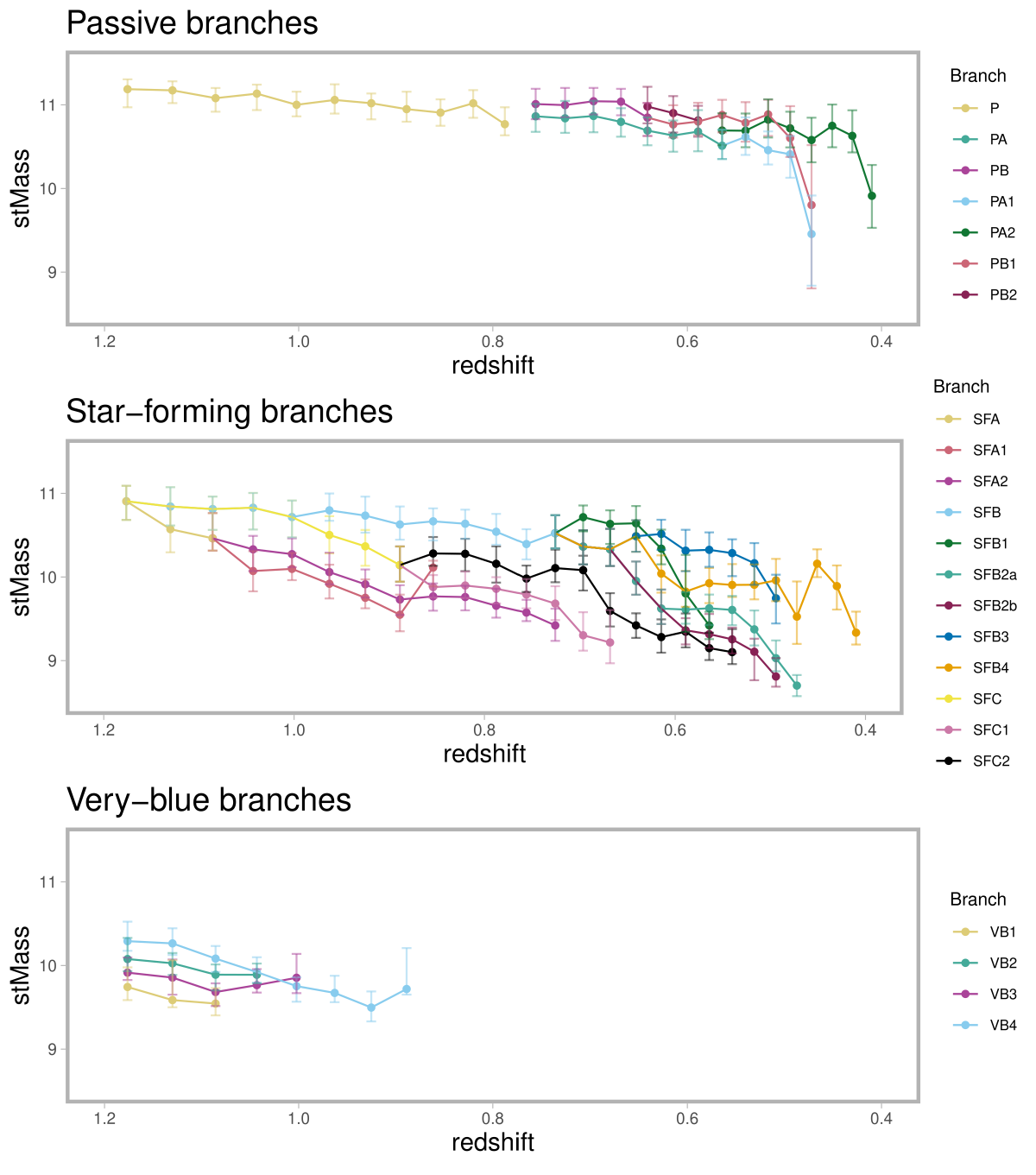}
    \caption{The distribution of the median values of stellar mass among the classes in the P, SF and VB branches. The error bars show the 25\% and 75\% quantiles.}
    \label{fig:StellarMass}
\end{figure}

Links between classes of successive epochs were constructed with the k-nearest-neighbours algorithm (k-NN) in order to follow the evolution of a given class over cosmic time, from 5 Gyr (z=1.2) to 9 Gyr (z=0.4) after the Big Bang. k-NN is a supervised classification algorithm that attributes a class to an object based on the majority vote of the $k$ closest (in terms of Euclidean distance) objects for which the class is known. Here we chose $k=\sqrt{n}$ where $n$ is the number of spectra of the corresponding sample. The Euclidean distance between two spectra $s$ and $t$ is defined as $d^2_{st}=\sum_{l=1}^p\left(f^2_{ls}-f^2_{lt}\right)$ where $p$ is the number of wavelengths in the common spectral range and $f_{ls}$ the monochromatic flux of spectrum $s$ at wavelength index $l$.
By applying this method to the average spectrum of a class $K_i$ at epoch $T_i$, and comparing it to all the spectra at epoch $T_{i-1}$ we are able to find which class $K_{i-1}$ is the closest to $K_i$. In other words, the mean spectrum of a class $K_i$ is compared to the $k$ closest spectra of the previous epoch $T_{i-1}$, and the connection is made with the class $K_{i-1}$ having the largest number of spectra within these $k$ closest ones (highest vote). Naturally there is no such value for the first bin at the highest redshift. This method implies that the class $K_i$ has one and only one precursor class, while class $K_{i-1}$ can have from zero to several successor classes. 
In this study, epoch $T_i$ corresponds to a later time than epoch $T_{i-1}$ and thus to a lower redshift. Repeating this process on subsequent bins creates a divergent tree-like structure (Fig.~\ref{fig:tree}) that highlights the change of galaxy population flow found automatically by the algorithm. 

The tree is robust in the sense that 16\% of the majority votes in k-NN are lower than 50\%, 6\% below 40\% and 2\% below 30\%. Table~\ref{table:bins} provides the fraction of classes within each bin which are connected to the previous epoch with a vote score higher than 50\%. The low-majority connections do not show any trend on the tree with respect to dead branches or redshift (hence bin size). These dead branches occur when a class at epoch $T_{i-1}$ is not associated (i.e. elected by k-NN) with any class at $T_i$. This can happen when the successor is not present in the sample (incompleteness), or not in sufficient number (for the majority vote) or because the previous association was not "correct" enough (due to noise for instance). 
In this section, we validate whether the algorithm has found the physically-driven pattern and thus the evolutionary paths of galaxy populations.

The structure visible in Fig.~\ref{fig:tree} can be divided into three independent main sub-structures which do not share any common ancestor. Our nomenclature follows the main physical properties of the galaxy populations within each branch, that are discussed below: i) the main branch of star-forming galaxies, hereafter SF-tree, ii) the right branch of passive galaxies, hereafter P-tree, and iii) a smallest branch of the bluest galaxies at the left, hereafter VB-tree.  The two main branches, SF-tree and P-tree, become more complex (i.e. the new subpopulations are emerging) when moving to lower redshift. Such a pattern could be connected to the increase in the diversity of galaxy population as time passes but this is probably mainly created by the building method of the tree (see above). Over cosmic time, galaxies can undergo various processes, including mergers, interactions, and quenching of star formation, which can be reflected by a more diverse mix of galaxy types when transiting to the lower-redshift population. Interestingly, the VB-tree does not present evolutionary sequence and disappears at z = 0.9, i.e. existing only for a ~1 Gyr at the epoch when the Universe was half of its age (but see Sects. \ref{section:VB} and \ref{section:limitations}).

The relative population of each tree in our sample, epoch by epoch (Fig.~\ref{fig:tree_share}), shows that at higher redshifts, the VB-tree is the most populated, but its size quickly diminishes down to zero at $z=0.9$. The SF-tree, on the other hand, sees its population increase at the same pace as that of the VB-tree decreases, suggesting the population of galaxies found in VB at higher redshifts are transferred to the SF-tree at lower redshifts. Finally, the share of P-tree galaxies remain quite constant around 25\%.

The mean stacked spectra computed for each of the three trees (Fig.~\ref{fig:subtree_spectra}) justify our nomenclature and the nature of the galaxy populations. The stacked spectrum of P-tree galaxies (see the top panel in Fig.~\ref{fig:subtree_spectra}) do not have any prominent emission lines, and is characterised by a redder continuum and absorption lines, typical of passive galaxies. A strong D4000\footnote{The D4000 break is a feature in the spectrum associated with the Balmer break, often used as an indicator of the stellar population age.} suggests an older stellar population, possibly indicating an ageing or evolved stellar population. The presence of old stellar population is further strengthen by the presence of the strong $\rm H\delta$, an indicator of the recent star formation history of a galaxy. The presence of NaD absorption lines at 5896\r{A} suggests the existence of cool, neutral gas along the line of sight. This could be indicative of the interstellar medium within these galaxies. The presence of Fe lines (around 4200\r{A} and 5100\r{A}) in absorption indicating a relatively high-metallicity and the other features (strong D4000, $\rm H\delta$ in absorption) suggest that P-tree population gathers galaxies that are currently quiescent in terms of the recent star formation. The presence of non-negligible OII emission may indicate some low-level, ongoing star formation, but it  does not dominated the stacked spectrum.
 
The mean spectrum of the SF-tree galaxies (see the middle panel in Fig.~\ref{fig:subtree_spectra}) contains galaxies with a rather blue continuum and pronounced emission lines. Strong OII, OIII, and $\rm H\beta$ lines indicate that these galaxies are actively forming stars.  The VB tree, although consisting of 4 independent structures, gathers very blue spectra with extremely intense emission features. Similarly to the SF-tree, these galaxies are characterised by prominent OII emission line, but we cannot compare OIII and $\rm H\beta$, which are already outside the optical windows at this redshift. However, there are visible differences between the spectra of the VB-tree in respect to SF-tree, such as: stronger OII emission, prominent and broad NeIII emission, $\rm H \gamma$ in emission. The differences can be visualised on the spectra at the highest redshift  (z=1.1763, Fig.~\ref{fig:vipers_classes_bin26}): the classes for VB1, VB2, VB3 and VB4 are respectively E4, A6, A1 and A2, while the class for SF is C6. Both stacked spectra (Fig.~\ref{fig:stacked_spectra}) show MgII in absorption suggesting the presence of cooler, lower-density gas in the interstellar medium of galaxies in VB and SF trees. The spectral differences (OIII strength and NeIII emission) could be indicative of variations in the underlying physical processes, such as the degree of star formation activity or the presence of different ionisation sources, possibly including contributions from an active galactic nucleus (AGN) in galaxies within VB-tree. The $\rm [NeIII]\lambda3869$ is present in high excitation galaxies and might by boost by non-thermal, obscured AGN~\citep[e.g.][]{Levesque2014}. The presence of VB-tree might be an important key in understanding the evolution in the ionisation conditions of the gas-phase medium between $\rm z \sim 2$ and $\rm z = 0$~\citep[e.g.][]{Steidel2014, Sanders2016, Sanders2018, Strom2017A}.

Within these three trees lie many intricate structures. In this paper we illustrate the power of the tree-like structure with a few example branches, as highlighted in Fig.~\ref{fig:tree}. Structurally, the VB-tree is the simplest and can be synthesised by 4 branches that will be referred to as VB1-4. Next, the P-tree has a more complex tree-like structure. It takes root with a common ancestor at $z=1.2$ and a single main branch (P) up until $z=0.76$, where it divides into two branches (PA and PB) which then separate again into two branches each at $z=0.54$ (PA1 and PA2) and $z=0.64$ (PB1 and PB2). Lastly, the SF-tree is significantly more complex with three main branches (SFA, SFB and SFC) emerging very early on: at z=1.13 for SFA and z=0.93 for SFB and SFC. Similarly to the P-tree, each of these branches then divides into several sub-branches. SFA is the smallest branch, with only 2 sub-branches emerging from it (SFA1 and SFA2), which appear at z=1.04 and disappear respectively at z=0.82 and z=0.70. Similarly, SFC divides into 2 sub-branches (SFC1 and SFC2) at z=0.85, which die out at z=0.64 and z=0.52. Lastly, the SFB branch divides into a larger number of sub-branches (SFB1, SFB2a, SFB2b, SFB3 and SFB4) at redshifts ranging from z=0.61 to z=0.70. The mean spectra of the all these branches (Fig.~\ref{fig:stacked_spectra}) show that the branches bring a finer degree of spectral discrimination with their own specificities.

The construction of the tree thus simplifies the classification results by connecting similar classes present at different epochs. The initial 754 classes are reduced to a total of 3 trees and 14 branches, allowing the analysis of the full dataset and the interpretation of the classification results in an evolutionary perspective.

\section{Discussion}
\label{section:discussion}

\subsection{Physical properties of the branches}
\subsubsection{Passive tree}

The spectra of the galaxies gathered in the P branches have a red continuum, a strong D4000 break and mostly no emission lines, indicative of old stellar populations. 
\citet{Kauffmann2003May} show that there is a bimodal distribution of D4000 within the SDSS spectra, which separates galaxies with old stellar populations (D4000$>1.5$) from those with more recent star formation (D4000$<1.5$). As suggested by the mean spectra, the vast majority of classes from the P-tree have D4000 values typical of galaxies populated by old stellar populations(Fig.~\ref{fig:D4000_3main}).
The first fork PA-PB at z=0.76 separates two populations of galaxies with distinct D4000 values (Fig.~\ref{fig:D4000_3main}). In PA, we find galaxies with D4000 mostly ranging from 1.5 to 1.7, while those with a 4000\AA\ break above 1.7 are found in PB. Interestingly, \citet{haines_vimos_2017} observe that many massive blue galaxies within the VIPERS dataset are being quenched around $z\approx0.7$, which is possibly what is seen in the PA branch. Further down the PA branch, two paths emerge. One of them (PA1) shows a decrease in D4000 break suggesting that this branch can gather galaxies that experienced merging processes and the inflow of the cool gas has triggered the recent star formation. The galaxies PA1, besides D4000 reaching value typical for blue star-forming galaxies, are characterised by bluer continuum and noticeable [OII] emission with still H$\delta$ absorption. In fact, the D4000 in PA1 shows a consistent decrease over several successive epochs. This branch shows characteristics of post-starburst galaxies~\citep[e.g.][]{Setton2022, Setton2023}, and could correspond to the quenched massive blue galaxies mentioned above. The other sub-branch, PA2, does not share these features and instead reaches D4000 levels comparable to the PB branch, as well as an overall redder spectrum. PA2, PB1, and PB2 are very similar, and mostly segregate different levels of ageing of the stellar populations, with PA1 being the younger, and PB2 the oldest (except for the last redshift bins). This may be explained by varying degrees of dust attenuation, but further insights are given below through the evolution of the properties along the branches. At the last stages of the life of PA2, PB1, and PA1 that occur at $\rm z\sim 0.44-0.5$ (Fig.~\ref{fig:D4000_3main}) the D4000 drops significantly suggesting that these galaxies have experienced a recent star formation activity due to the tidal stripping, merger activity or AGN activity that have reshaped their properties. However, we stress that the ends of branches typically contain only a few galaxies with a lot of dispersion, and may not really corresponding to the physical processes driving their change. 

The NUVrK colour diagram (Fig.~\ref{fig:NUVrK}) provides information relative to the star formation activity. It is in agreement with the above interpretation: (i) the classes of the P-tree are in majority located in the red region of the diagram and (ii) the PA1 branch evolved from the green region to blue region with decreasing redshift, (iii) the ends of branches PA1, PB1 and PA2 follow the conclusion of recent star formation activity by moving to bluer colours. 

The stellar mass evolution (Fig.~\ref{fig:StellarMass}) where the high redshift galaxies are associated with higher stellar mass (with $\rm log(M_{star}/M_{\odot}\sim 11)$), can be naturally explained by selection bias in a magnitude limited survey: at high redshifts only massive galaxies get targeted, and low-mass galaxies are only targeted at low redshifts. This also explains the anti-correlation between SFR and stellar mass: a higher SFR makes the galaxy appear brighter and gets targeted despite low stellar mass.

\subsubsection{Star-forming tree}

Galaxies found in the SF-tree show more complex spectral features than in the P-tree, typically found in star-forming galaxies. In addition to a much bluer continuum and a small D4000 break (see Fig.~\ref{fig:D4000_3main}), we observe several significant emission lines (OII; $\rm H\gamma$; $\rm H\beta$; OIII) in all the SF-tree, with varying intensities and ratios across the branches (see Fig.~\ref{fig:stacked_spectra}). The SFA branches are the ones with the most intense emission lines and the bluest continuum, and both SFA1 and SFA2 have very similar features. Nonetheless, SFA2 contains slightly bluer galaxies with more intense [OII] emission than SFA1. In comparison, the galaxies in the SFC branches have less strong emission features, a slightly redder continuum, and also show strong H$\delta$ absorption. We observe the same characteristics in SFBs but to a greater extreme; the emission lines get much fainter, and the spectra redder. In particular, the spectra of galaxies found in SFB1 and SFB3 resemble that of the least quiescent P-branch, PA1.

The classes of the SF-tree are, as opposed to the P-tree, characterised by a D4000 systematically smaller than 1.5 (Fig.~\ref{fig:D4000_3main}), typical of galaxies with recent star formation events. There is nonetheless a significant amount of diversity within that tree. While the uncertainty on the estimations of star formation rate are too high to observe any significant differences between the branches, the NUVrK diagram (Fig.~\ref{fig:NUVrK}) is quite informative in that regard: the SFA and SFC branches are associated to the higher end of the star formation rate, whereas the PB branches are more so located in the intermediate region. In particular, the spectra of galaxies found in SFB1 and SFB3 resemble that of the PA1 branch, which was interpreted as a class of currently quenching galaxies.

Interestingly, we observe that overall, SF galaxies see their stellar mass decrease at smaller redshifts (Fig.~\ref{fig:StellarMass}). Additionally, the branches isolate distinct ranges of stellar masses, with a clear anti-correlation between star-formation rate and stellar mass. The SFA branches, which are the most star-forming, are also the less massive, while the SFB which are the least star-forming, are the most massive. This observation can likely be attributed to 'downsizing' \citep{cowie_new_1996}. Generally, SF galaxies are also significantly less massive than the P galaxies.

\subsubsection{Very-Blue branches}
\label{section:VB}

Structurally, the VB branches are different from Ps and SFs as they do not originate from the same ancestor. They are actually 4 independent branches that span through several redshifts down to $z=0.9$, and while smaller than the P and SF trees, the VB branches still gather a significant amount of galaxies, especially at higher redshifts (Fig.~\ref{fig:tree_share}). It is to be noted that despite not sharing a common ancestor, the four VB branches show very similar characteristics.: the four branches separate different levels of blueness of the continuum (VB1 being the bluest, and VB4 the less blue) and line intensity (in the same order).

At first glance, these branches can be seen as 'extreme' versions of the SF-tree. They are characterised by a particularly blue continuum, low D4000 values (Fig.~\ref{fig:D4000_3main}), and very intense emission lines (Fig.~\ref{fig:stacked_spectra}). They are all located on the region of the NUVrK diagram associated to the highest star formation rate (Fig.~\ref{fig:NUVrK}). This interpretation might be supported by the fact that the VB branches are well integrated in the reverse tree even serving as precursors of some SF branches (Fig.~\ref{fig:reversetree}, see Sect.~\ref{section:limitations}). However, we note a significant NeIII emission, which is usually associated to nuclei activity. It is difficult to draw clear conclusions on that aspect, since MgII and NeV emissions would also be expected in stronger intensity than NeIII, but is not really observed here.

The trend between stellar mass and star formation rate observed for the P and SF galaxies also holds for the VB galaxies: they are the most star-forming, and also the less massive at a given redshift (Fig.~\ref{fig:StellarMass}). Contrary to the SF-tree, however, it is difficult to really see a trend with redshift, since the VB tree branches die out at $z\approx0.9$.

On a last note, the fact that this tree disappears at $z=0.9$ is more likely to be a consequence of the representativeness of the dataset rather than an actual evolutionary meaning. It is possible that there is somewhat of a selection bias which limits the amount of VB galaxies in the dataset at lower redshift. This also limits its correlation with AGNs as most of VIPERS AGNs are rather populating lower redshifts ($\rm z< 0.7$; see Fig. 6 in \citealt{Siudek2023AGN}). It is also probable that the disappearance of VB spectra at z=0.9 might be due to the lost connection between the VB and SF branches when the O II line is masked.

\subsection{Comparison with local galaxies}
\label{chpt:vipers-sect:discussion-sub:comparison}

\begin{figure*}
    \centering
    \includegraphics[width=0.99\textwidth]{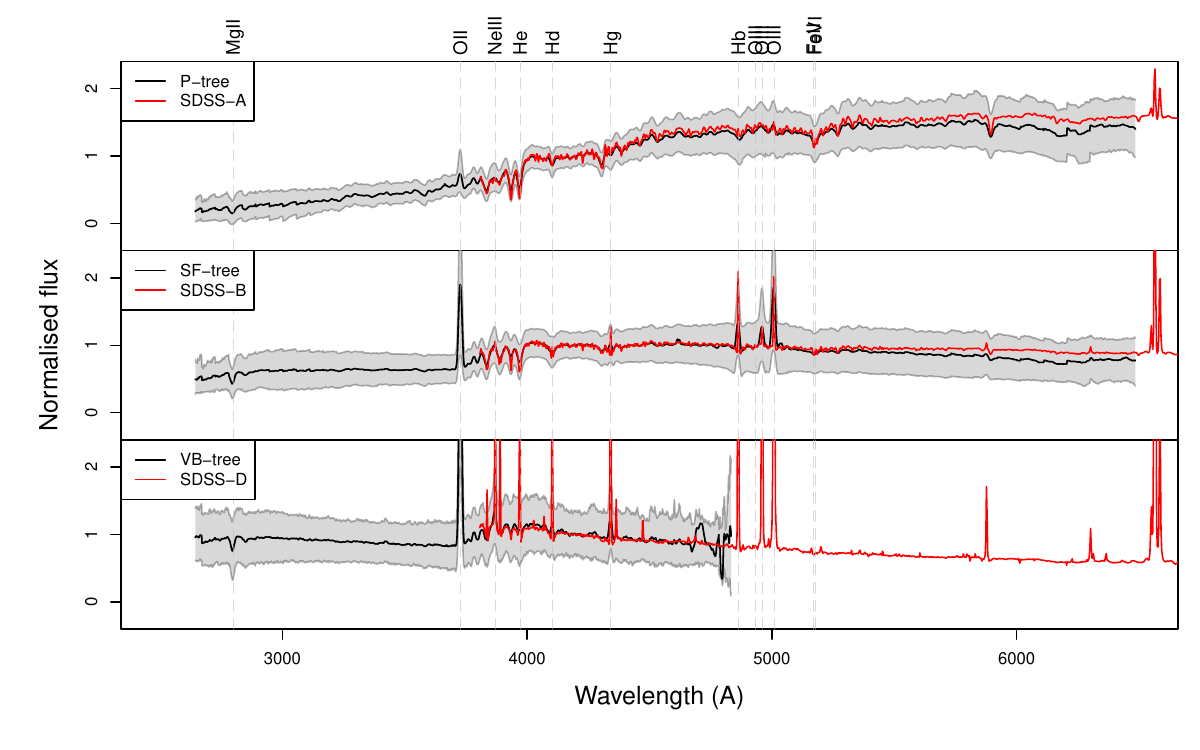}
    \caption{Same figure as in Fig.~\ref{fig:subtree_spectra} with an additional layer showing the mean spectra of the SDSS main classes A, B and D. The SDSS spectra were renormalised in the same spectral region than the VIPERS spectra to make the comparison possible.}
    \label{fig:comparison_SDSS}
\end{figure*}

\citet{Fraix-Burnet2021} classified the SDSS dataset into 4 main classes, which were then subclassified resulting in their final classification with 86 classes. These four main classes essentially consist of: (A) a class of red quiescent galaxies, (B) a class of galaxies with bluer continuum and some emission lines, (C) a diverse class with a lot of dispersion both in continuum shapes and line intensities, and (D) a small class of galaxies with a particularly blue continuum and very intense emission lines. These main classes are in a way reminiscent of the P, SF and VB categorisations that emerged from the construction of the evolutionary tree. 

After normalising the mean spectra of the 4 main SDSS classes on the same spectral region as for VIPERS, it is clear (Fig.~\ref{fig:comparison_SDSS}) that the P-tree and SF-tree match remarkably well with the SDSS-A and SDSS-B classes respectively, both from a continuum and a line intensity standpoint. The continuum slope of the VB-tree also matches perfectly with that of the SDSS-D class, although the latter has a lot more visible lines. Overall, this suggests that we are witnessing the same main categories of galaxy emerge from the analysis both in the $z$\textasciitilde$1$ vicinity and in the local Universe, similarly to what was found by~\cite{Turner2021} when comparing the evolution of galaxy populations found within SDSS and VIPERS datasets. Nonetheless, there is no equivalent of class SDSS-C, but it is difficult to draw any conclusions due to its diversity in continuum shapes. Still, it seems to be characterised by noticeable line broadening, which we do not retrieve within the VIPERS classification (the VIPERS broad-line AGN were excluded from this analysis).

This simple comparison could be improved by lowering the resolution of the SDSS mean spectra down to the VIPERS resolution (R$=220$). Additionally, the SDSS spectra appear to be generally redder, which could be explained by cosmic expansion. 
Still, the comparison portrayed here is good enough to conclude of a remarkable match between the SDSS main classes and the VIPERS sub-trees. In any case, this comparison could be refined with the SDSS subclasses and the k-NN procedure used in the present work, but this would be more rewarding by adding intermediate redshift data between SDSS and VIPERS. This is our goal for future work.

\subsection{Limitations}
\label{section:limitations}

The difficulty in the study of spectra at different redshifts is the lack of a large common wavelength range. Binning is the only option but this reduces the size of the samples for the clustering analysis. Despite the large VIPERS dataset, its large redshift range imposes a significant number of bins and thus limits the diversity of spectra within each bin. This implies that some class present at a given redshift might be absent in neighbouring ones due to the limited sample size. This could have a non-negligible impact on the tree structure.

In a given sample, subclassification allows for a finer detection of typical spectra. Indeed, subclassification can be pursued at more than two levels like it has been performed in this work. The number of levels is mainly given by a sufficiently low dispersion within classes, but there is no objective criterion apart from the scientific goal and interpretation. The level of subclassification might affect the tree structure.

The tree was built using the Euclidean distance between the whole spectra with a majority vote through the k-NN algorithm. However, 
the Euclidean distance tends to favour the continuum over narrow features because the former are more redundant.
In addition, the spectra contain several hundreds of monochromatic fluxes while the distance metrics suffer from the curse of dimensionality. 

To optimise the overlap between the mean spectra of the classes, we were obliged to compare consecutive redshift bins. This does not allow complex structures to emerge, such as branch merging, or links that skip an epoch. Making this possible, in addition to changes in the distance computation, would likely help reduce the number of dead branches, which are very numerous in the current version of the tree. Some of them may have an actual physical meaning, being that this type of galaxy is no longer observed at lower redshifts. But so far, many of the dead branches end up in small classes of outliers spectra that seem to contain calibration issues or much noise (see for instance class D in Fig.~\ref{fig:vipers_classes_bin19}, the dead end of VP4). 

All these limitations have solutions for improvement that we will investigate in future works. However, it is important to note that the tree established in this paper is not an evolutionary tree of galaxies in the sense of the phylogeny \citep[e.g.][]{Fraix2012,StatsRef2017,Fraix-Burnet2019} but rather a chronological evolution of the classes of galaxy spectra. For that one would need properties that characterise the evolutionary status of galaxies and can be given at any redshift. This might prove difficult using optical spectra only since the number of common features in the dataset tends toward zero with an increasing redshift range.

The simple approach used in this paper cannot build a converging tree. This means that we could not expect to get the exact reverse tree of Fig.~\ref{fig:tree} beginning the links from the lower redshifts as reference instead of the higher redshifts as we show in the present paper  (Fig.~\ref{fig:reversetree}). We find the interpretation of the reverse tree more difficult since, chronologically, the new “dead ends” occur from nothing. Of course this could probably better be explained by the sample incompleteness exactly as for the tree in Fig.~\ref{fig:tree}. Nevertheless, the branches identified on the tree (Fig.~\ref{fig:tree}) in this paper are rather nicely recovered on the reverse tree (Fig.~\ref{fig:reversetree}) with two main branches, one for the P and one for the SF, the VB appearing as “precursors” of SF.

\section{Conclusion}
\label{section:conclusion}

In this work, we successfully used the discriminant latent subspace Gaussian mixture model clustering algorithm Fisher-EM on a large dataset of 79~224 optical spectra of galaxies with redshift of 0.4 < z < 1.2 from the VIPERS DR2. We split it into 26 samples by bins of redshift. A classification was obtained on each individual sample, followed by a one-level subclassification, yielding in total 754 classes of spectra with 20 to 43 classes per sample except for one sample which has 8 classes only. Subsequently, links between the classes of the 26 samples were established using the k-Nearest Neighbour algorithm, resulting in a tree-like structure that highlights evolutionary paths of the classes.

Three main categories of classes are highlighted by the evolutionary tree through three sub-trees: (i) red passive galaxies, (ii) blue star forming galaxies and (iii) very blue galaxies with intense emission features. The mean spectra of these three types match remarkably well what that of the SDSS main classes A, B and D from \citet{Fraix-Burnet2021} at $z<0.25$. 

As demonstrated in a previous paper \citep{dubois_unsupervised_2022}, the classes obtained with Fisher-EM do contain physical specificities. Here, we characterise the classes according to properties such as D4000 break, colour, star formation rate, and stellar mass, and visualise their evolution through our tree. 
Indeed, the three sub-trees show complex branching structures that highlight different physical and evolutionary behaviours. For instance, different intensities of star-formation are distinguished within the SF-tree, and quenching events might be identified in a sub-branch of the P-tree. Also, while the VB-tree is indubitably associated to very high star formation rates, some spectral features could indicate the potential presence of AGNs within the classes, but it remains inconclusive due to the absence of Mg and NeV emission lines. 

The present work has innovated in two ways: firstly in the first unsupervised classification of galaxy spectra at redshifts around 1, and secondly in the construction of a tree to depict the evolution of the classes by their similarity. While the classification synthetises the diversity of spectra of galaxies, the tree summarises the diversification of galaxies through epochs.

This novel approach to galaxy classification can easily be extended to higher redshifts with future surveys. Most notably, it will be possible to use the JWST spectroscopic observations of very high redshift galaxies, hence contributing to shedding light on the question that are left surrounding the intricate mechanisms governing galaxy formation and evolution.

\begin{acknowledgements}
We warmly thank the referee for very insightful comments that helped clarify some important points.
This paper uses data from the VIMOS Public Extragalactic Redshift Survey (VIPERS). VIPERS has been performed using the ESO Very Large Telescope, under the "Large Programme" 182.A-0886. The participating institutions and funding agencies are listed at http://vipers.inaf.it. We are grateful to Thibaud Moutard for allowing us to use some unpublished data from the VIPERS survey.
This work has been supported by the Polish National Agency for Academic Exchange (Bekker grant BPN/BEK/2021/1/00298/DEC/1), the European Union's Horizon 2020 Research and Innovation programme under the Maria Sklodowska-Curie grant agreement (No. 754510).
\end{acknowledgements}

\bibliographystyle{aa}
\bibliography{bibliography}
%
%

\begin{appendix}
    \section{Mean spectra of all bin classifications}
    \label{appendix:spectra}

\newpage 

\begin{figure}
    \centering
    \includegraphics[width=\hsize]{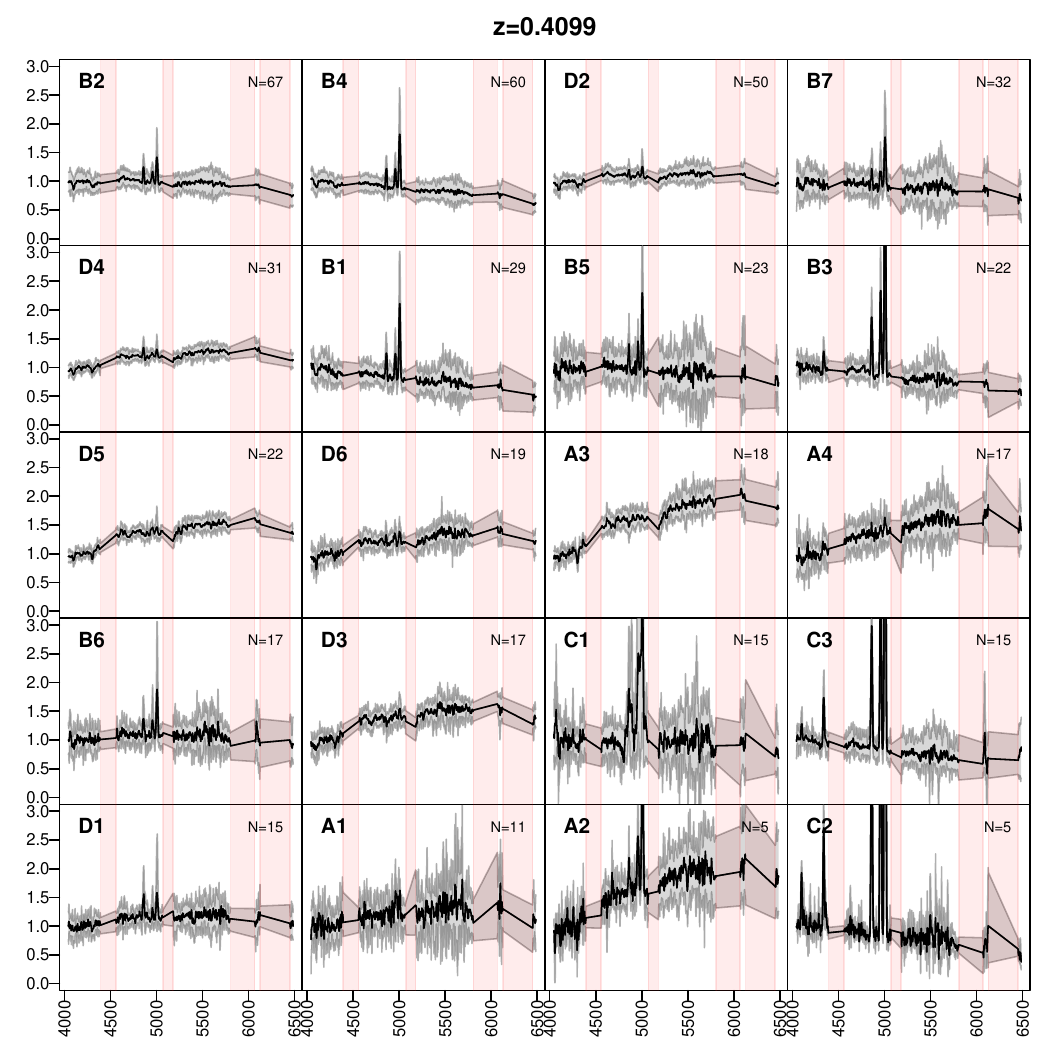}
    \caption{Mean spectra (black) of the classes obtained on the first bin, at $z=0.41$. The dispersion (10\% and 90\% quantiles) are shown in grey, and the masks in red. Each panel shows one class, whose label is displayed in the top left corner and size in the top right one. The letter in the class label corresponds to the first run of the algorithm on this bin, and the number to the subclassification of these main classes.}
    \label{fig:vipers_classes_bin1}
\end{figure}

\begin{figure}
    \centering
    \includegraphics[width=\hsize]{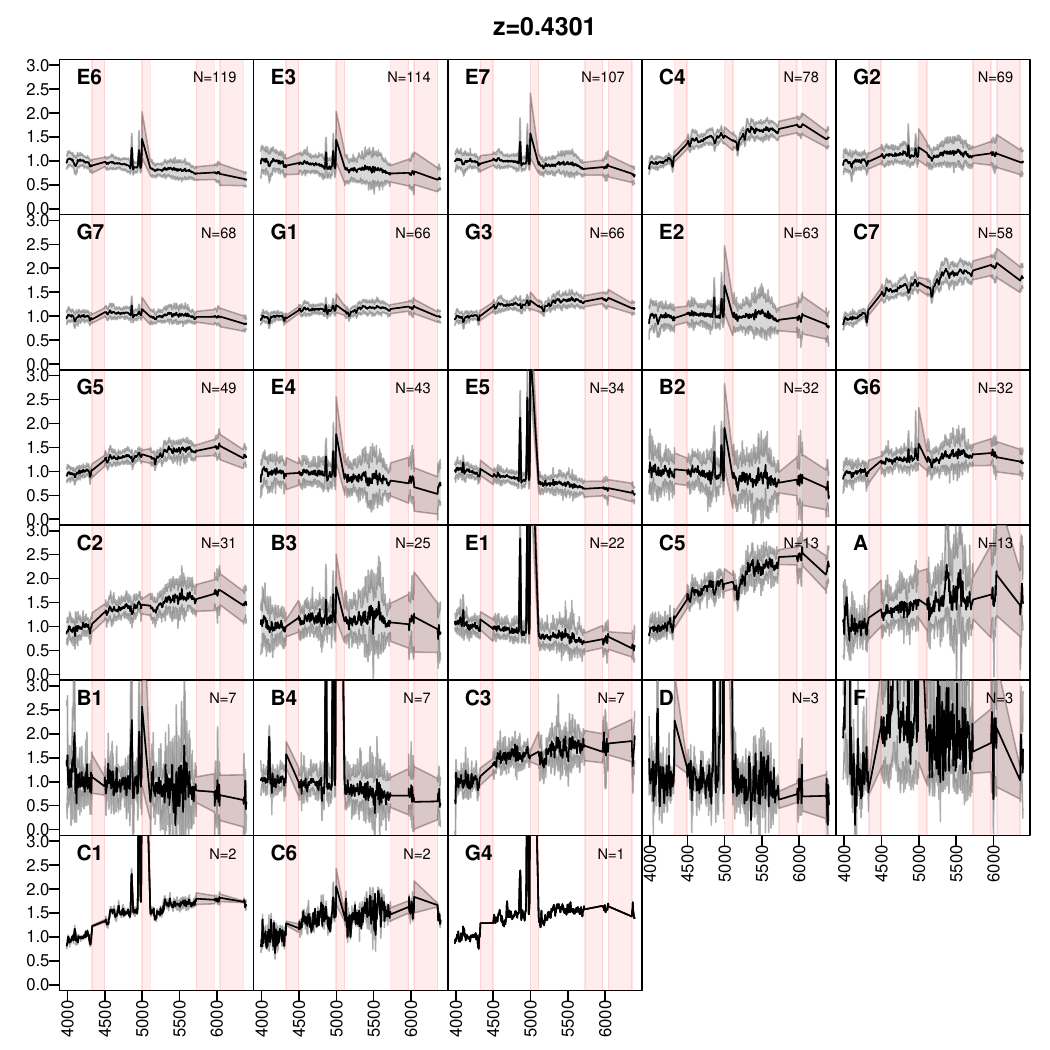}
    \caption{Stacked spectra of the classes of bin 2 (see Fig.~\ref{fig:vipers_classes_bin1} for further information)}
    \label{fig:vipers_classes_bin2}
\end{figure}

\begin{figure}
    \centering
    \includegraphics[width=\hsize]{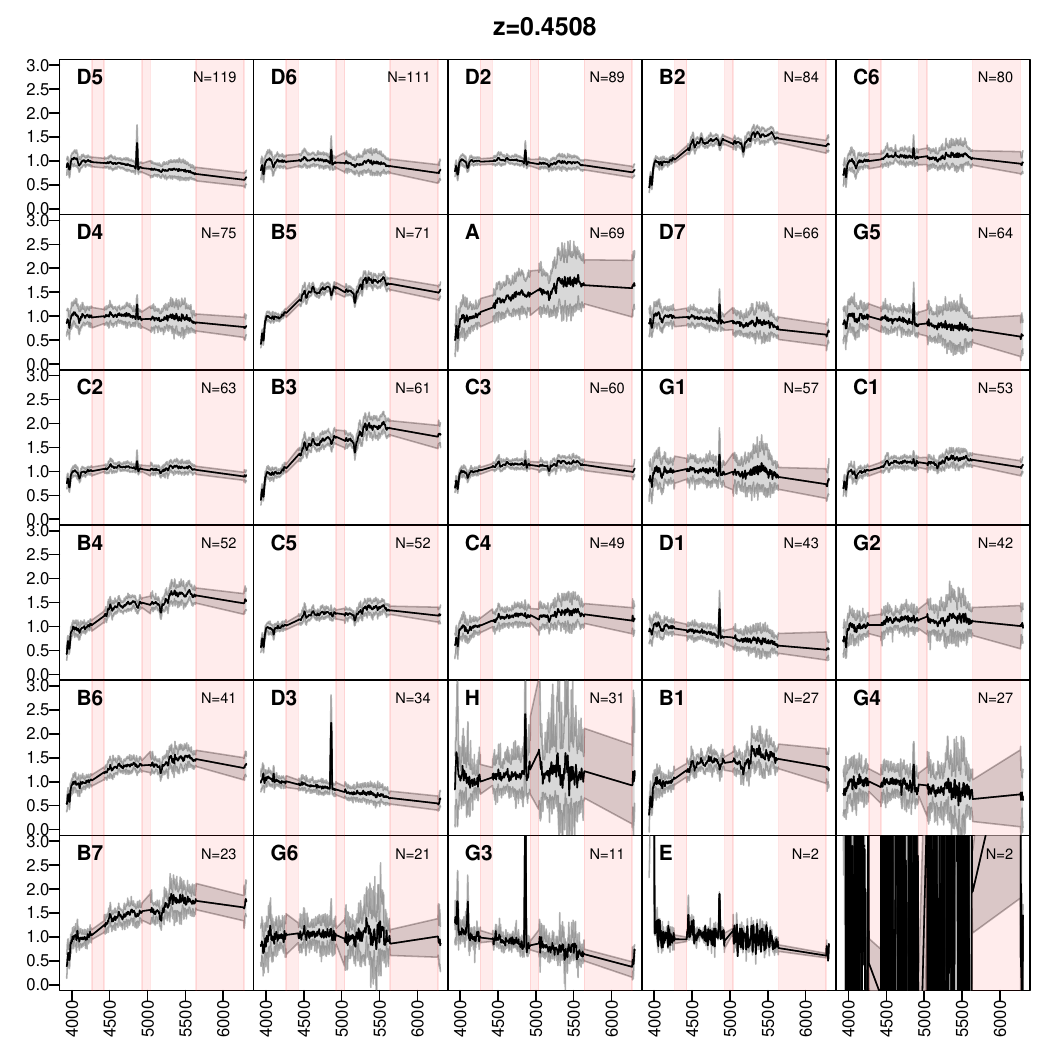}
    \caption{Stacked spectra of the classes of bin 3 (see Fig.~\ref{fig:vipers_classes_bin1} for further information)}
    \label{fig:vipers_classes_bin3}
\end{figure}

\begin{figure}
    \centering
    \includegraphics[width=\hsize]{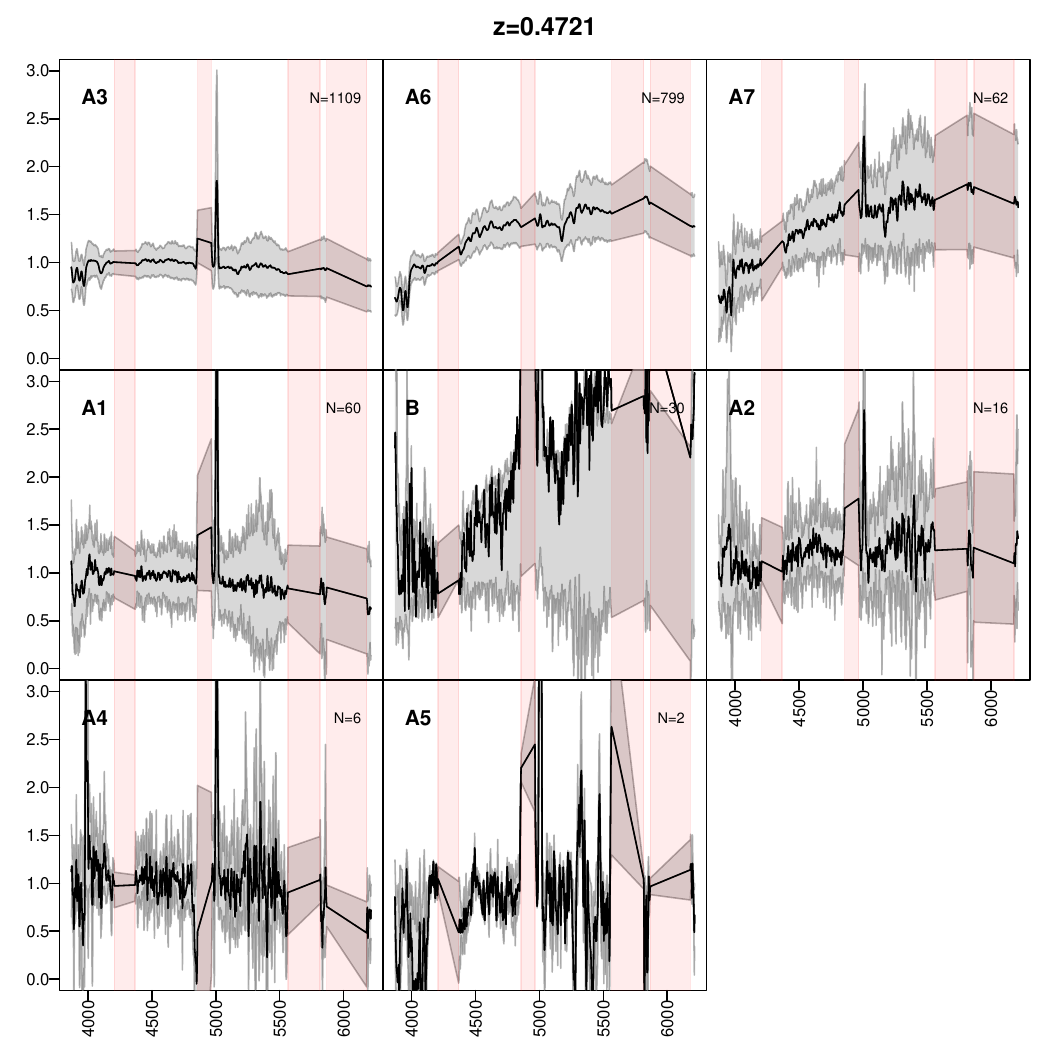}
    \caption{Stacked spectra of the classes of bin 4 (see Fig.~\ref{fig:vipers_classes_bin1} for further information)}
    \label{fig:vipers_classes_bin4}
\end{figure}

\begin{figure}
    \centering
    \includegraphics[width=\hsize]{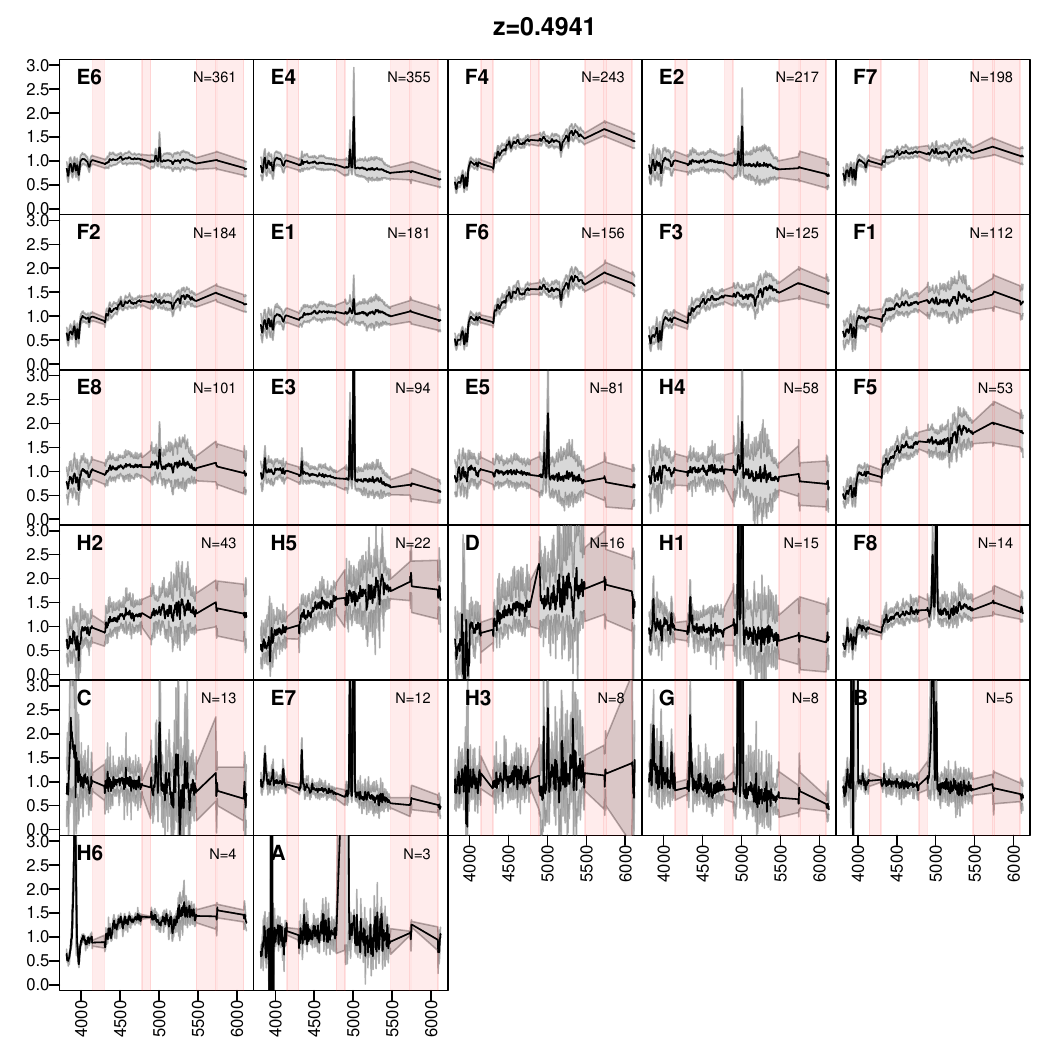}
    \caption{Stacked spectra of the classes of bin 5 (see Fig.~\ref{fig:vipers_classes_bin1} for further information)}
    \label{fig:vipers_classes_bin5}
\end{figure}

\begin{figure}
    \centering
    \includegraphics[width=\hsize]{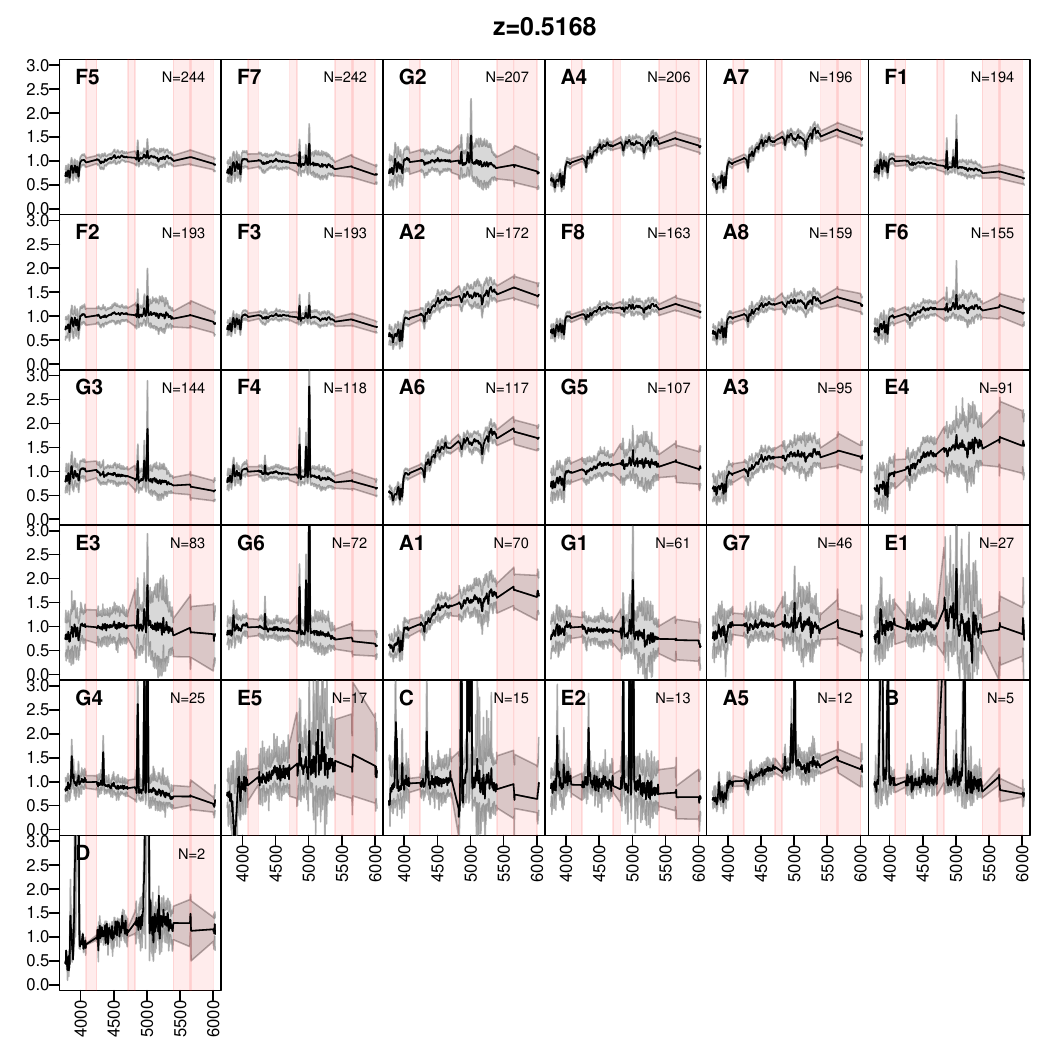}
    \caption{Stacked spectra of the classes of bin 6 (see Fig.~\ref{fig:vipers_classes_bin1} for further information)}
    \label{fig:vipers_classes_bin6}
\end{figure}

\begin{figure}
    \centering
    \includegraphics[width=\hsize]{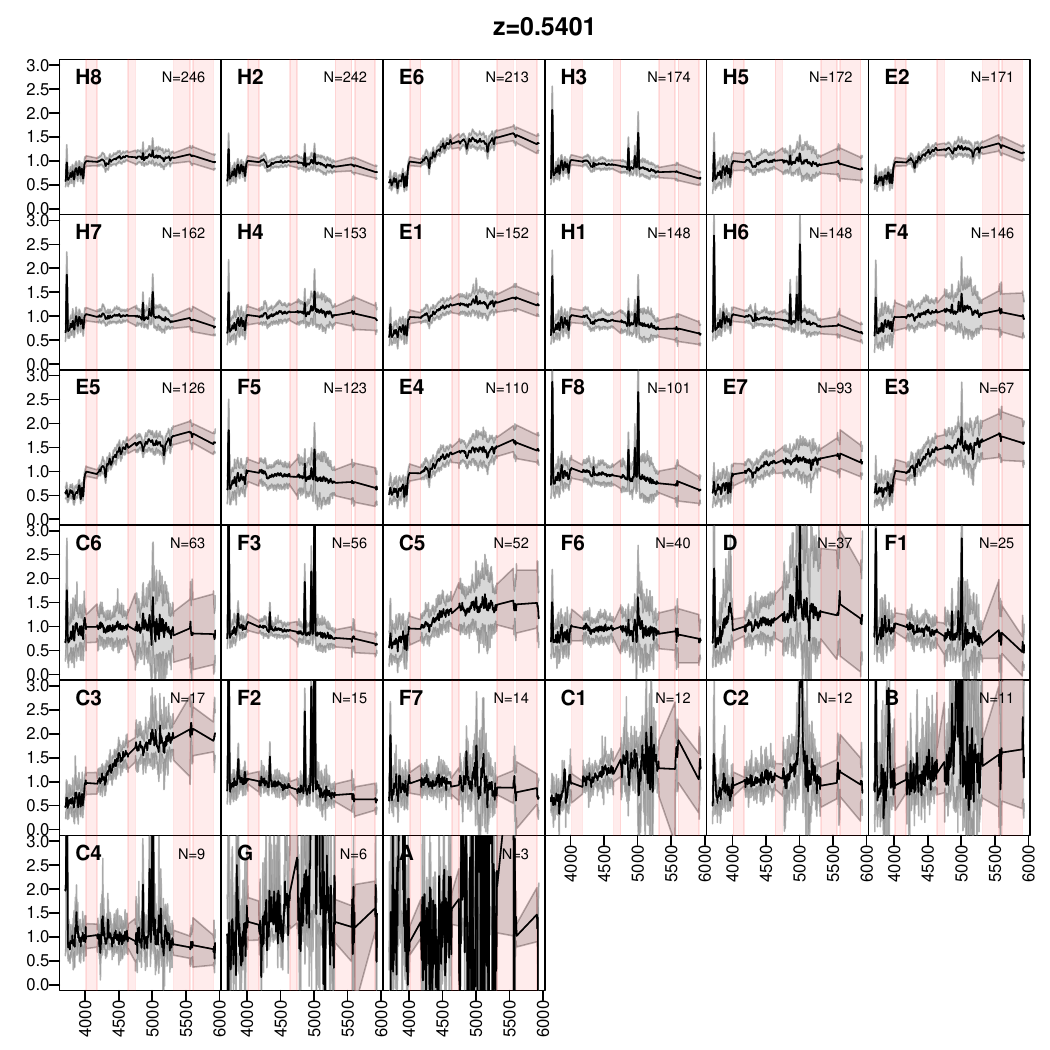}
    \caption{Stacked spectra of the classes of bin 7 (see Fig.~\ref{fig:vipers_classes_bin1} for further information)}
    \label{fig:vipers_classes_bin7}
\end{figure}

\begin{figure}
    \centering
    \includegraphics[width=\hsize]{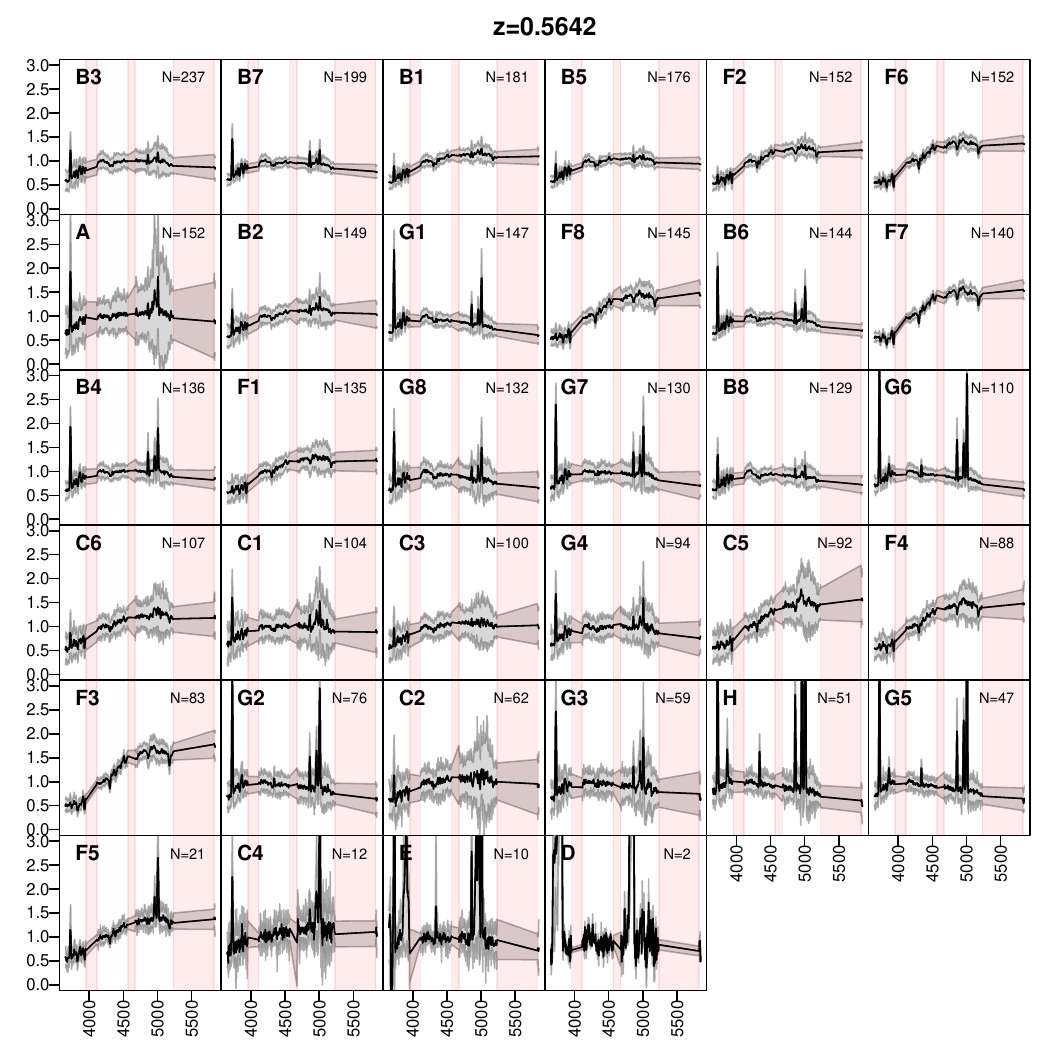}
    \caption{Stacked spectra of the classes of bin 8 (see Fig.~\ref{fig:vipers_classes_bin1} for further information)}
    \label{fig:vipers_classes_bin8}
\end{figure}

\begin{figure}
    \centering
    \includegraphics[width=\hsize]{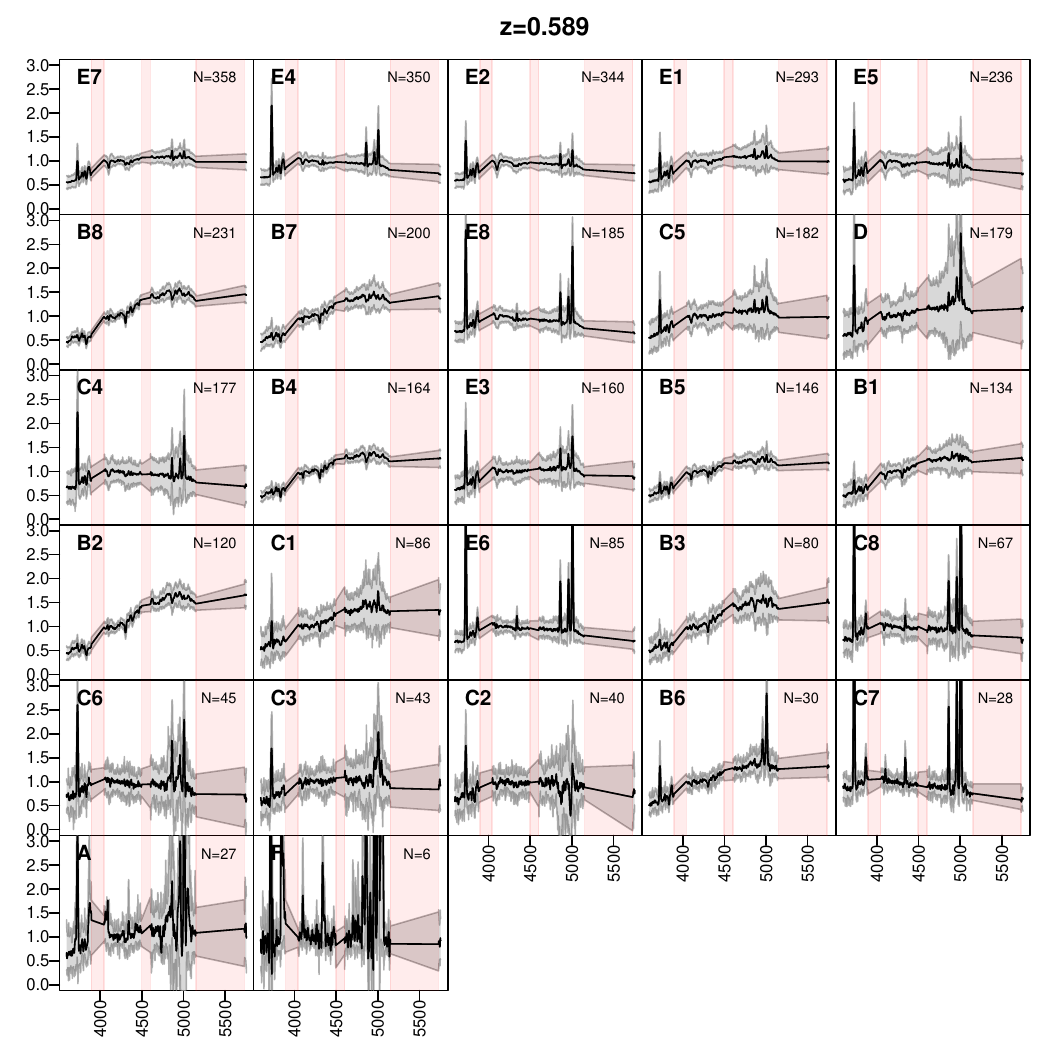}
    \caption{Stacked spectra of the classes of bin 9 (see Fig.~\ref{fig:vipers_classes_bin1} for further information)}
    \label{fig:vipers_classes_bin9}
\end{figure}

\begin{figure}
    \centering
    \includegraphics[width=\hsize]{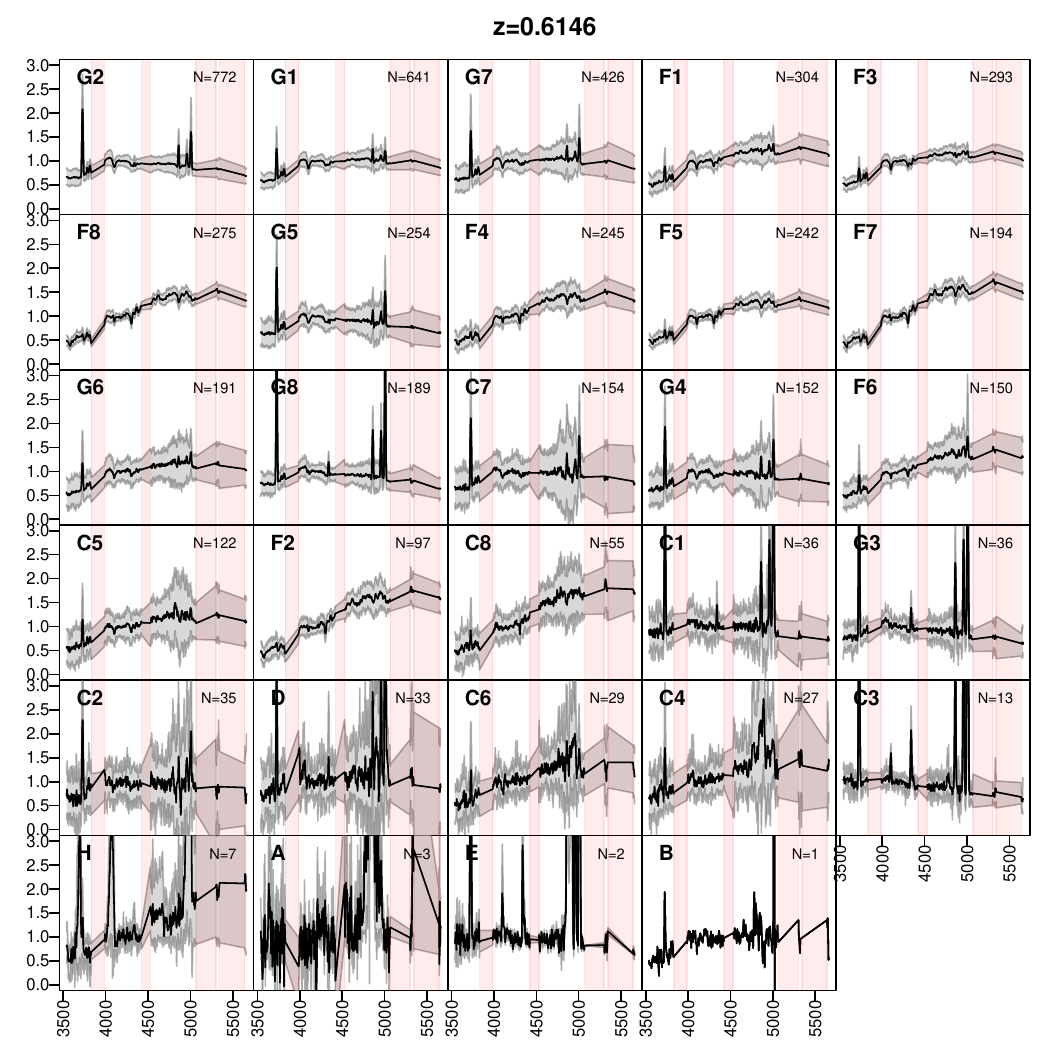}
    \caption{Stacked spectra of the classes of bin 10 (see Fig.~\ref{fig:vipers_classes_bin1} for further information)}
    \label{fig:vipers_classes_bin10}
\end{figure}

\begin{figure}
    \centering
    \includegraphics[width=\hsize]{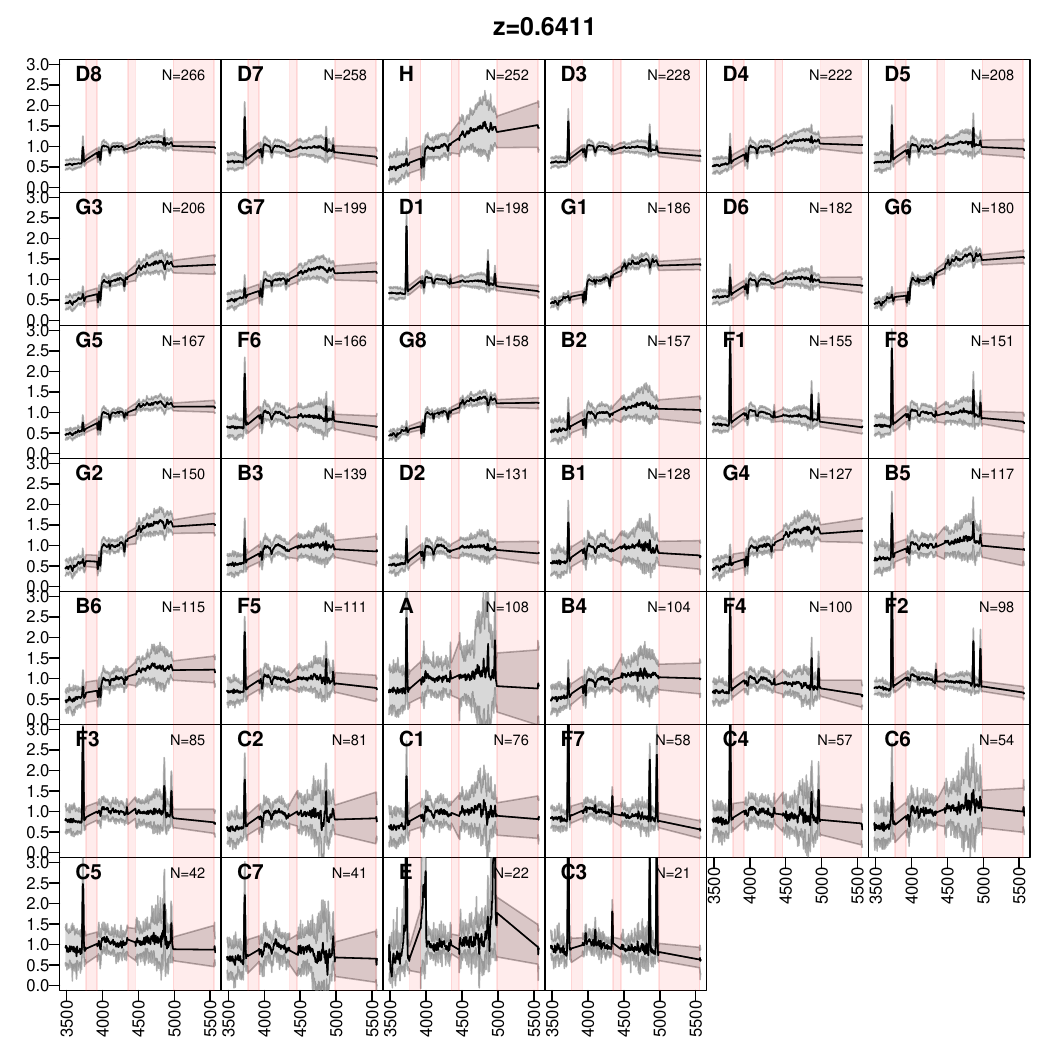}
    \caption{Stacked spectra of the classes of bin 11 (see Fig.~\ref{fig:vipers_classes_bin1} for further information)}
    \label{fig:vipers_classes_bin11}
\end{figure}

\begin{figure}
    \centering
    \includegraphics[width=\hsize]{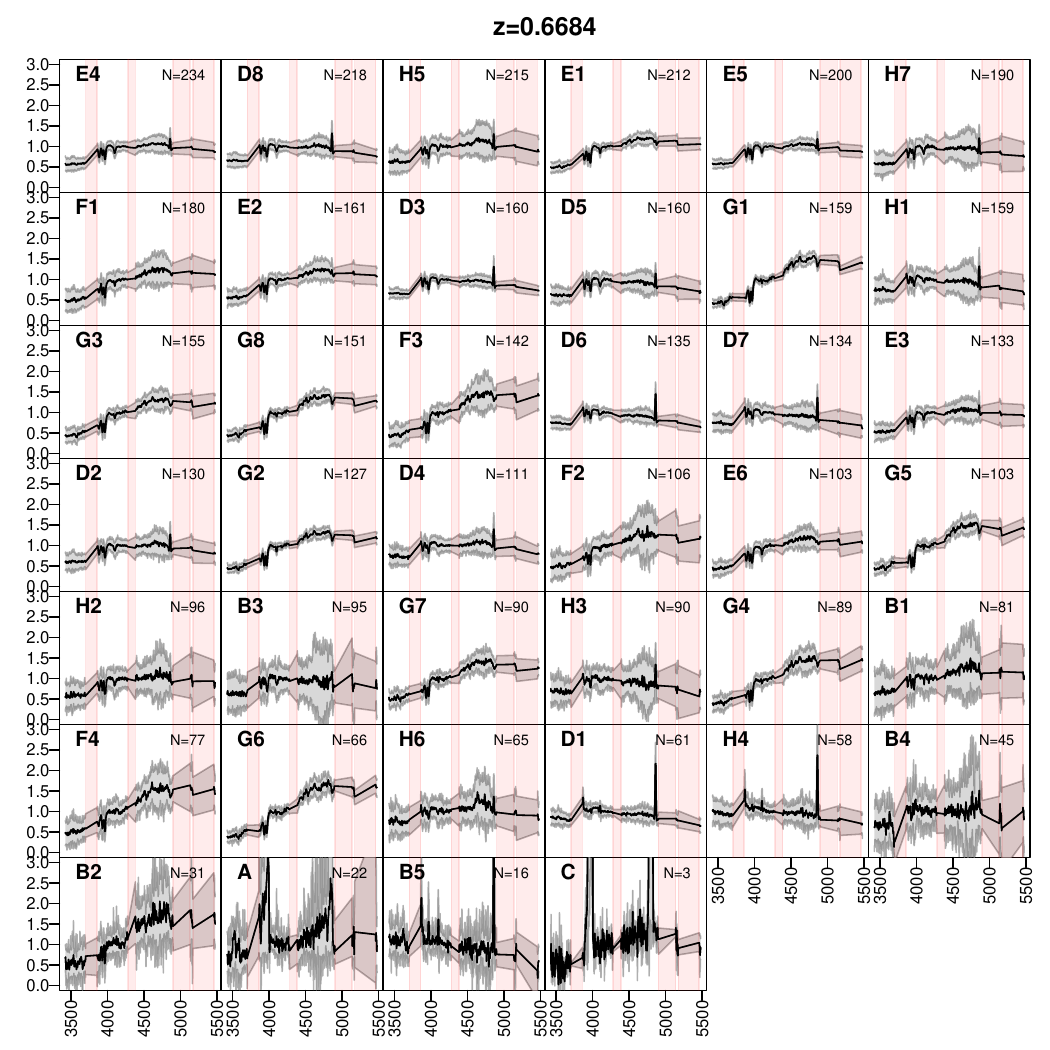}
    \caption{Stacked spectra of the classes of bin 12 (see Fig.~\ref{fig:vipers_classes_bin1} for further information)}
    \label{fig:vipers_classes_bin12}
\end{figure}

\begin{figure}
    \centering
    \includegraphics[width=\hsize]{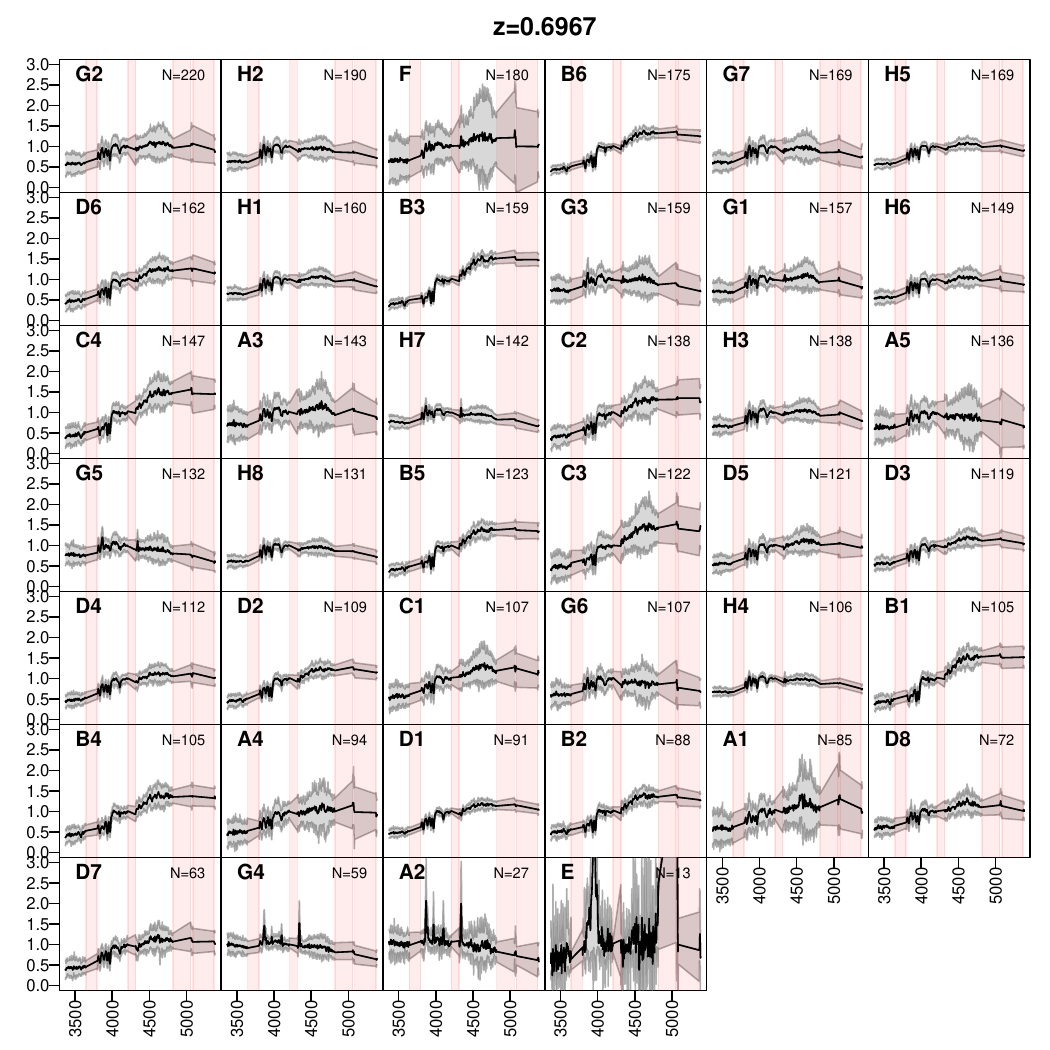}
    \caption{Stacked spectra of the classes of bin 13 (see Fig.~\ref{fig:vipers_classes_bin1} for further information)}
    \label{fig:vipers_classes_bin13}
\end{figure}

\begin{figure}
    \centering
    \includegraphics[width=\hsize]{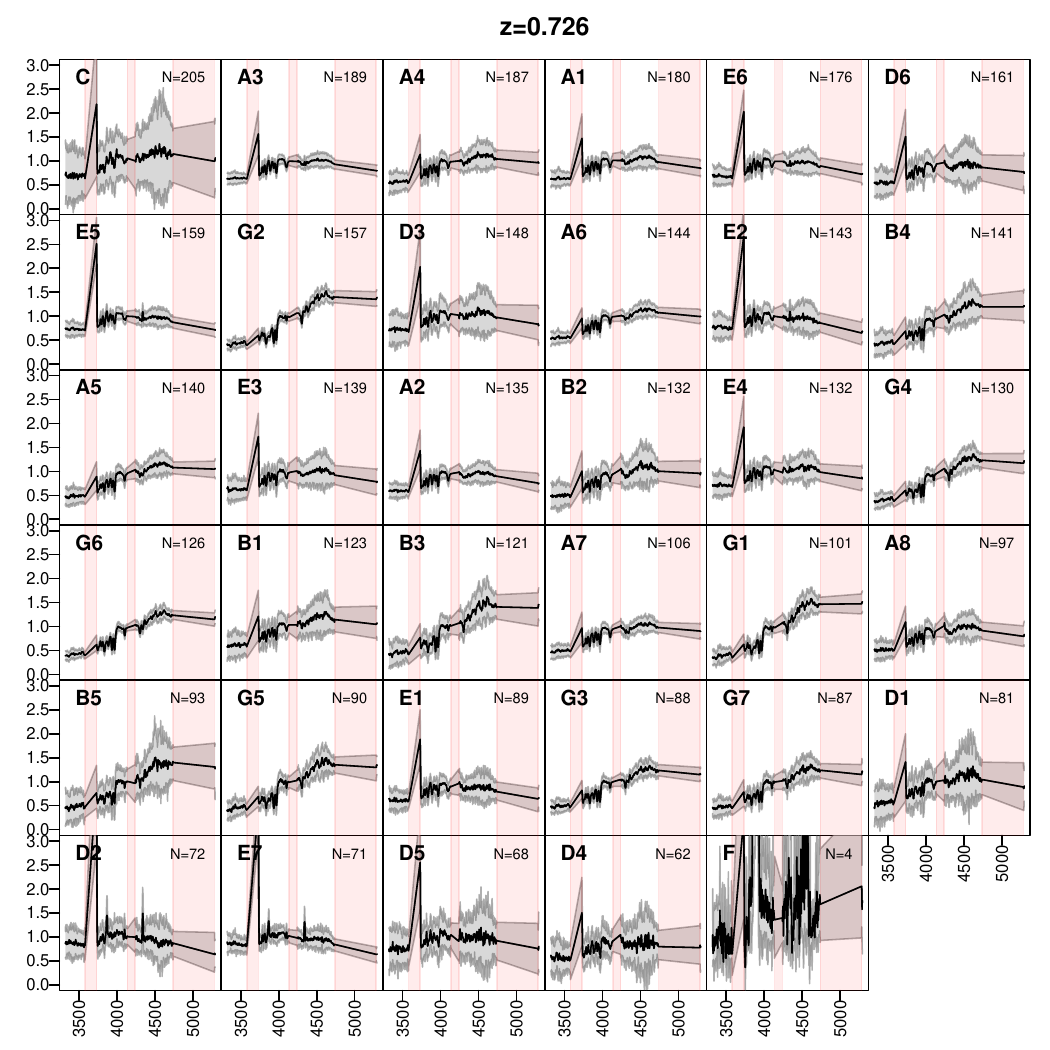}
    \caption{Stacked spectra of the classes of bin 14 (see Fig.~\ref{fig:vipers_classes_bin1} for further information)}
    \label{fig:vipers_classes_bin14}
\end{figure}

\begin{figure}
    \centering
    \includegraphics[width=\hsize]{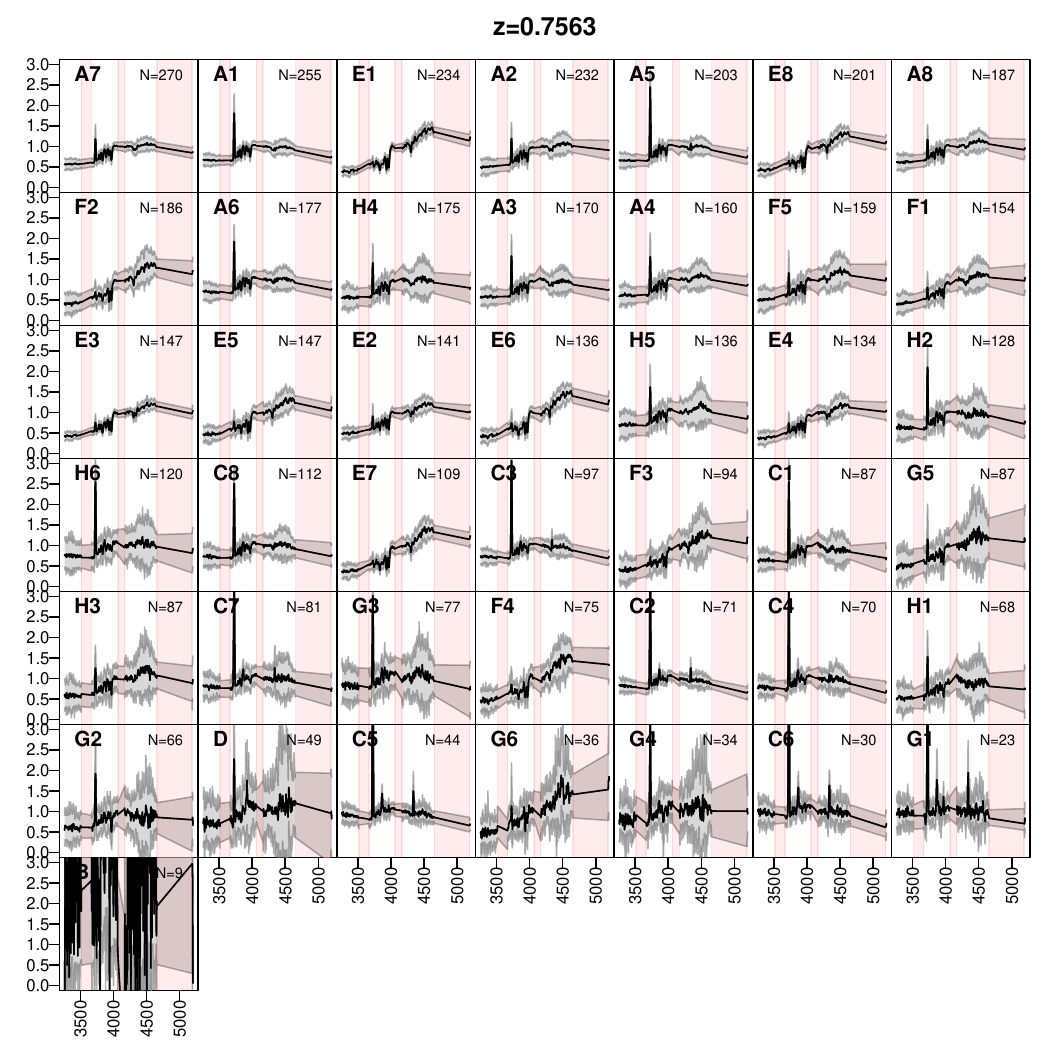}
    \caption{Stacked spectra of the classes of bin 15 (see Fig.~\ref{fig:vipers_classes_bin1} for further information)}
    \label{fig:vipers_classes_bin15}
\end{figure}

\begin{figure}
    \centering
    \includegraphics[width=\hsize]{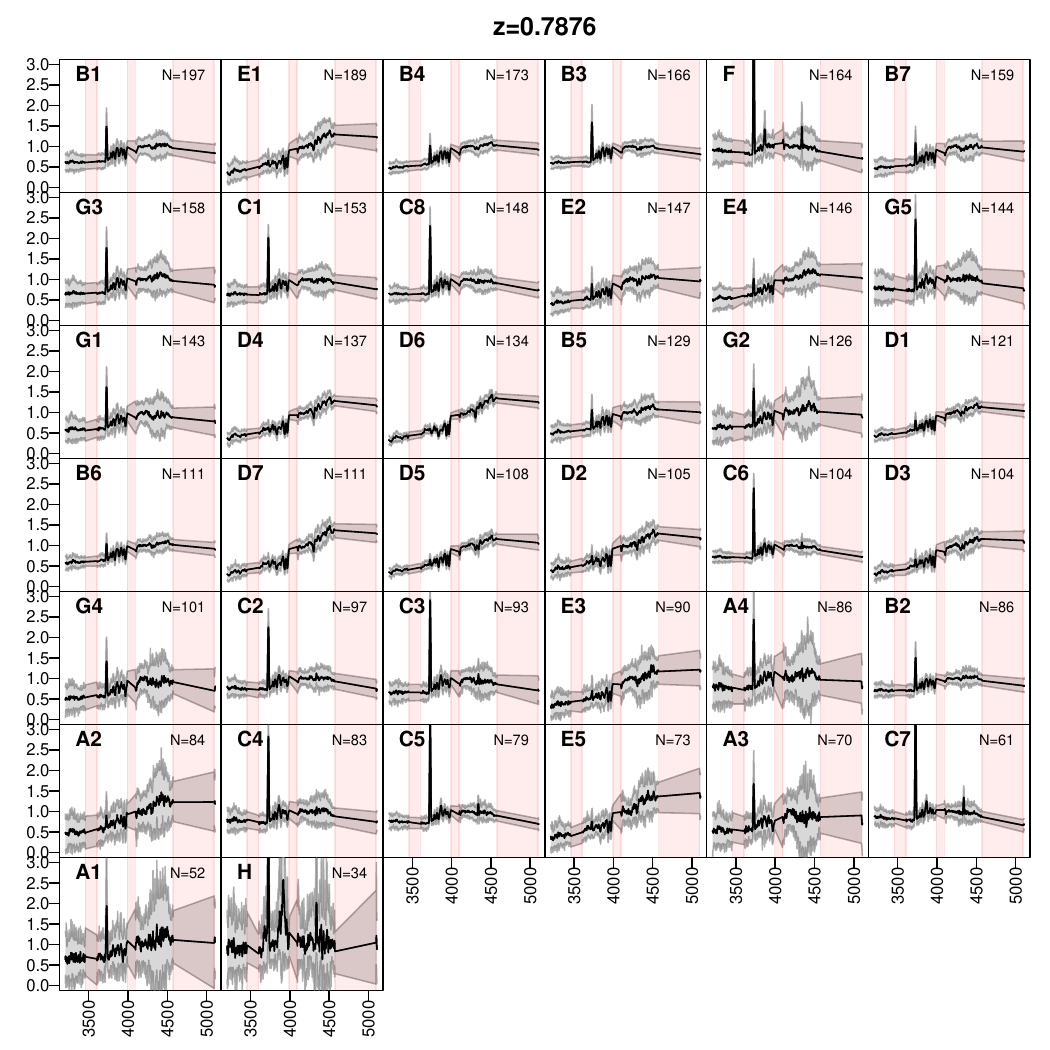}
    \caption{Stacked spectra of the classes of bin 16 (see Fig.~\ref{fig:vipers_classes_bin1} for further information)}
    \label{fig:vipers_classes_bin16}
\end{figure}

\begin{figure}
    \centering
    \includegraphics[width=\hsize]{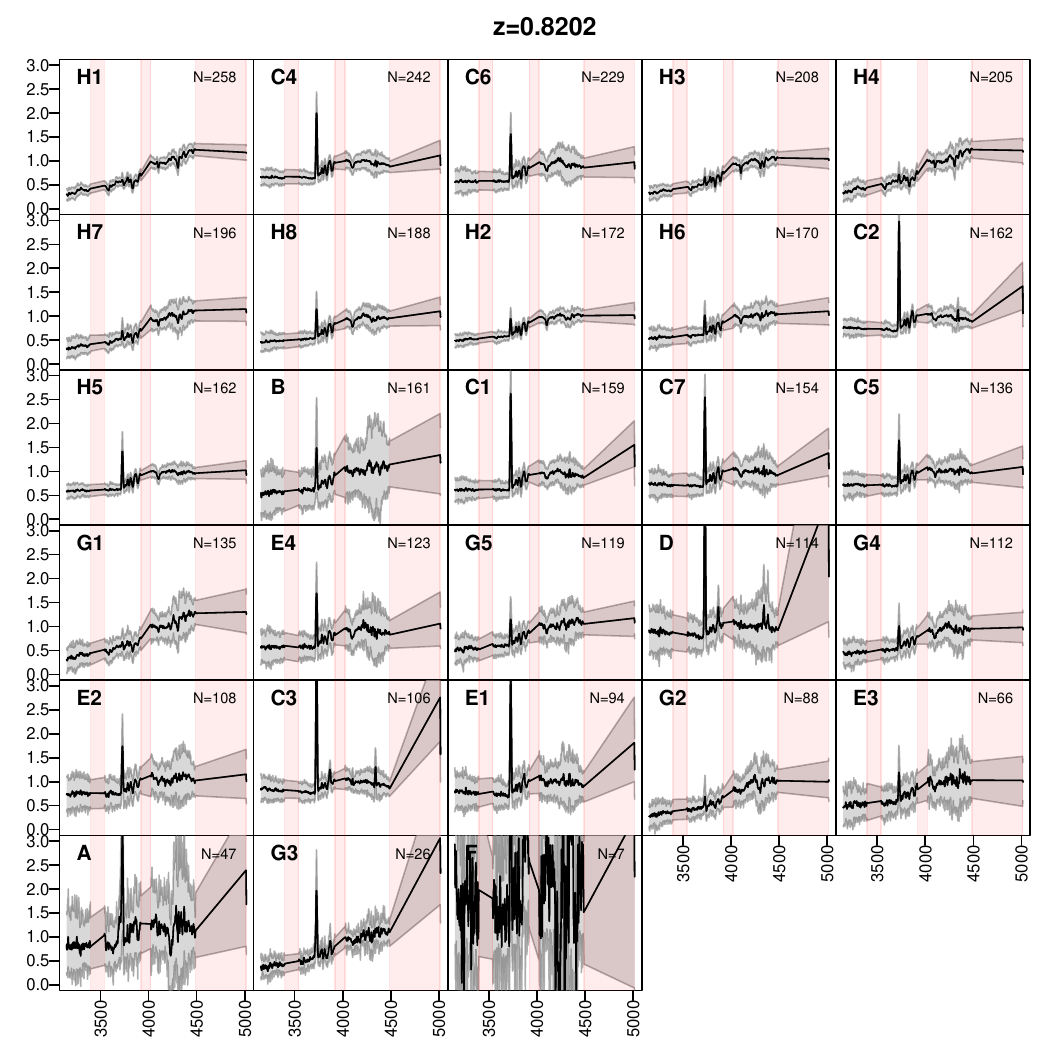}
    \caption{Stacked spectra of the classes of bin 17 (see Fig.~\ref{fig:vipers_classes_bin1} for further information)}
    \label{fig:vipers_classes_bin17}
\end{figure}

\begin{figure}
    \centering
    \includegraphics[width=\hsize]{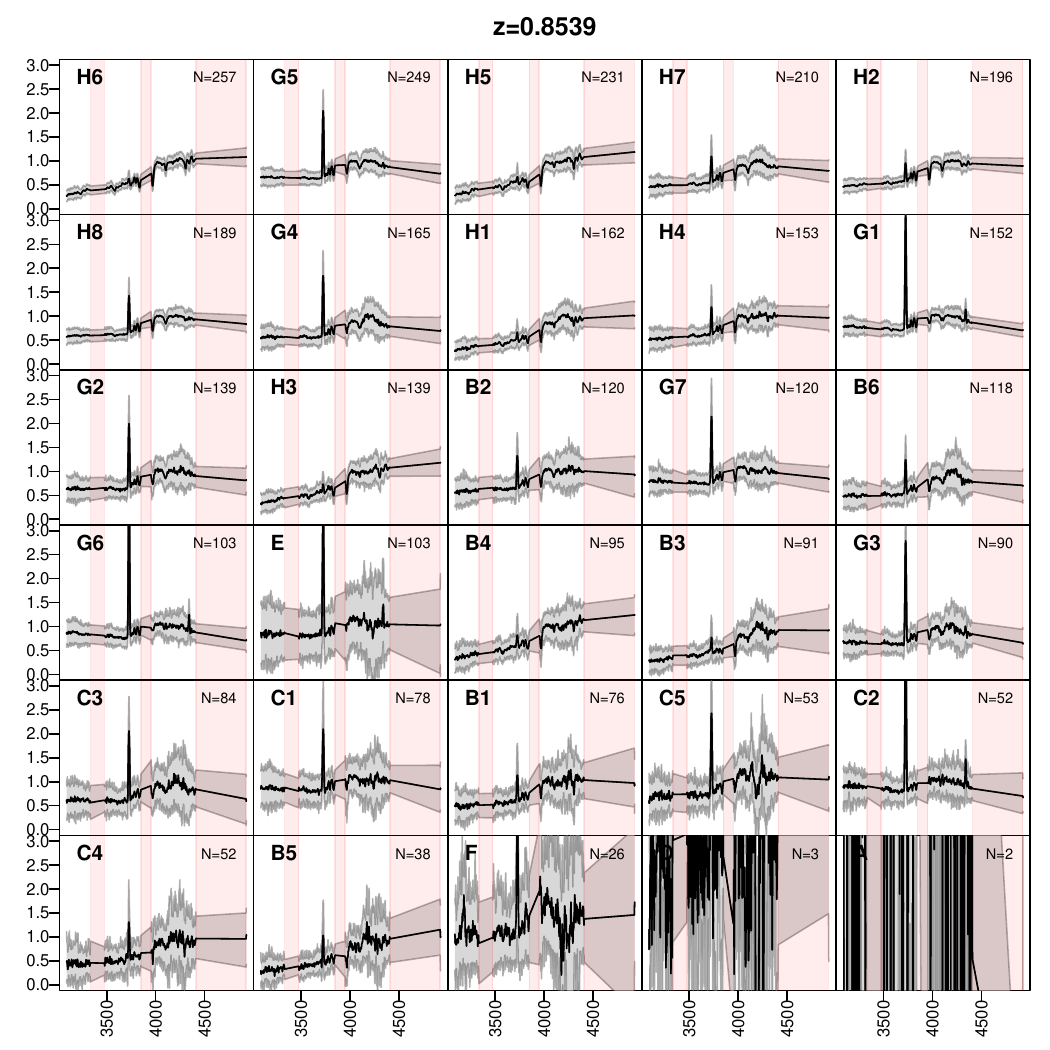}
    \caption{Stacked spectra of the classes of bin 18 (see Fig.~\ref{fig:vipers_classes_bin1} for further information)}
    \label{fig:vipers_classes_bin18}
\end{figure}

\begin{figure}
    \centering
    \includegraphics[width=\hsize]{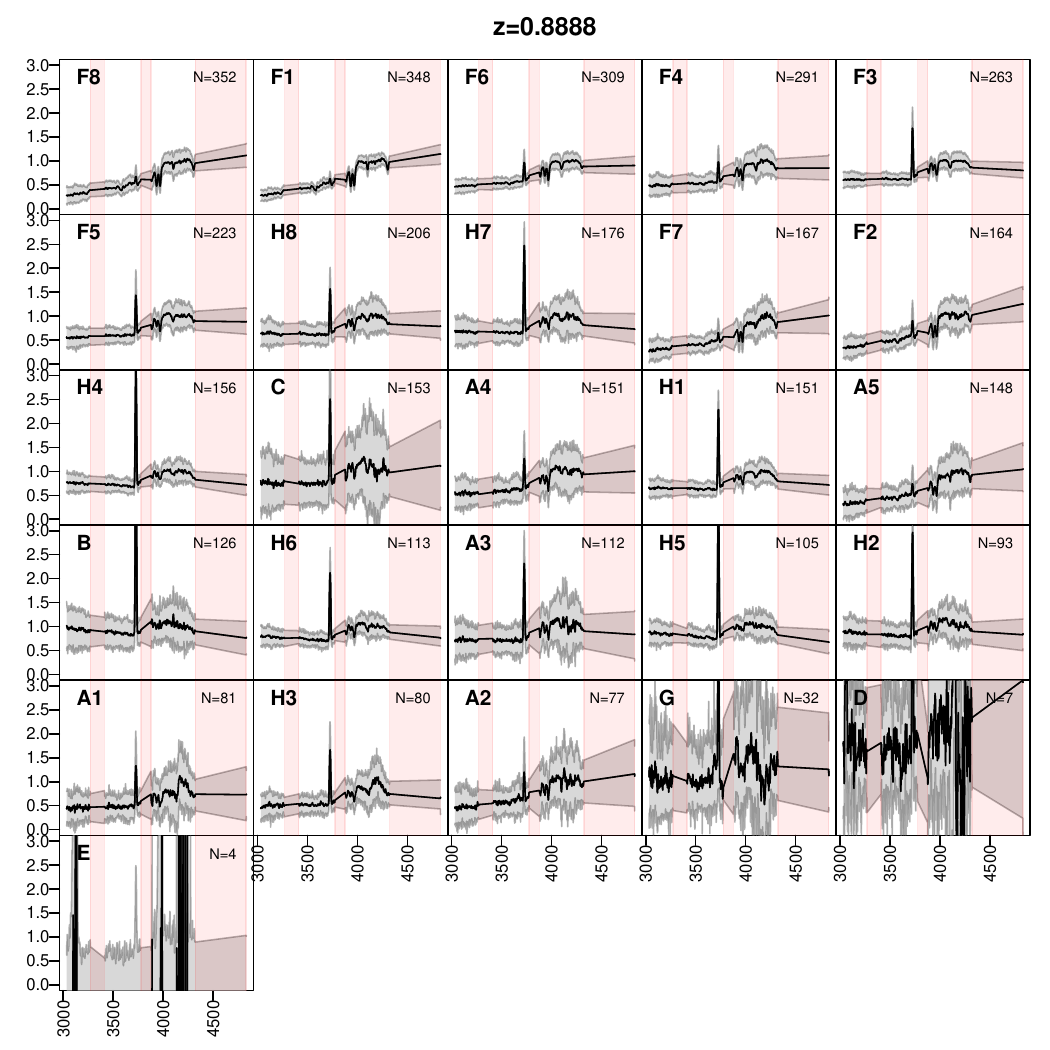}
    \caption{Stacked spectra of the classes of bin 19 (see Fig.~\ref{fig:vipers_classes_bin1} for further information)}
    \label{fig:vipers_classes_bin19}
\end{figure}

\begin{figure}
    \centering
    \includegraphics[width=\hsize]{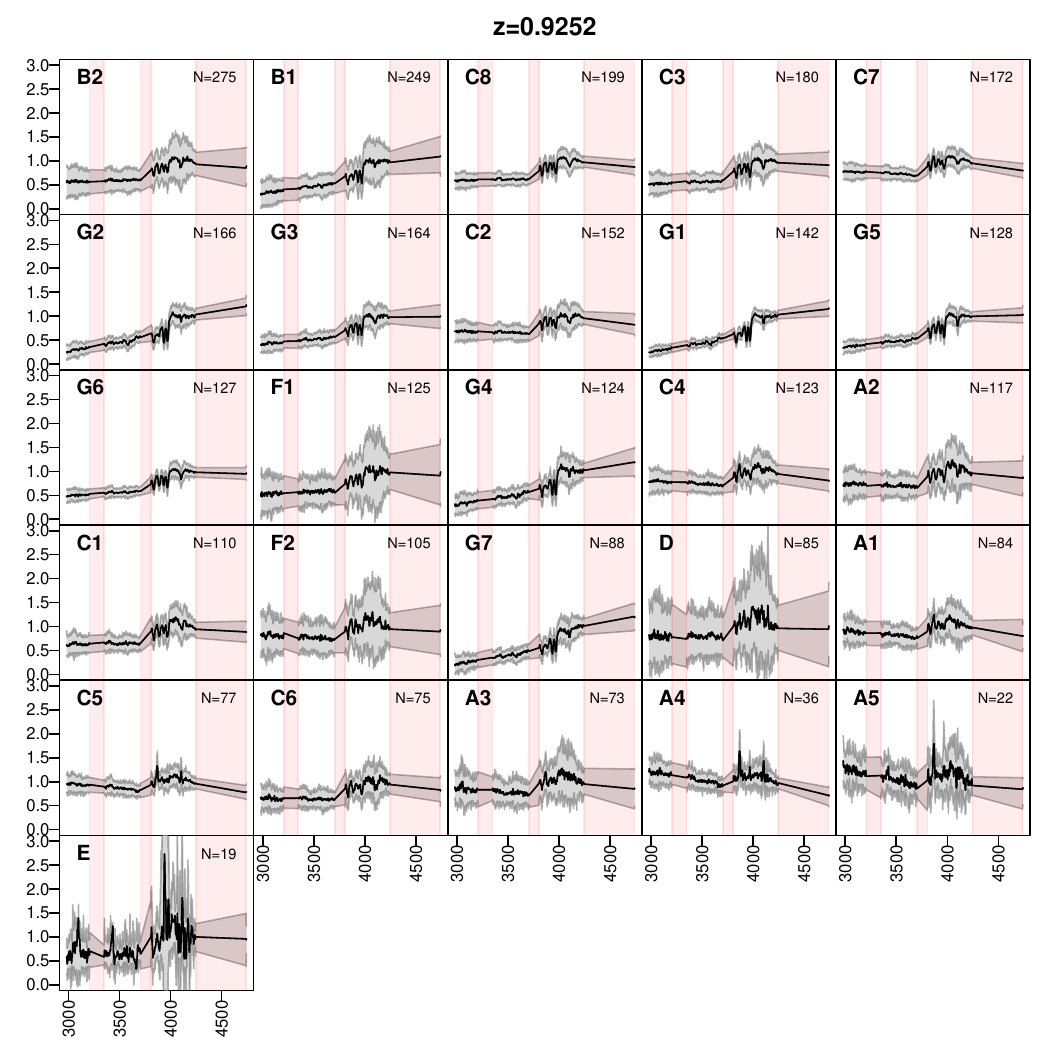}
    \caption{Stacked spectra of the classes of bin 20 (see Fig.~\ref{fig:vipers_classes_bin1} for further information)}
    \label{fig:vipers_classes_bin20}
\end{figure}

\begin{figure}
    \centering
    \includegraphics[width=\hsize]{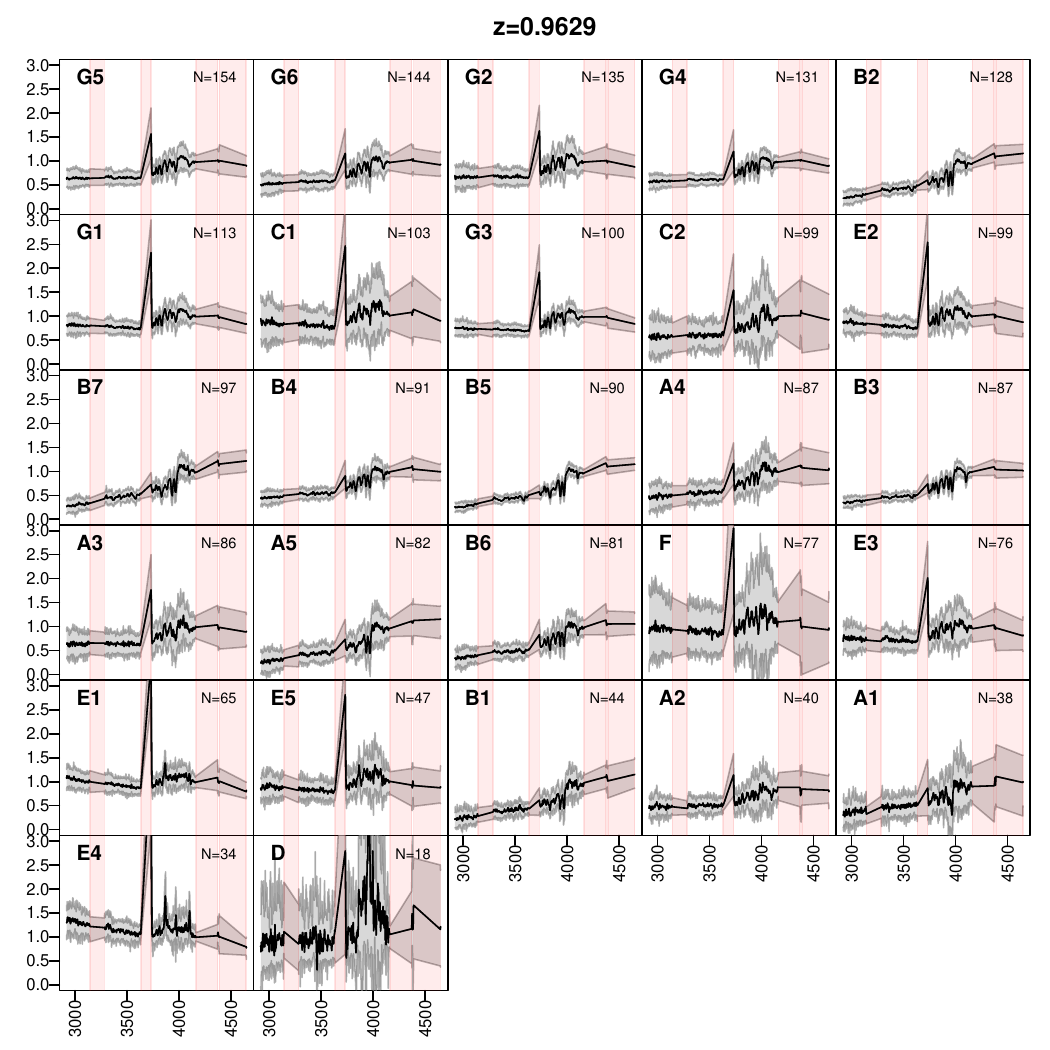}
    \caption{Stacked spectra of the classes of bin 21 (see Fig.~\ref{fig:vipers_classes_bin1} for further information)}
    \label{fig:vipers_classes_bin21}
\end{figure}

\begin{figure}
    \centering
    \includegraphics[width=\hsize]{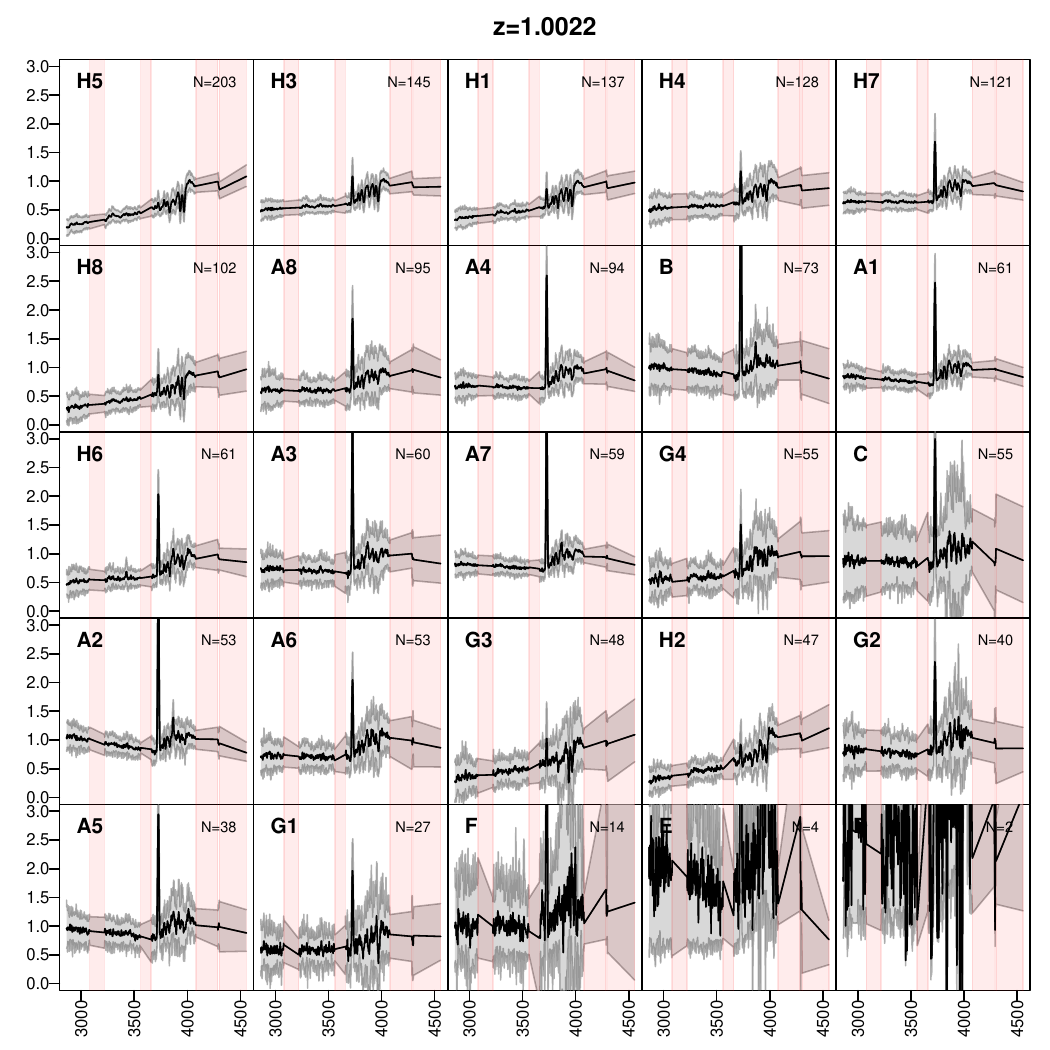}
    \caption{Stacked spectra of the classes of bin 22 (see Fig.~\ref{fig:vipers_classes_bin1} for further information)}
    \label{fig:vipers_classes_bin22}
\end{figure}

\begin{figure}
    \centering
    \includegraphics[width=\hsize]{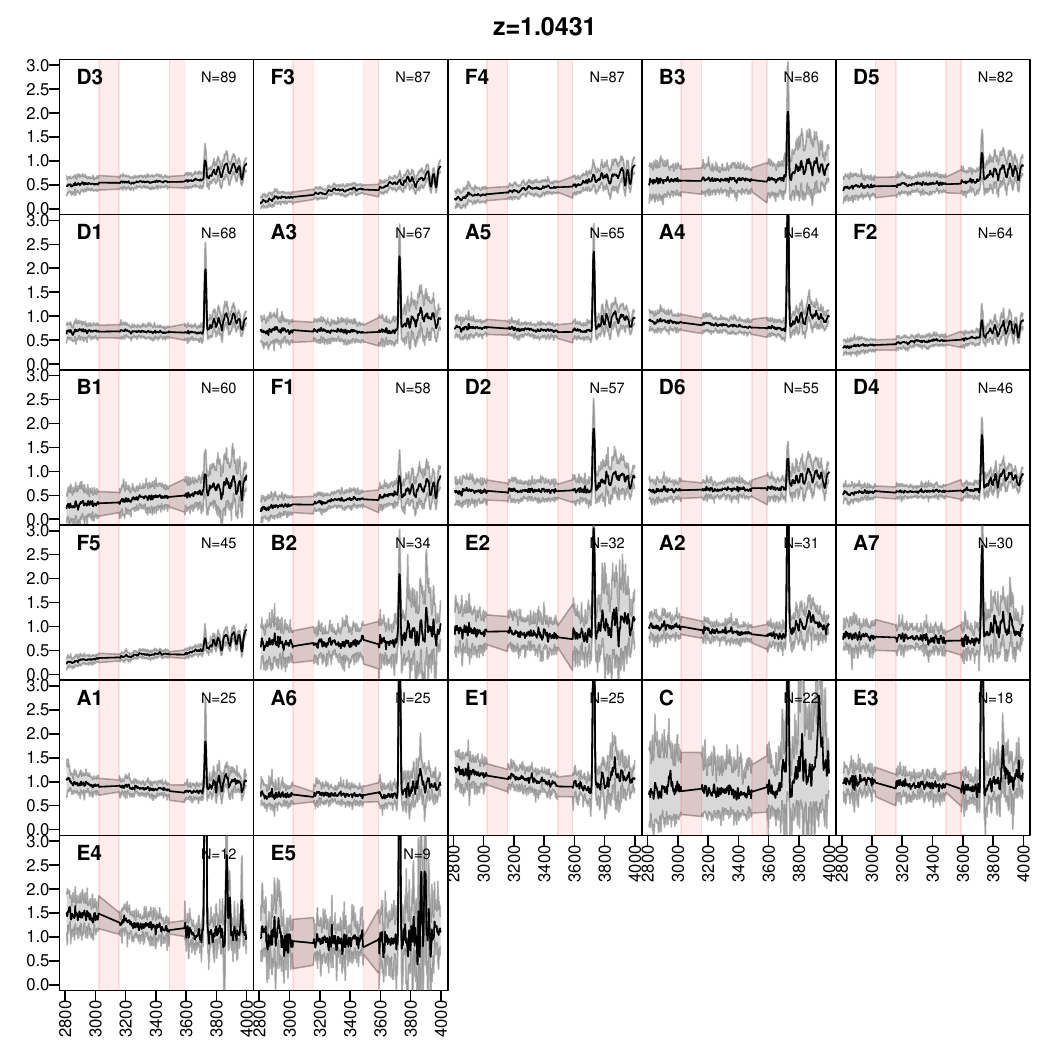}
    \caption{Stacked spectra of the classes of bin 23 (see Fig.~\ref{fig:vipers_classes_bin1} for further information)}
    \label{fig:vipers_classes_bin23}
\end{figure}

\begin{figure}
    \centering
    \includegraphics[width=\hsize]{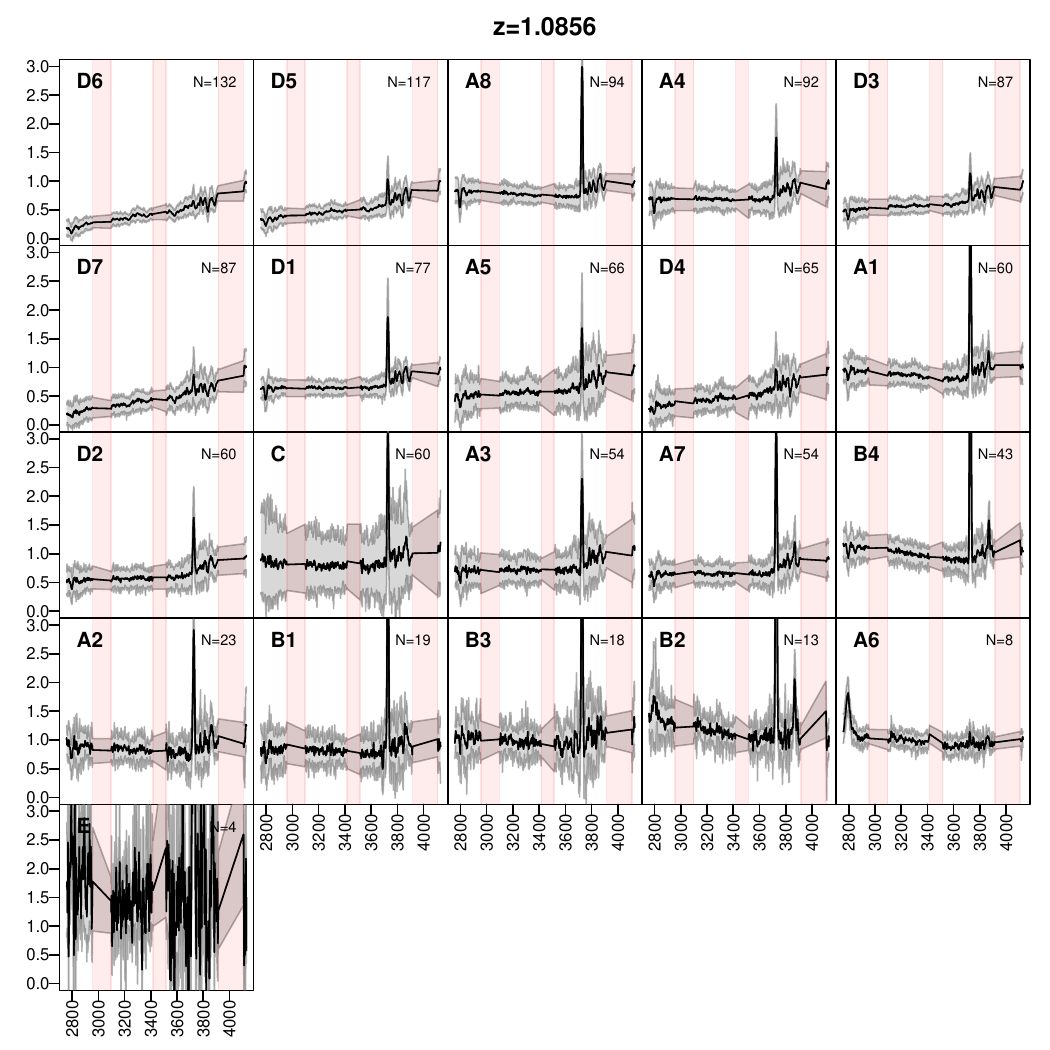}
    \caption{Stacked spectra of the classes of bin 24 (see Fig.~\ref{fig:vipers_classes_bin1} for further information)}
    \label{fig:vipers_classes_bin24}
\end{figure}

\begin{figure}
    \centering
    \includegraphics[width=\hsize]{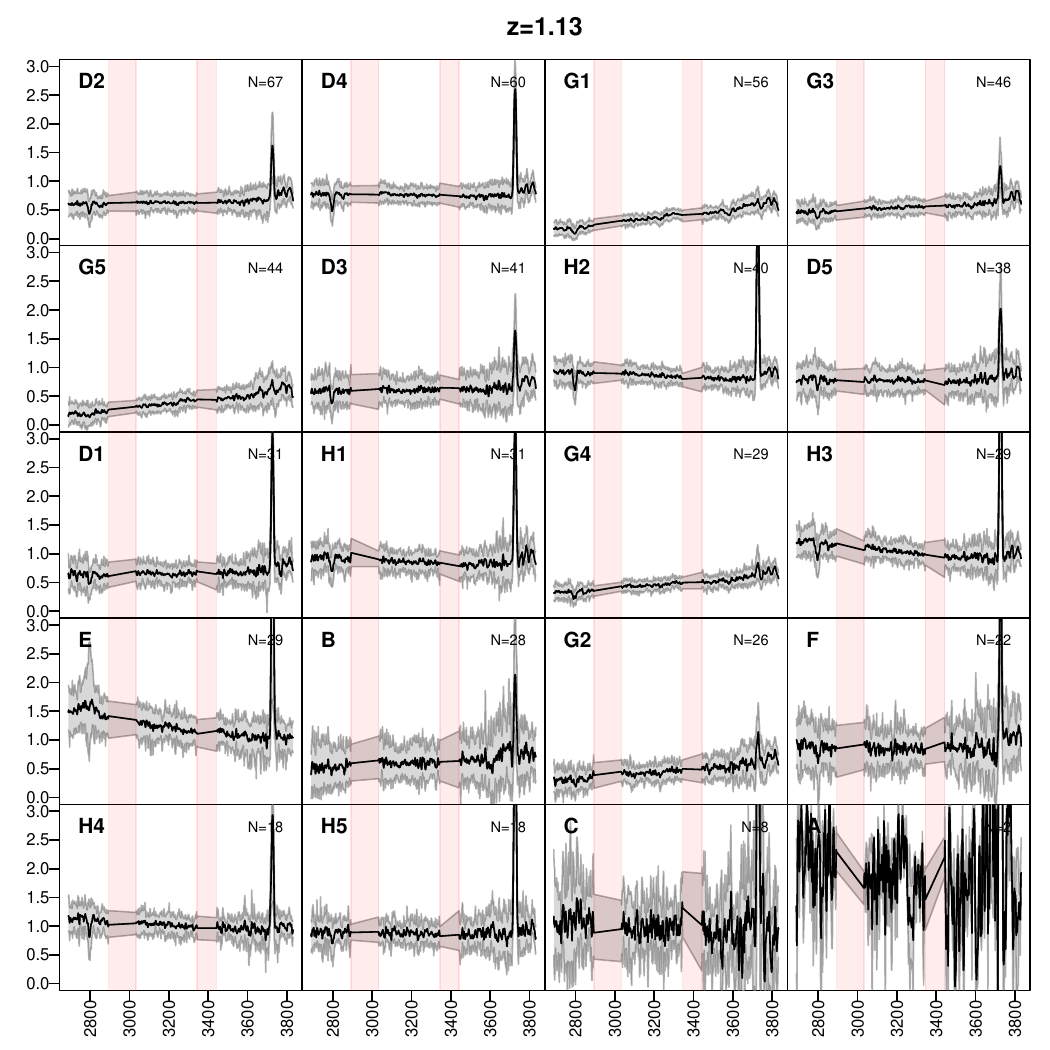}
    \caption{Stacked spectra of the classes of bin 25 (see Fig.~\ref{fig:vipers_classes_bin1} for further information)}
    \label{fig:vipers_classes_bin25}
\end{figure}

\begin{figure}
    \centering
    \includegraphics[width=\hsize]{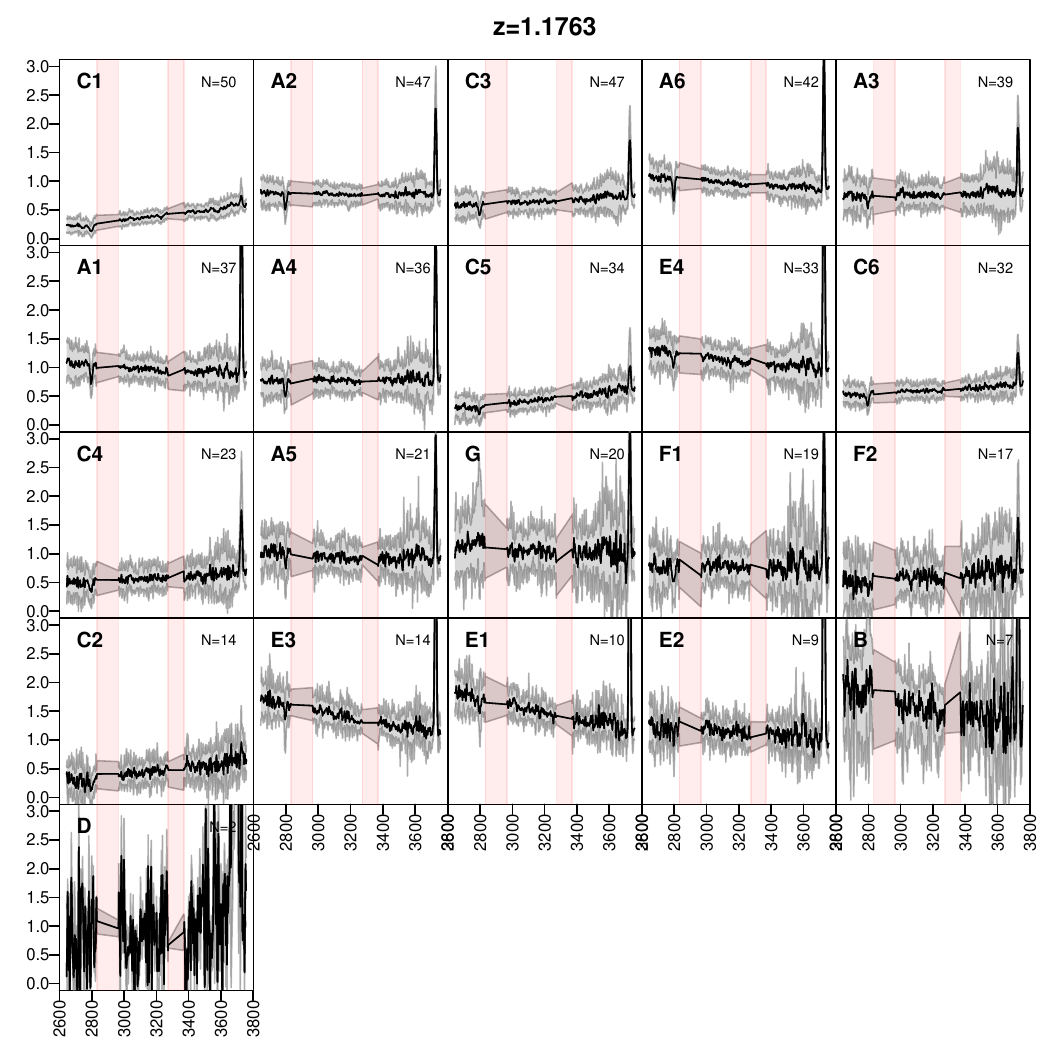}
    \caption{Stacked spectra of the classes of bin 26 (see Fig.~\ref{fig:vipers_classes_bin1} for further information)}
    \label{fig:vipers_classes_bin26}
\end{figure}

\clearpage

    \section{MEx diagrams}
\label{appendix:MEx}
\newpage

\begin{figure}
    \centering
    \includegraphics[width=\hsize]{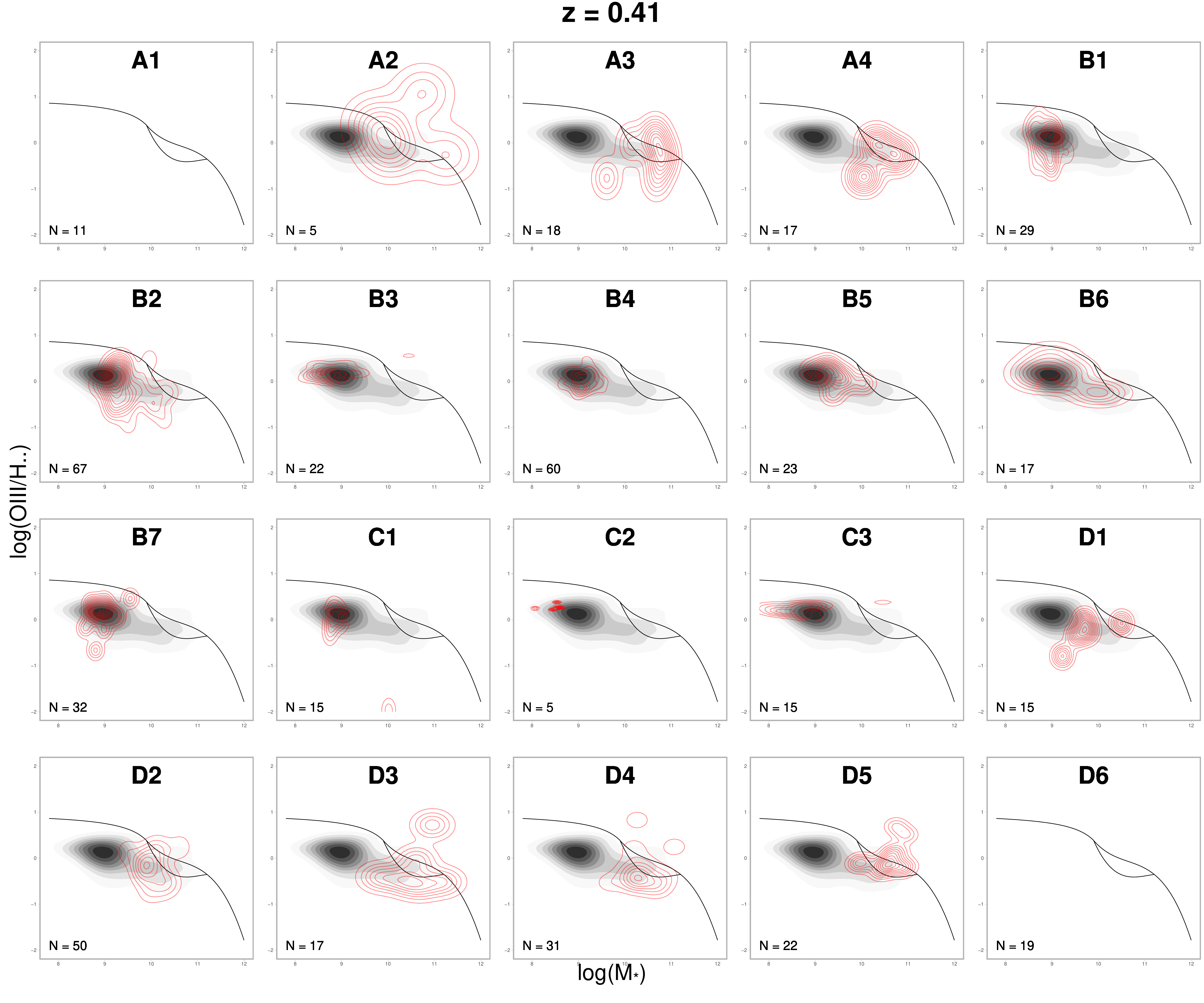}
    \caption{MEx diagrams of the 20 classes from the first bin, at $z=0.41$. Each panel shows one class, whose name and sizes are displayed at the top and bottom left corner. The black line delimits the star-forming region (bottom), AGN region (top), and intermediate region (centre). The black contours show the distribution of the bin, and the red contours show the distribution of the class. }
    \label{fig:vipers_MEx_bin1}
\end{figure}

\begin{figure}
    \centering
    \includegraphics[width=\hsize]{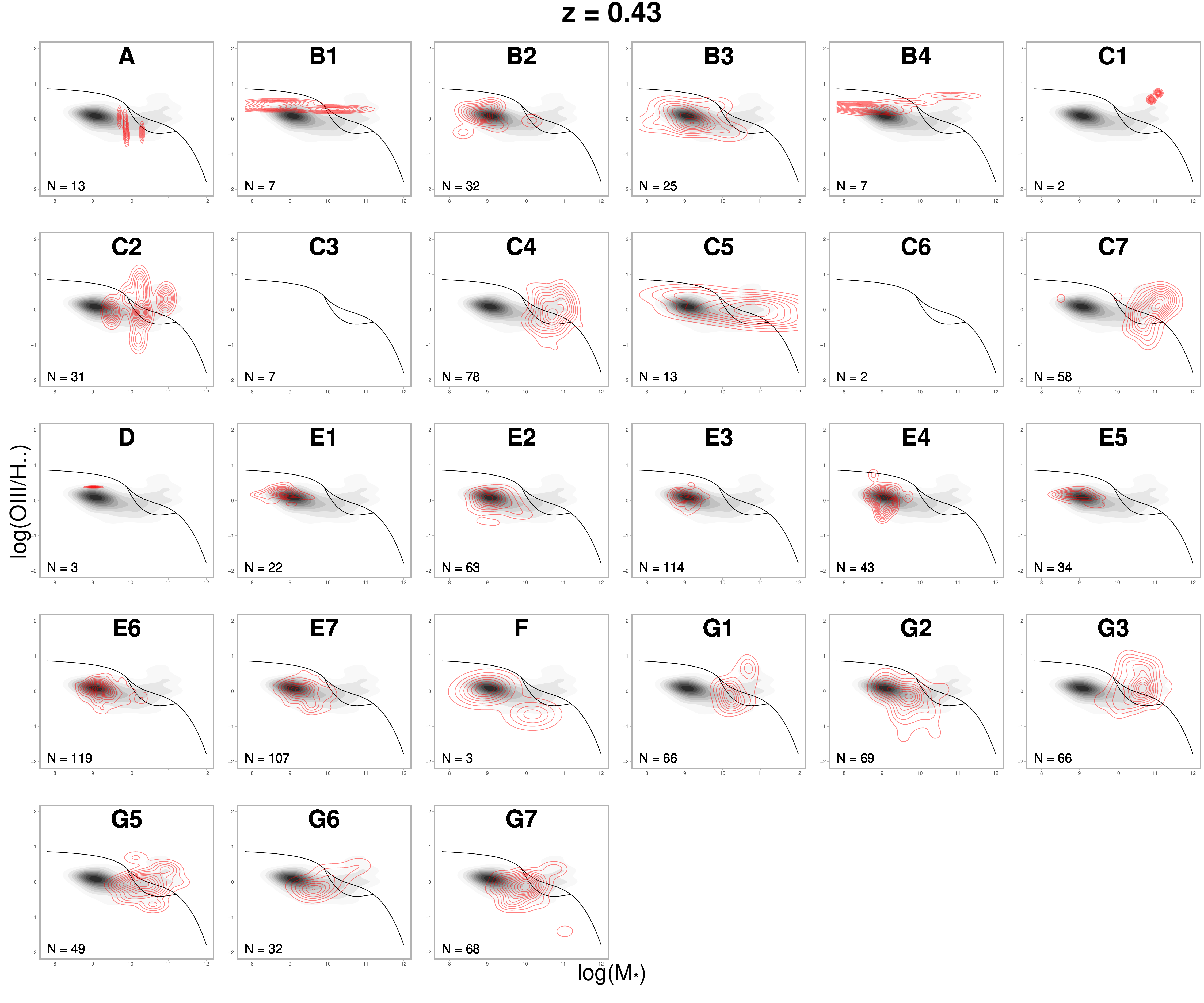}
    \caption{MEx diagram of the classes of bin 2 (see Fig.~\ref{fig:vipers_MEx_bin1} for further information)}
    \label{fig:vipers_MEx_bin2}
\end{figure}

\begin{figure}
    \centering
    \includegraphics[width=\hsize]{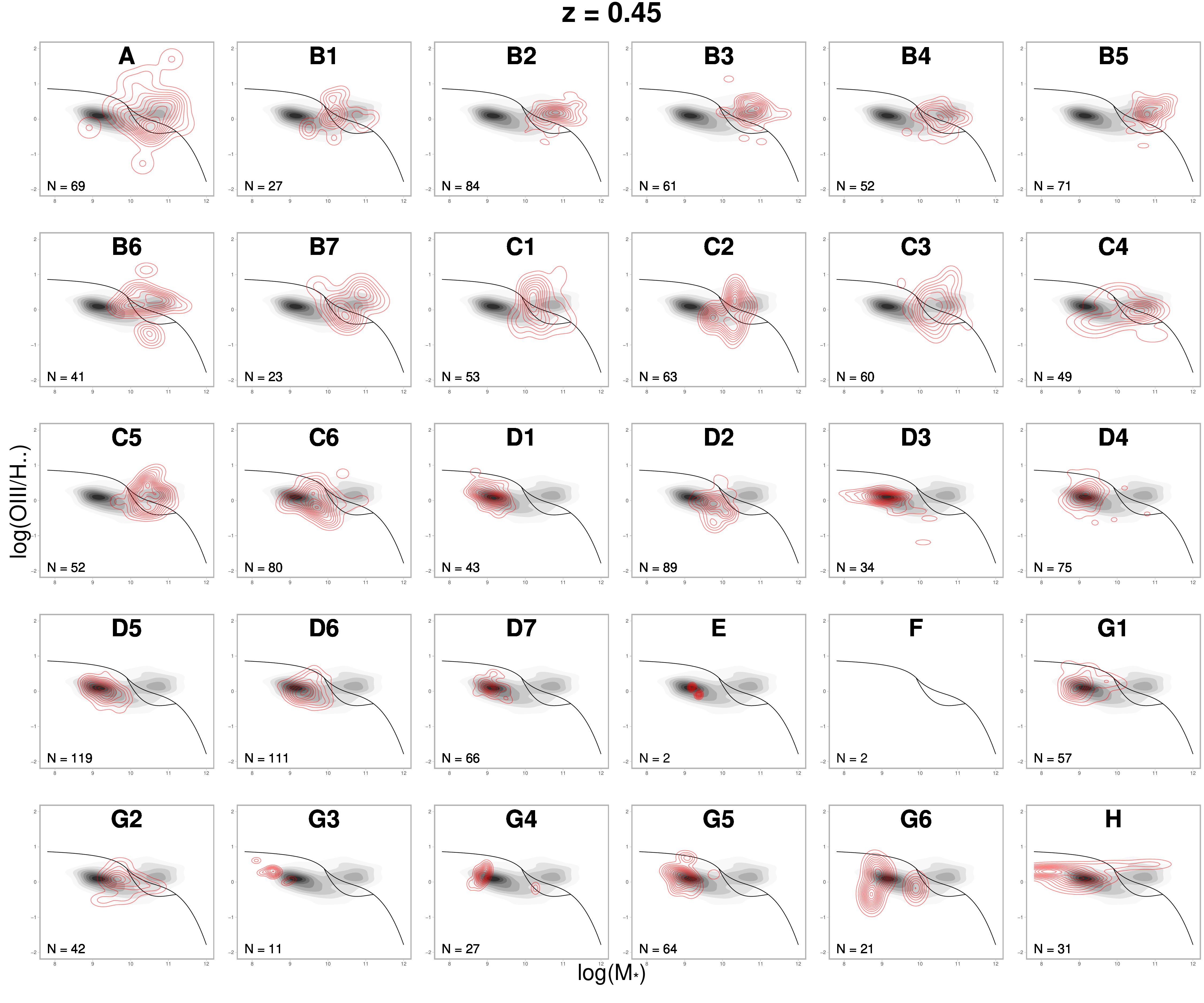}
    \caption{MEx diagram of the classes of bin 3 (see Fig.~\ref{fig:vipers_MEx_bin1} for further information)}
    \label{fig:vipers_MEx_bin3}
\end{figure}

\begin{figure}
    \centering
    \includegraphics[width=\hsize]{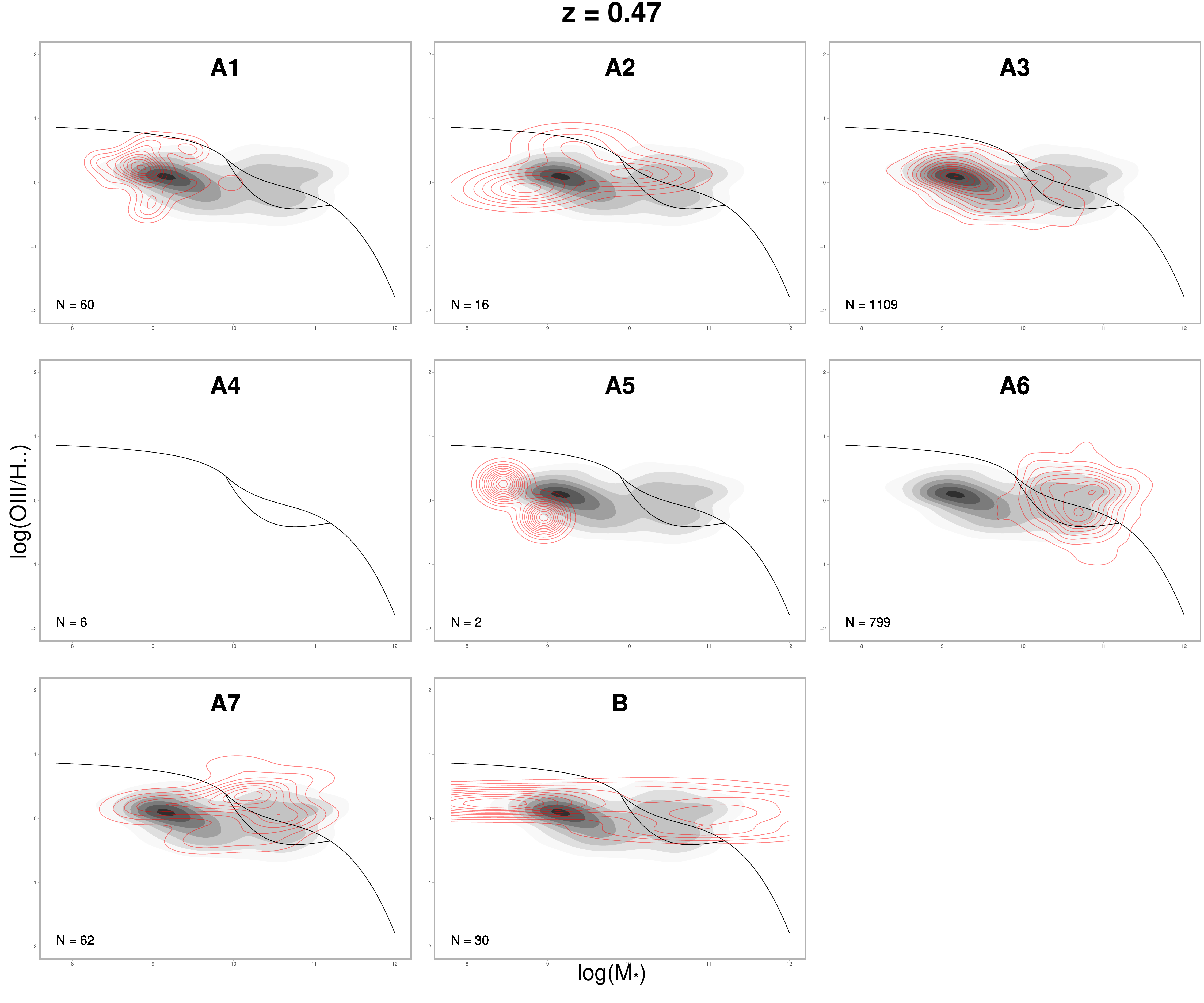}
    \caption{MEx diagram of the classes of bin 4 (see Fig.~\ref{fig:vipers_MEx_bin1} for further information)}
    \label{fig:vipers_MEx_bin4}
\end{figure}

\begin{figure}
    \centering
    \includegraphics[width=\hsize]{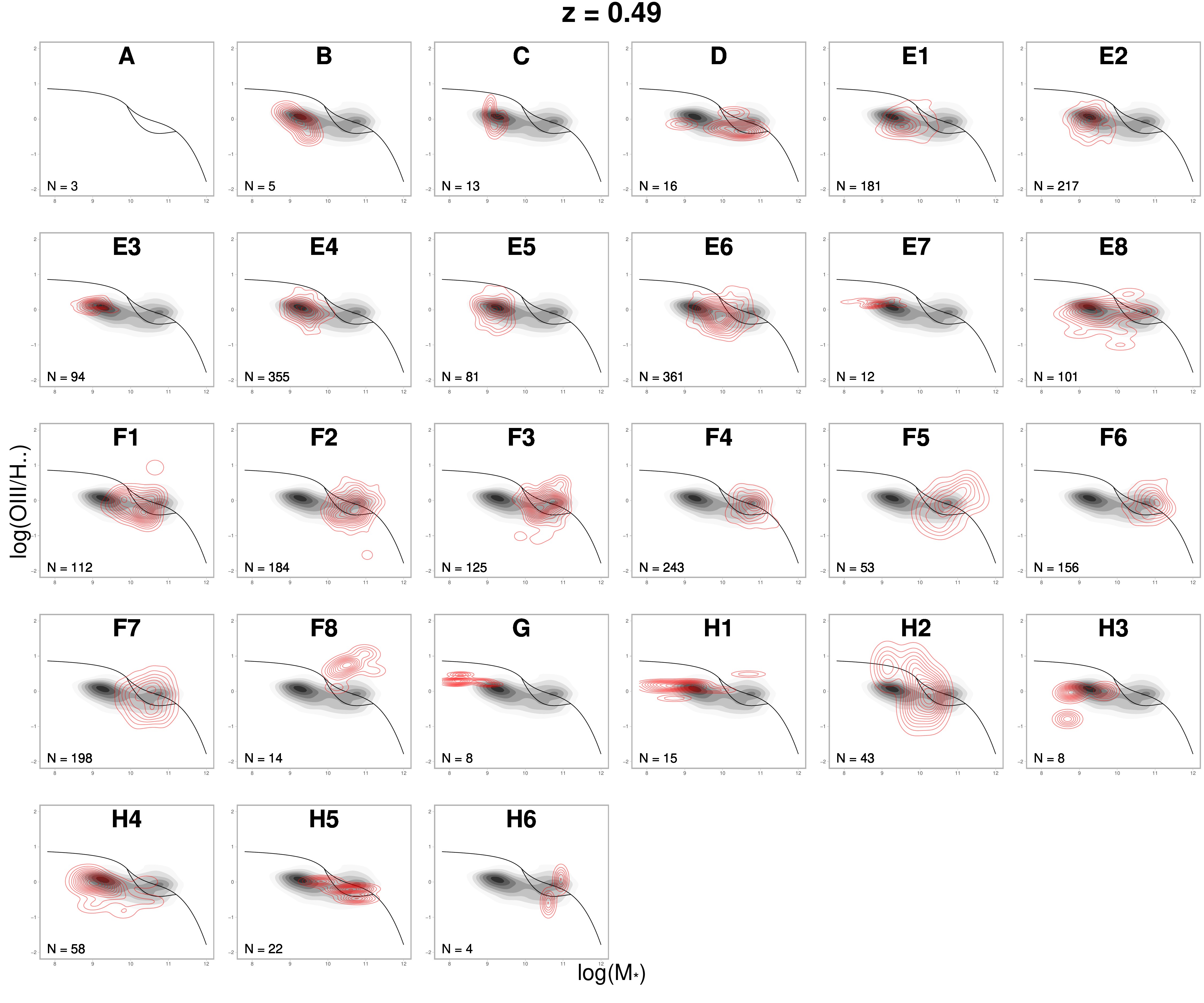}
    \caption{MEx diagram of the classes of bin 5 (see Fig.~\ref{fig:vipers_MEx_bin1} for further information)}
    \label{fig:vipers_MEx_bin5}
\end{figure}

\begin{figure}
    \centering
    \includegraphics[width=\hsize]{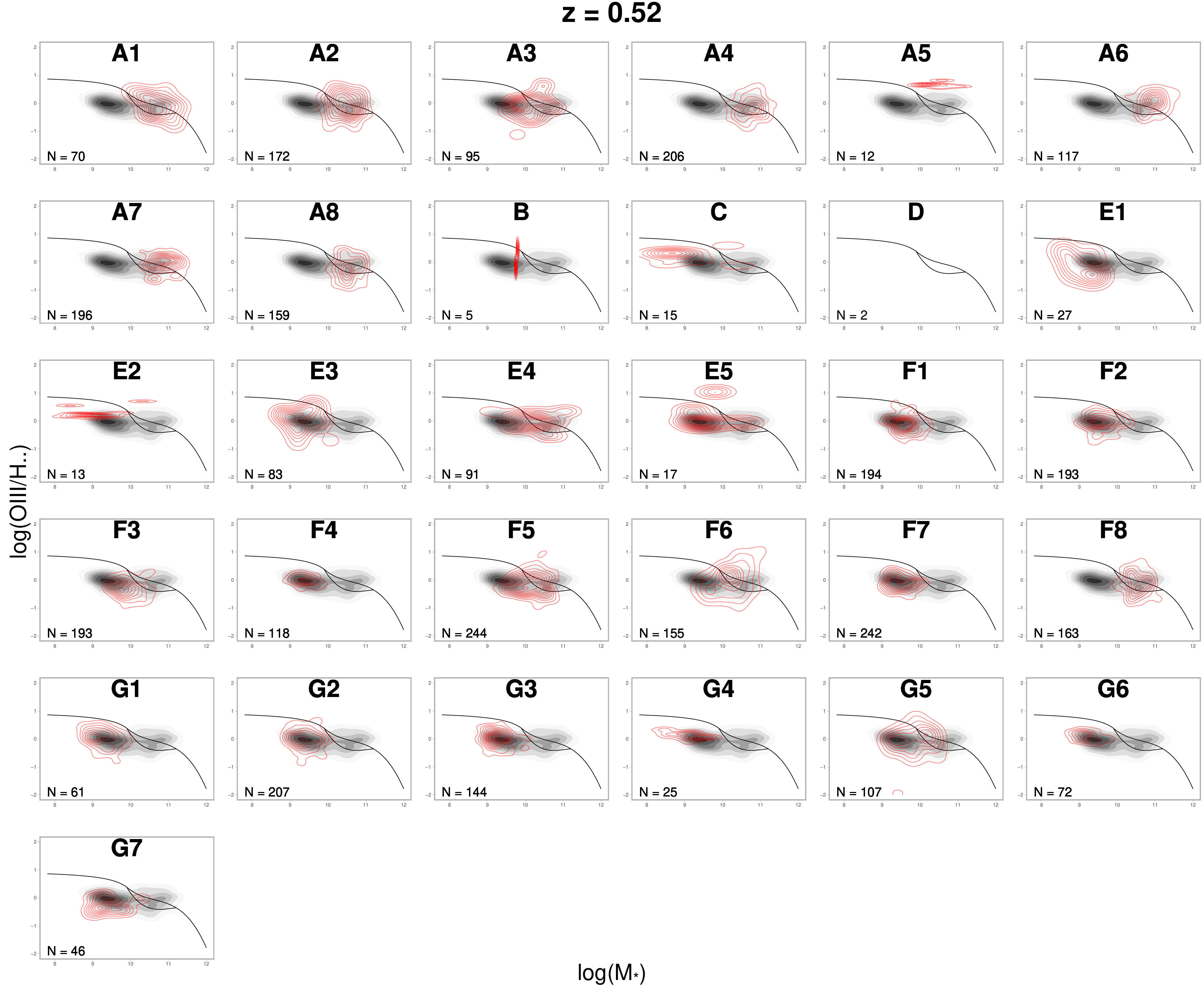}
    \caption{MEx diagram of the classes of bin 6 (see Fig.~\ref{fig:vipers_MEx_bin1} for further information)}
    \label{fig:vipers_MEx_bin6}
\end{figure}

\begin{figure}
    \centering
    \includegraphics[width=\hsize]{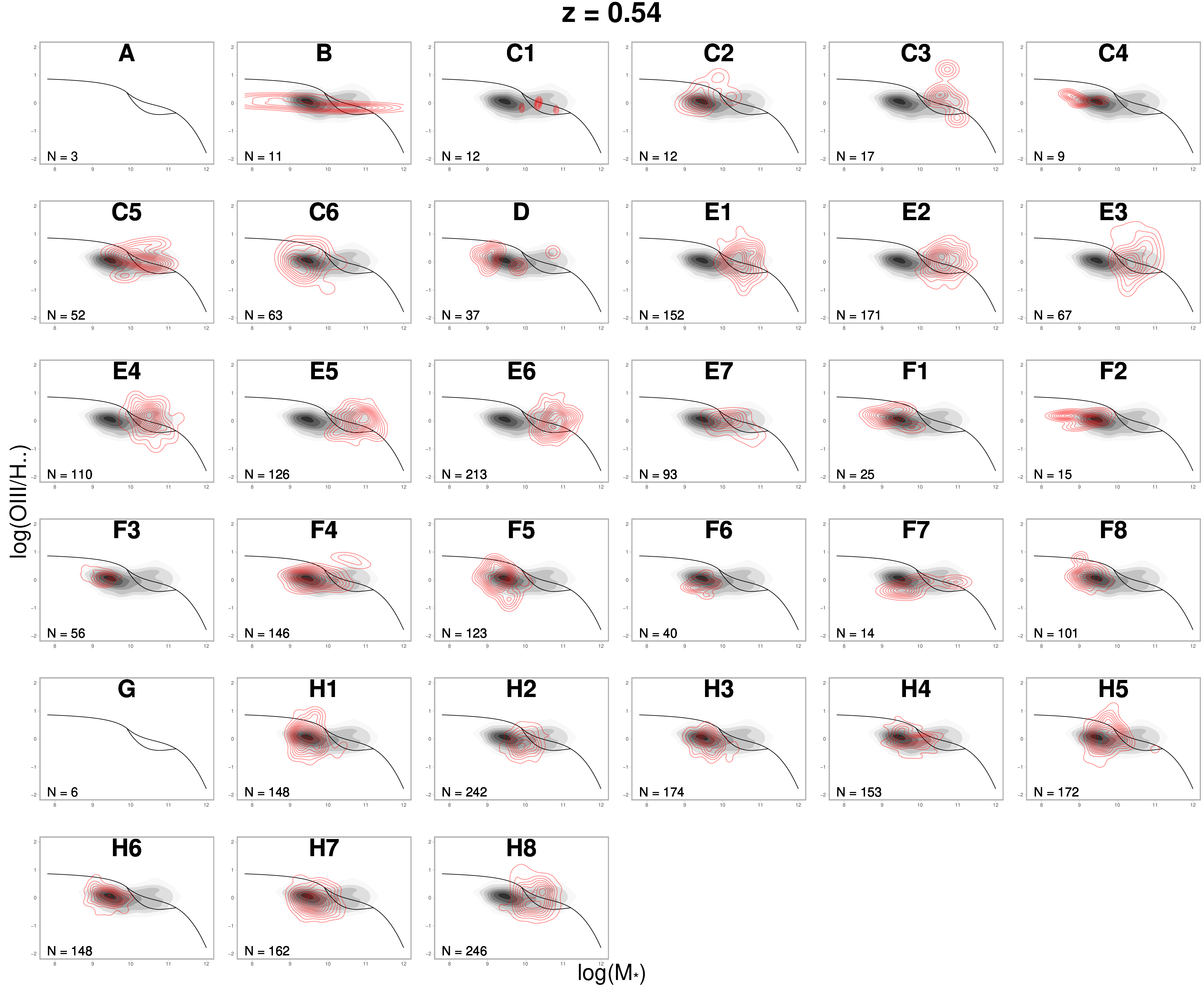}
    \caption{MEx diagram of the classes of bin 7 (see Fig.~\ref{fig:vipers_MEx_bin1} for further information)}
    \label{fig:vipers_MEx_bin7}
\end{figure}

\begin{figure}
    \centering
    \includegraphics[width=\hsize]{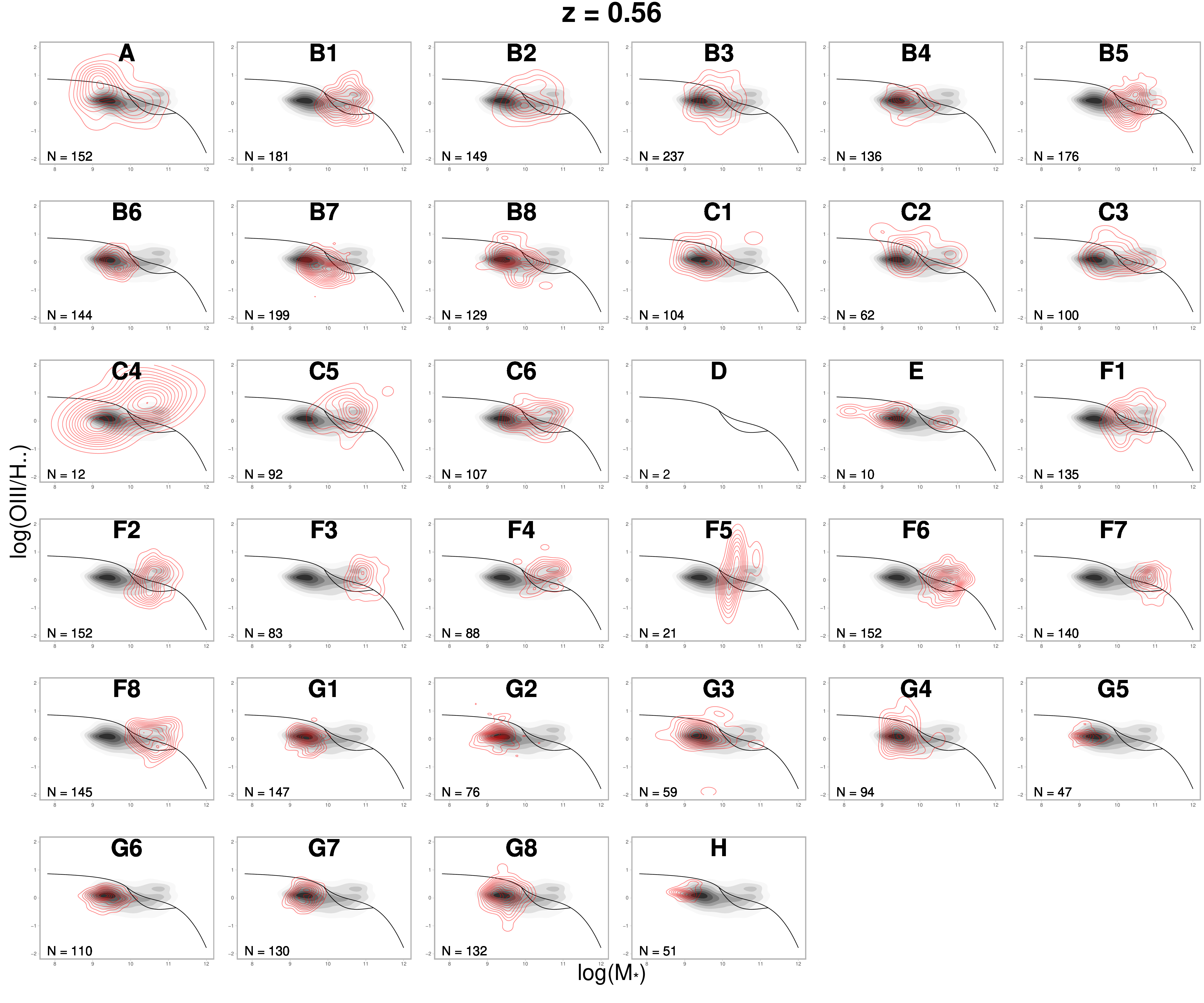}
    \caption{MEx diagram of the classes of bin 8 (see Fig.~\ref{fig:vipers_MEx_bin1} for further information)}
    \label{fig:vipers_MEx_bin8}
\end{figure}

\begin{figure}
    \centering
    \includegraphics[width=\hsize]{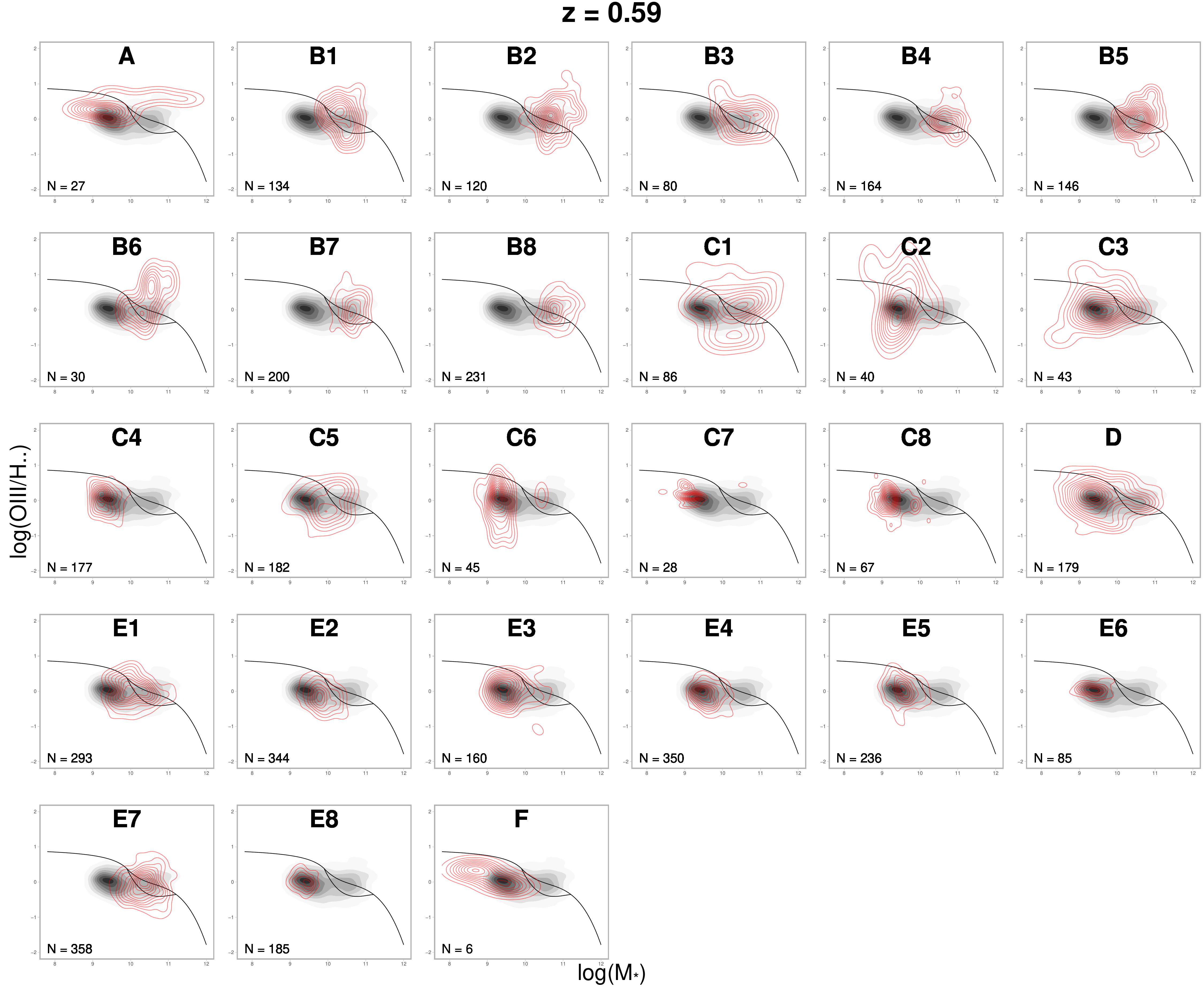}
    \caption{MEx diagram of the classes of bin 9 (see Fig.~\ref{fig:vipers_MEx_bin1} for further information)}
    \label{fig:vipers_MEx_bin9}
\end{figure}

\begin{figure}
    \centering
    \includegraphics[width=\hsize]{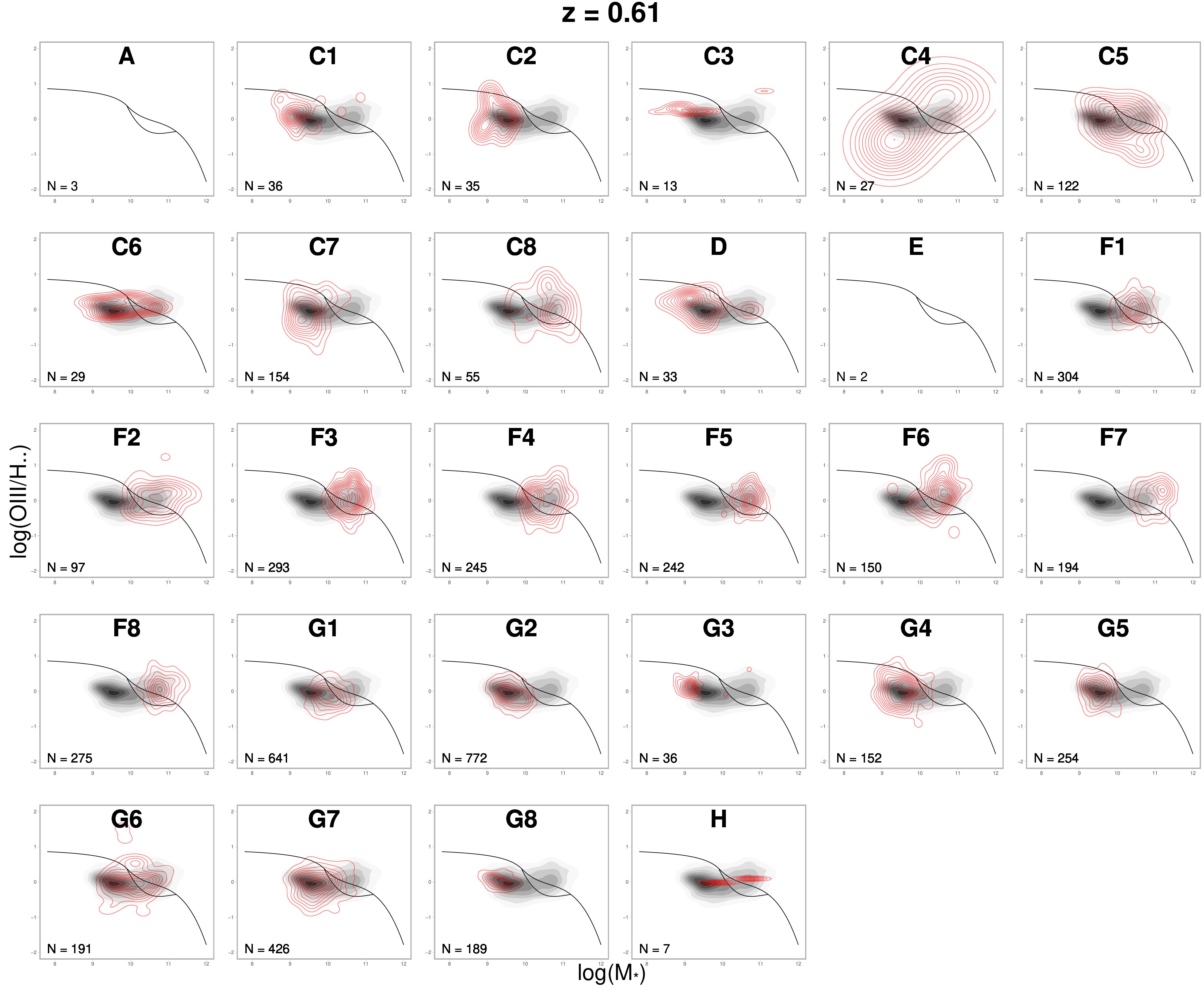}
    \caption{MEx diagram of the classes of bin 10 (see Fig.~\ref{fig:vipers_MEx_bin1} for further information)}
    \label{fig:vipers_MEx_bin10}
\end{figure}

\begin{figure}
    \centering
    \includegraphics[width=\hsize]{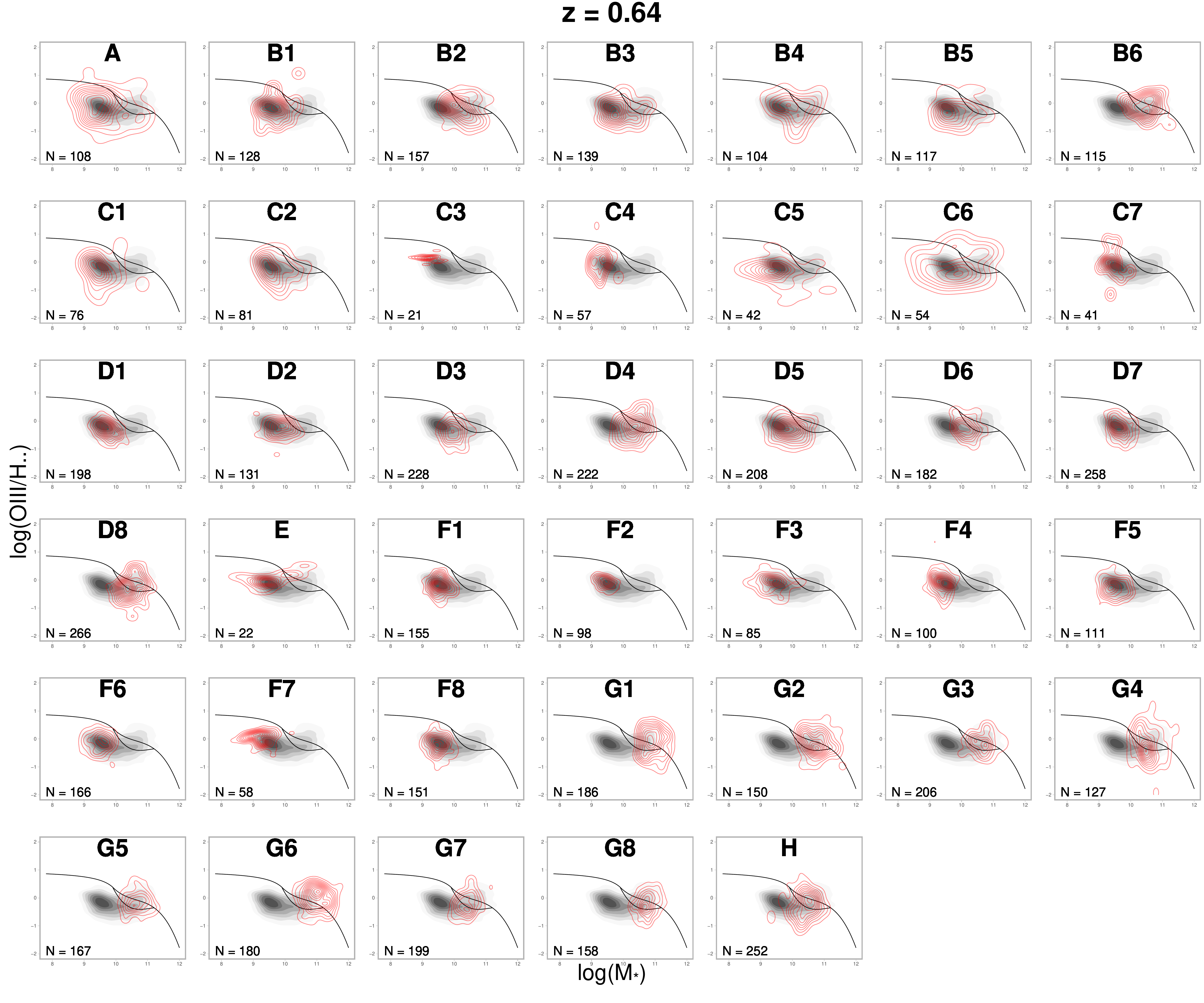}
    \caption{MEx diagram of the classes of bin 11 (see Fig.~\ref{fig:vipers_MEx_bin1} for further information)}
    \label{fig:vipers_MEx_bin11}
\end{figure}

\begin{figure}
    \centering
    \includegraphics[width=\hsize]{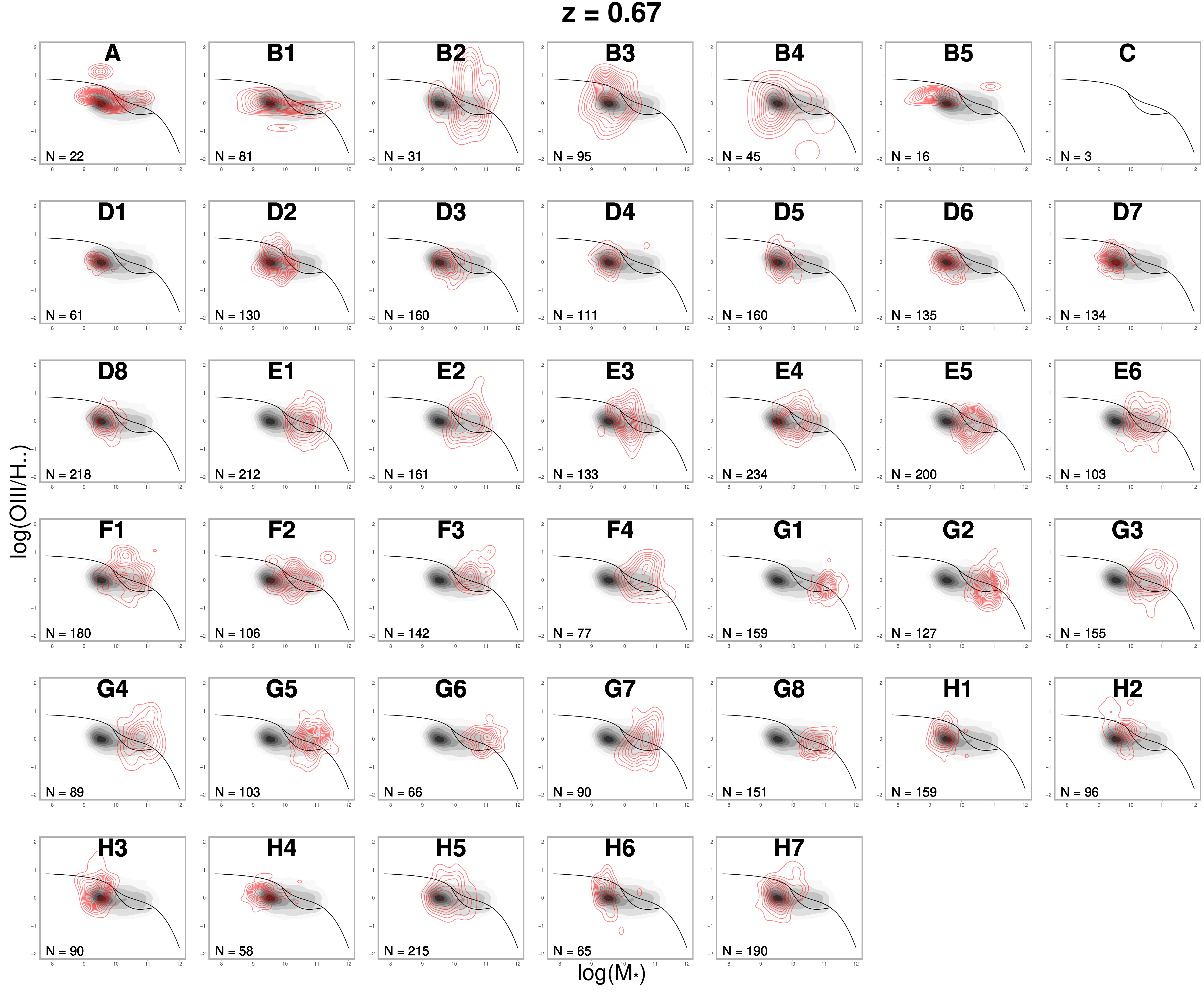}
    \caption{MEx diagram of the classes of bin 12 (see Fig.~\ref{fig:vipers_MEx_bin1} for further information)}
    \label{fig:vipers_MEx_bin12}
\end{figure}

\begin{figure}
    \centering
    \includegraphics[width=\hsize]{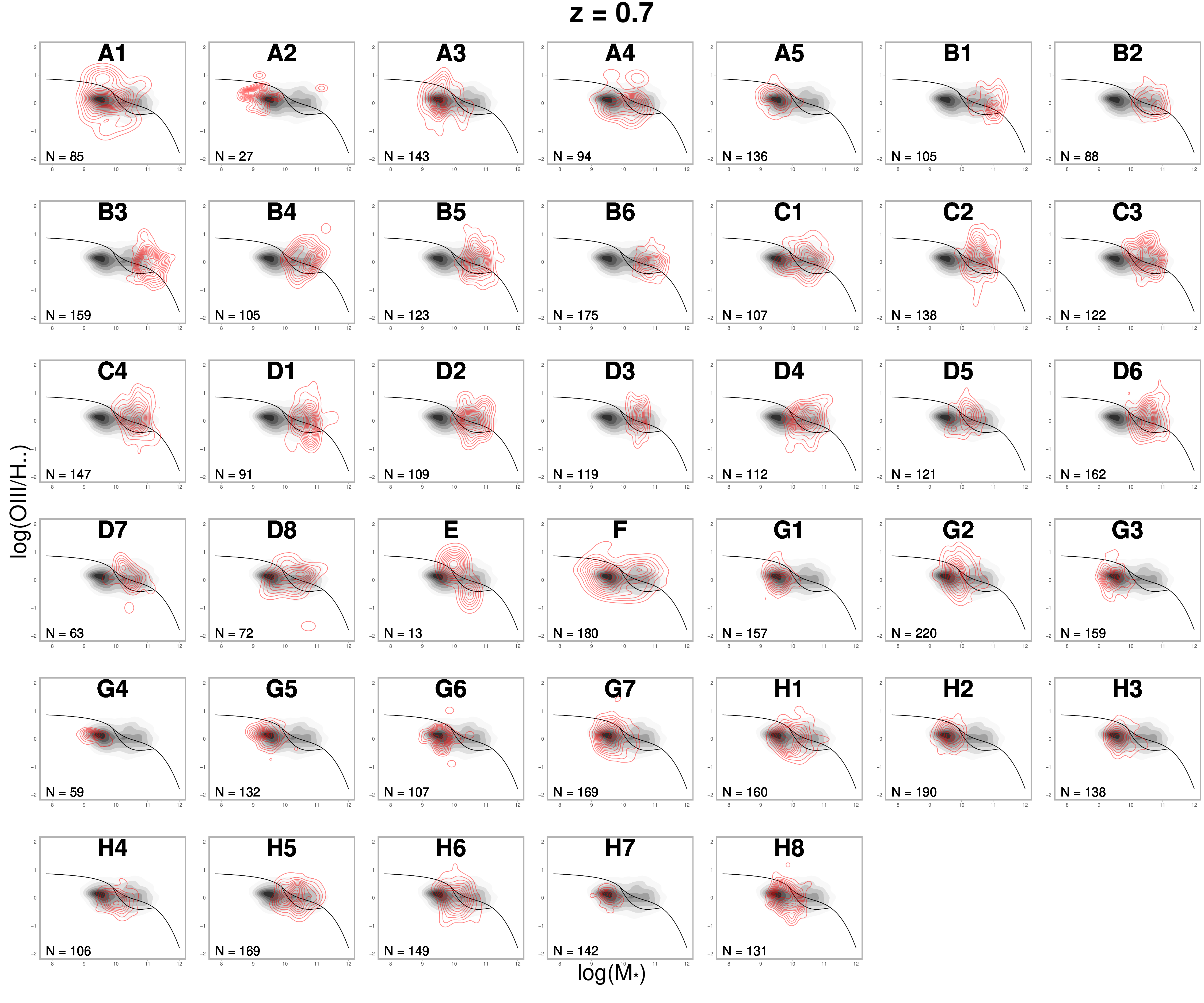}
    \caption{MEx diagram of the classes of bin 13 (see Fig.~\ref{fig:vipers_MEx_bin1} for further information)}
    \label{fig:vipers_MEx_bin13}
\end{figure}

\begin{figure}
    \centering
    \includegraphics[width=\hsize]{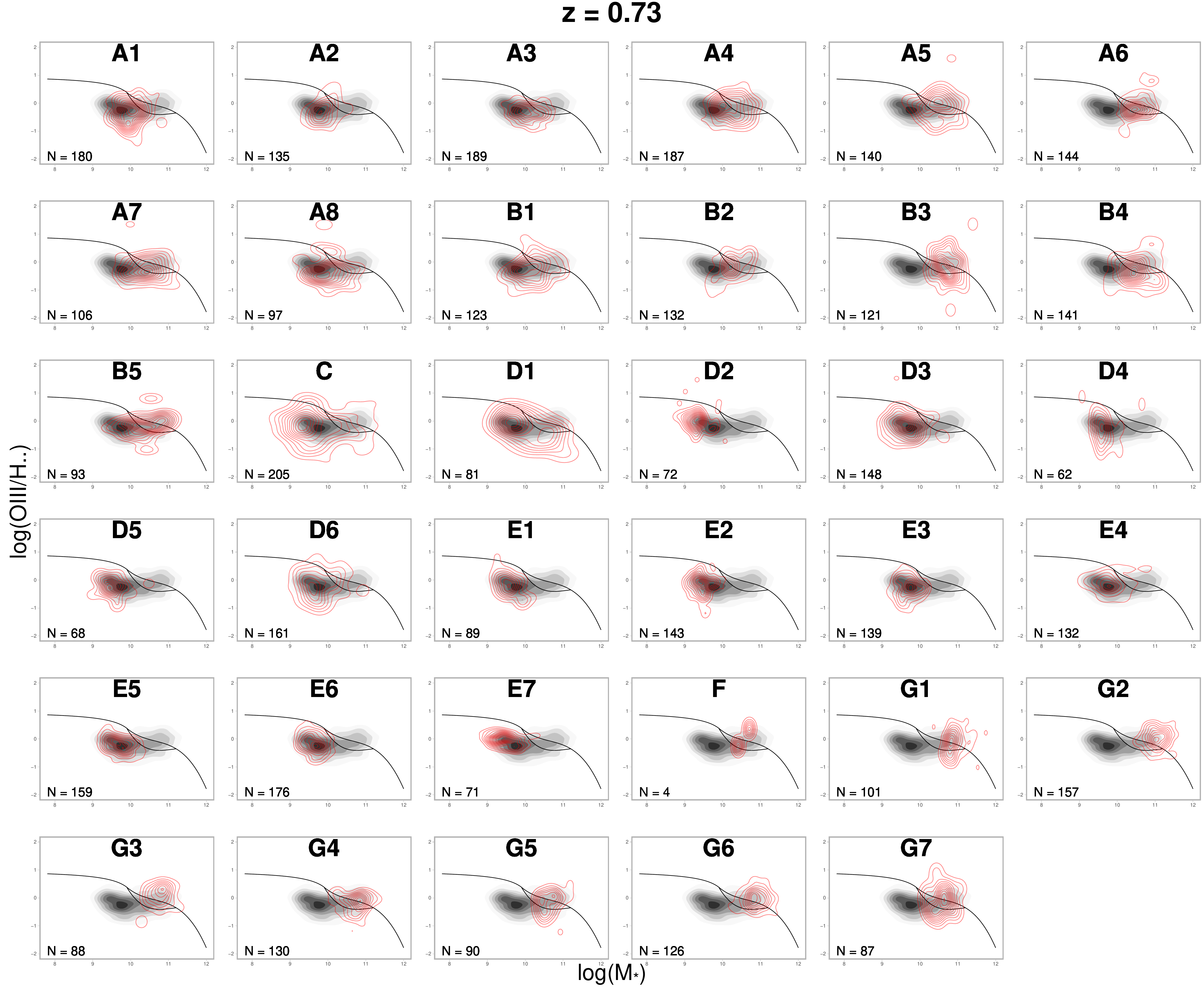}
    \caption{MEx diagram of the classes of bin 14 (see Fig.~\ref{fig:vipers_MEx_bin1} for further information)}
    \label{fig:vipers_MEx_bin14}
\end{figure}

\begin{figure}
    \centering
    \includegraphics[width=\hsize]{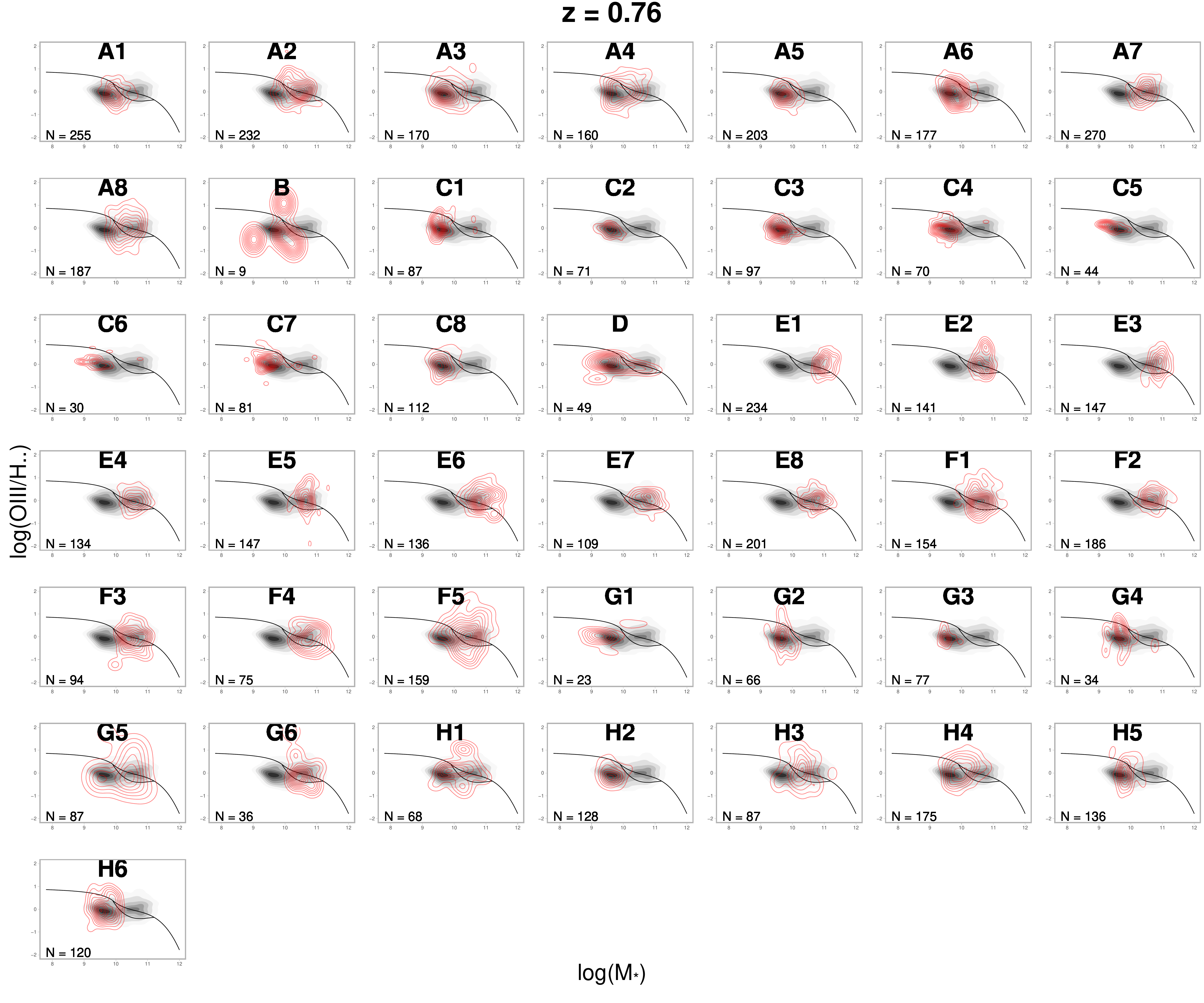}
    \caption{MEx diagram of the classes of bin 15 (see Fig.~\ref{fig:vipers_MEx_bin1} for further information)}
    \label{fig:vipers_MEx_bin15}
\end{figure}

\begin{figure}
    \centering
    \includegraphics[width=\hsize]{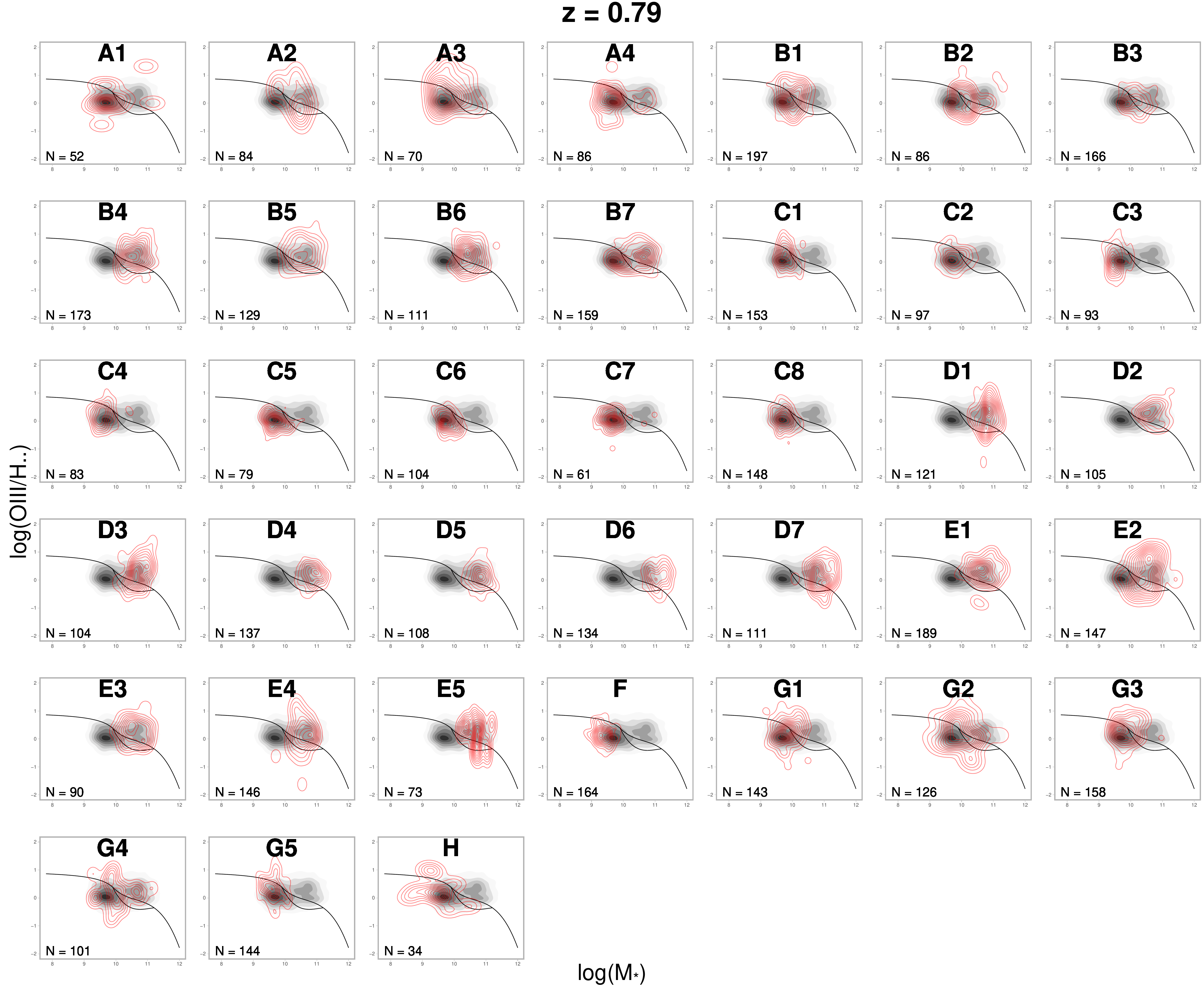}
    \caption{MEx diagram of the classes of bin 16 (see Fig.~\ref{fig:vipers_MEx_bin1} for further information)}
    \label{fig:vipers_MEx_bin16}
\end{figure}

\begin{figure}
    \centering
    \includegraphics[width=\hsize]{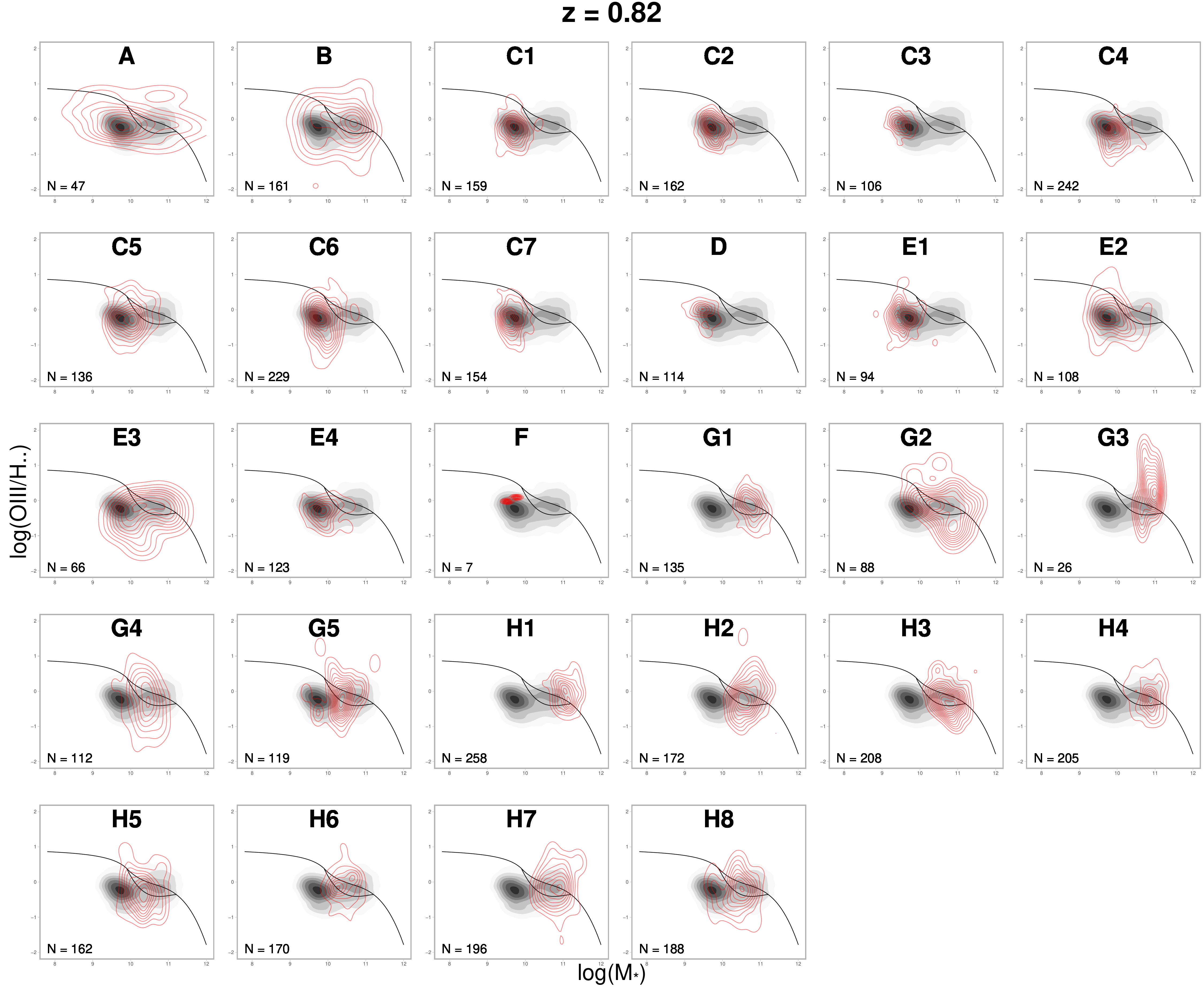}
    \caption{MEx diagram of the classes of bin 17 (see Fig.~\ref{fig:vipers_MEx_bin1} for further information)}
    \label{fig:vipers_MEx_bin17}
\end{figure}

\begin{figure}
    \centering
    \includegraphics[width=\hsize]{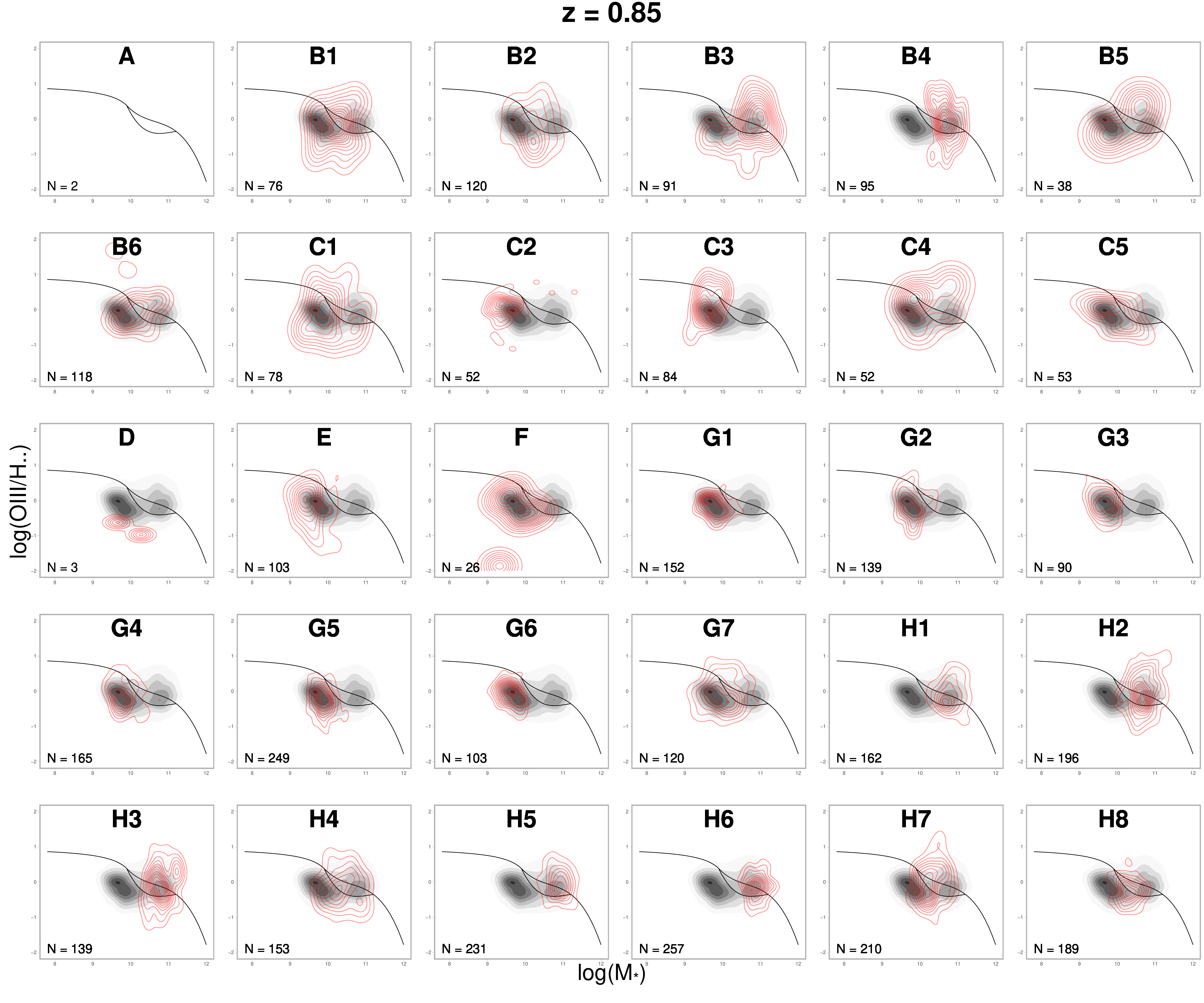}
    \caption{MEx diagram of the classes of bin 18 (see Fig.~\ref{fig:vipers_MEx_bin1} for further information)}
    \label{fig:vipers_MEx_bin18}
\end{figure}

\begin{figure}
    \centering
    \includegraphics[width=\hsize]{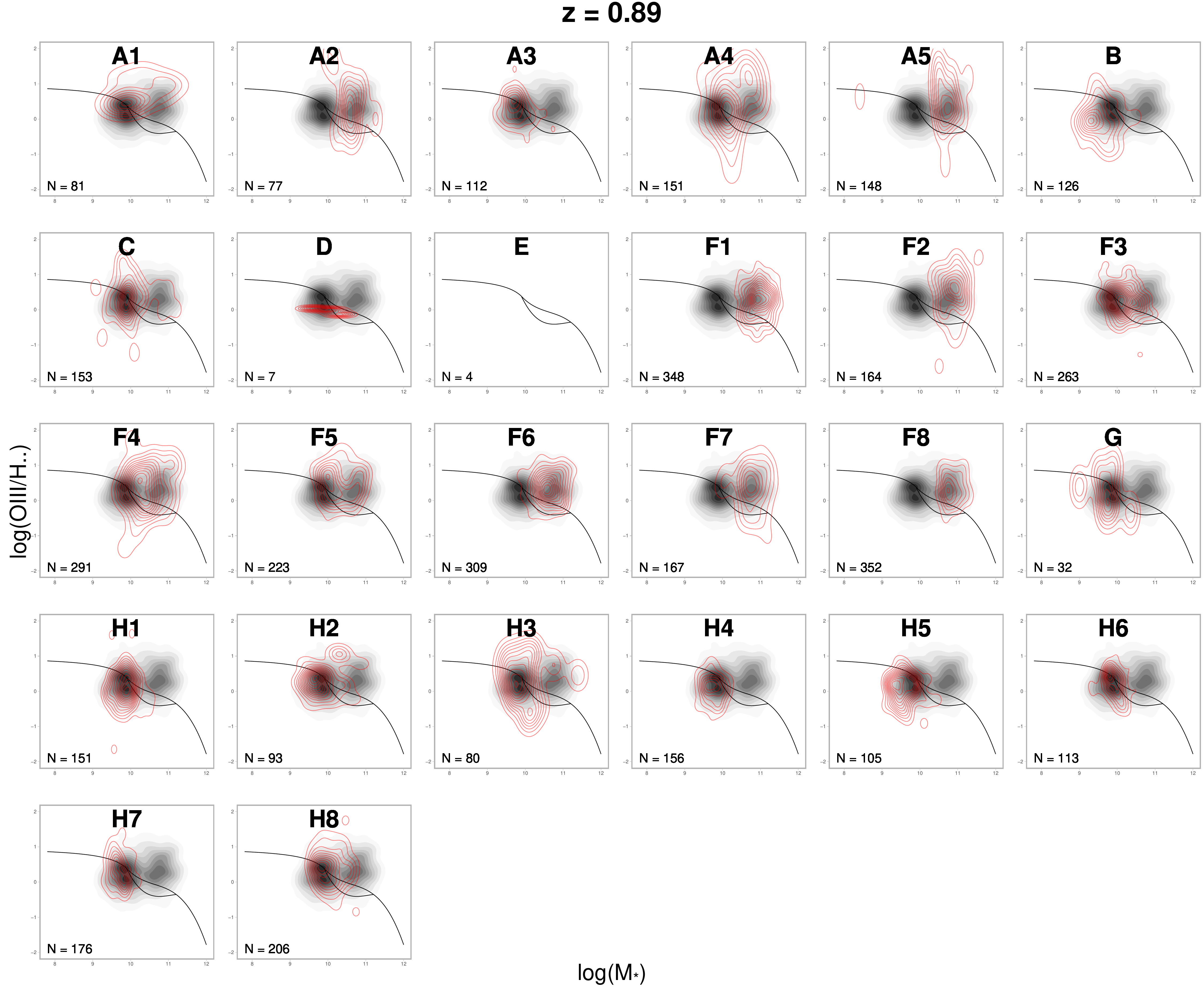}
    \caption{MEx diagram of the classes of bin 19 (see Fig.~\ref{fig:vipers_MEx_bin1} for further information)}
    \label{fig:vipers_MEx_bin20}
\end{figure}

\clearpage

    \section{Reverse tree}
\label{appendix:reverse}
\newpage

\begin{figure*}
    \centering
    \includegraphics[width=\linewidth]{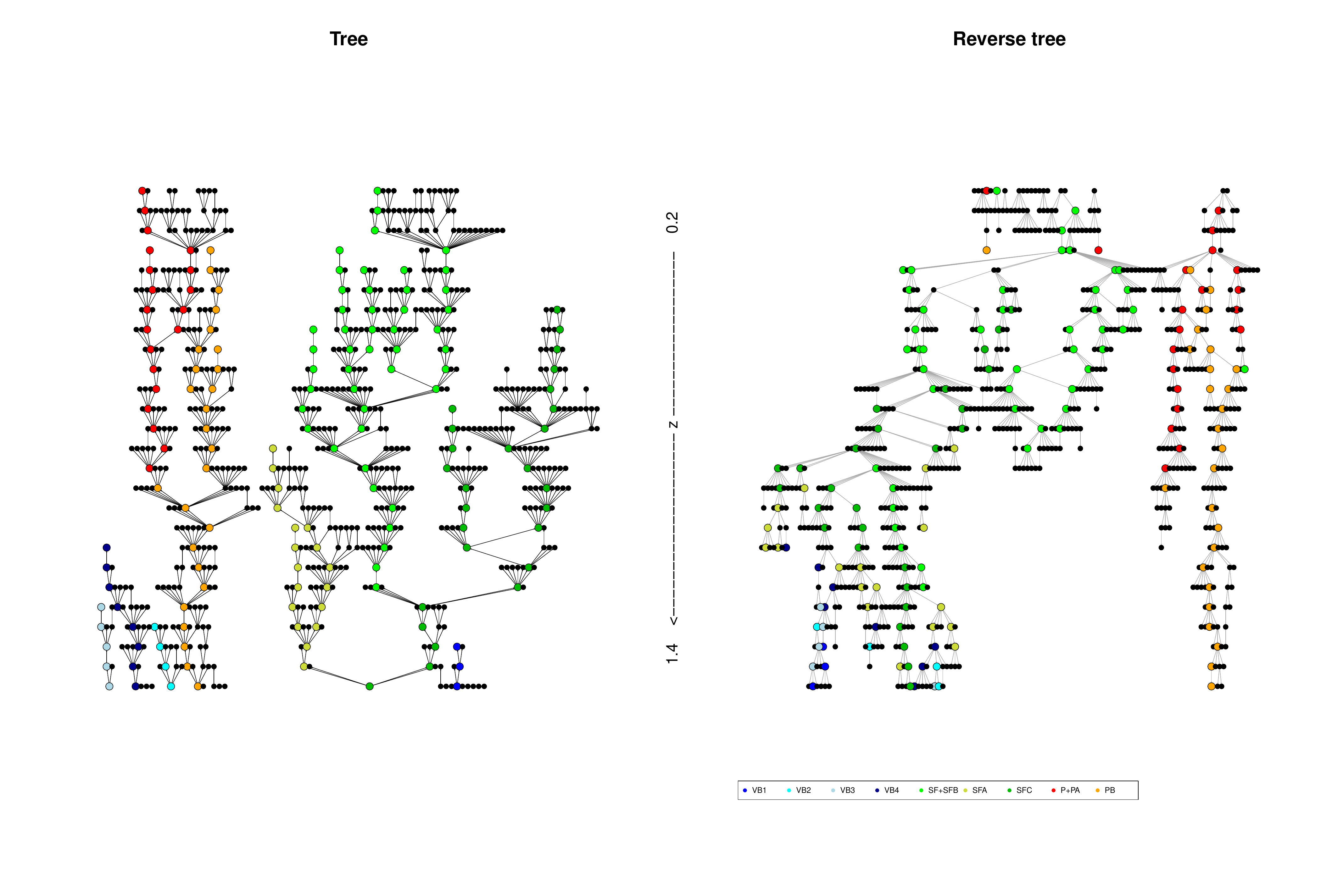}
    \caption{\textit{Left}: same tree as in  Fig.~\ref{fig:tree} with same example branches identified by colours of nodes. \textit{Right}: Reverse tree obtained by starting from lowest redshift (see Sect.~\ref{section:limitations}) with same branches as to the left.}
    \label{fig:reversetree}
\end{figure*}

\end{appendix}

\end{document}